\documentclass[aip,jmp,reprint,amssymb,amsmath,amsfonts,groupedaddress,onecolumn]{revtex4-1} 

\usepackage{titlesec}

\providecommand{\abs}[1]{\lvert#1\rvert}
\providecommand{\ket}[1]{\lvert#1\rangle}
\providecommand{\bra}[1]{\langle#1\rvert}

\usepackage{tabularx}
\usepackage{xspace}

\newcommand{\SU}{\mathrm{SU}}

\newcommand{\R}{\mathbb R}
\newcommand{\C}{\mathbb C}
\newcommand{\idperm}{\varepsilon}
\newcommand{\tr}{\mathrm{Tr}}
\renewcommand{\vec}[1]{\mathrm{vec}(#1)}
\usepackage{adjustbox}

\usepackage{graphicx} 
\usepackage{float}
\usepackage{bbold}
\usepackage{bm}
\usepackage{amsthm}
\usepackage{amsmath}

\usepackage[bookmarks=false,pdfstartview={FitH}]{hyperref}
\usepackage{lmodern}

\usepackage{afterpage}

\usepackage{multirow}
\usepackage{bigdelim}

\usepackage{ytableau}

\newcommand{\TTab}[1]{\ytableaushort{#1}}

\begin{document}


\title{Symmetry-adapted decomposition of tensor operators and the visualization of coupled spin systems} 



\author{David Leiner}
\email[]{david.leiner@tum.de}
\author{Robert Zeier}
\email[]{zeier@tum.de}
\author{Steffen J. Glaser}
\email[]{glaser@tum.de}
\affiliation{Technical University of Munich, Department of Chemistry,\\ 
Lichtenbergstrasse 4, 85747 Garching, Germany}
\date{24 September 2018}


\begin{abstract}
We study the representation and visualization of finite-dimensional quantum systems. 
In a generalized Wigner representation, multi-spin operators can be decomposed into 
a symmetry-adapted tensor basis and they are
mapped to multiple spherical plots that are each assembled from linear combinations of spherical harmonics.
We apply two different approaches based on explicit projection operators
and coefficients of fractional parentage in order to obtain this basis 
for up to six spins $1/2$ (qubits), for which various examples are presented.
An extension
to two coupled spins with arbitrary spin numbers (qudits) is provided, also
highlighting a quantum system of a spin~$1/2$ coupled to a spin~$1$ (qutrit).
\end{abstract}


\maketitle 



\ytableausetup{smalltableaux,centertableaux}

\section{Introduction}
Quantum systems exhibit an intricate structure
and numerous methods have been established for the visualization of their quantum state.
A two-level quantum system, such as a single spin $1/2$ (qubit), 
can always be faithfully represented by a three-dimensional vector (Bloch vector),
as shown in the seminal work of Feynman et al.\cite{Feynman_Vernon_57}
Applications of the Bloch vector are frequently found in the field of quantum physics, in particular in
magnetic resonance imaging,\cite{handbook_04,EBW87} spectroscopy,\cite{EBW87} and quantum optics.\cite{SchleichBook}
However, for systems consisting of coupled spins,  standard Bloch vectors can only 
partially represent
the density matrix, whereas important terms, such as  multiple-quantum coherence\cite{EBW87}
or spin alignment,\cite{EBW87} are not captured.
In this case, the complete density operator can be visualized 
by bar charts, in which the real and imaginary parts of each element of the density matrix is represented
a vertical bar, an approach which is commonly used to graphically display the experimental results of quantum 
state tomography.\cite{NC00}
Alternatively, energy-level diagrams can 
can be illustrate populations
by circles on energy levels and coherences 
by lines between energy levels.\cite{SEL:1983}
Density operators can also be visualized by non-classical vector 
representations based on single-transition operators.\cite{EBW87,Donne_Gorenstein,Freeman97}
However, these techniques are inconvenient for larger spin systems and often do not provide an intuitive view of the spin dynamics.

Phase space representations,\cite{SchleichBook,Curtright-review,Zachos2005,Schroeck2013}
in particular Wigner functions,\cite{Wig32,SchleichBook,Curtright-review} which originally arise 
in the description of the infinite-dimensional quantum 
state of light,\cite{smithey1,smithey2,smithey3,leonhardt,paris_rehacek}
provide a powerful alternative approach for the characterization and visualization of finite-dimensional quantum systems.
One valuable class for the representation of finite-dimensional systems are discrete Wigner functions
\cite{Wooters87,leonhardt1996,Miquel,Miquel2,gibbons2004,FE09} but we will focus on continuous 
representations, which naturally reflect the inherent rotational symmetries of spins.
General criteria for defining continuous Wigner functions for finite-dimensional quantum systems
had been established in the work by Stratonovich\cite{Stratonovich}
and the case of single-spin systems has been studied in the 
literature.\cite{Agarwal81,VGB89,Brif98,Brif97,heiss2000discrete,klimov2002ExactEvolution,klimov2005classical,StarProd}
Extensions to multiple spins have been considered in 
Ref.~\onlinecite{SchleichBook,DowlingAgarwalSchleich,JHKS,PhilpKuchel,Harland},
but a general strategy for multiple coupled spins was still missing.\cite{PhilpKuchel,Harland}
Recently, Garon et al.\cite{Garon15} identified such a general strategy.
Subsequently, further approaches to phase-space representations 
have been developed,\cite{tilma2016,koczor2016,rundle2017,RTD17,koczor2017,koczor2018}
while rotated parity operators\cite{heiss2000discrete,tilma2016,rundle2017,RTD17,koczor2017,koczor2018}
and tomographic techniques\cite{rundle2017,davidtomo,koczor2017,LG18}
became further focal points.

We build in this work on the general Wigner representation for multiple coupled spins
introduced in Ref.~\onlinecite{Garon15}.
This Wigner representation is denoted as DROPS representation (discrete representation of
operators for spin systems).
It is based on mapping operators  to a finite set of spherical plots, which  
are each assembled from linear combinations of spherical harmonics\cite{Jac99} 
and which are denoted as {\it droplets} or {\it droplet functions}.\cite{Garon15}
These characteristic droplets preserve crucial symmetries of the quantum system. 
One particular version of this representation relies on a specific choice of a tensor-operator basis, 
the so-called LISA basis,\cite{Garon15} which characterizes tensors according to
their linearity, their set of involved spins, their permutation symmetries with respect to
spin permutations, and their rotation symmetries under rotations that operate uniformly on each spin.
These symmetry-adapted tensors can be constructed 
using explicit projection operators given as 
elements of the group ring of the 
symmetric group.\cite{James78,JK81,CSST10,Boerner67,Hamermesh62,Sagan01,tung1985group,Garon15}
We apply this approach to a larger number of coupled spins $1/2$ (qubits)
and also to two-spin systems with arbitrary spin numbers (qudits).
In addition,
we implement 
a second, alternative computational methodology 
that relies on 
so-called coefficients of fractional 
parentage (CFP)\cite{Racah65,EL57,Kaplan75,Silver76,Chisholm76,KJS81,JvW51}
in order to obtain 
the symmetry-adapted LISA basis.

Our contribution can also be put into a general context of 
symmetry-adapted decompositions of tensor operators. 
Symmetry-adapted (tensor) bases have a very long tradition in physics.
Important mathematical contributions were made by Weyl\cite{Wey27,Weyl31,Weyl50,Weyl46}
and Wigner,\cite{Wigner31,Wigner59} even though the corresponding group theory
was (at least in the beginning) not universally embraced in the physics community
(see p.~10-11 in Ref.~\onlinecite{CondonShortley}). Building on Ref.~\onlinecite{CondonShortley},
Racah\cite{Racah41,Racah42,Racah43,Racah49,Racah65,Fano59} developed
tensor-operator methods for the analysis of electron spectra.
These tensor methods have been widely studied\cite{edmonds1960,Griffith06,Judd98,Silver76}
and initiated an active exchange between group theory and 
physics.\cite{Weyl50,Wigner59,Hamermesh62,Boerner67,Miller72,tung1985group,Ludwig96}
Moreover, tensor operators (as well as coefficients of fractional parentage)
play an important role in applications to atomic and nuclear structure
for which an expansive literature exsists.\cite{EL57,Slater1960,ShalitTalmi63,Pauncz67,Wybourne70,Kaplan75,
Chisholm76,ED79,Condon1980,Rudzikas97,CH98,RW10}
In this context, we also mention the work of Listerud et al.\cite{Listerud_thesis,LGD93}
which partly motivated the approach taken in Ref.~\onlinecite{Garon15} and this work.

This paper is structured as follows. In Sec.~\ref{sec:Vis_summary}, we introduce
the symmetry-adapted tensor basis and its mapping to Wigner functions.
An overview of the construction process of this tensor basis using either 
explicit projection operators
or fractional parentage coefficients
is presented in Sec.~\ref{sec:sum_method}.
In Sec.~\ref{sec:mulitple_1_2}, the tensor-operator basis  is illustrated 
for up to six coupled spins $1/2$ by
examples and applications from quantum information and nuclear magnetic
resonance spectroscopy. Coupled two-spin systems with
arbitrary spin numbers are treated in Sec.~\ref{sec:arbitrary}.
The explicit construction of the tensor-operator basis is detailed
in Sec.~\ref{sec:construciondetails}.
Before we conclude,
challenges related to
the construction method that relies on explicit projection operators
are discussed 
in Sec.~\ref{sec:discussion}.
Additional illustrative examples for spins $1/2$ are presented in 
Appendix~\ref{sec:further_viz} and Appendix~\ref{CFP_app}
lists the employed values of the fractional 
parentage coefficients.

\section{Symmetry-adapted decomposition and visualization of operators of coupled spin systems\label{sec:Vis_summary}}

\addtocounter{footnote}{1}
\footnotetext[\value{footnote}]{Spherical harmonics
$Y_{jm}(\theta,\phi)=r(\theta,\phi) \exp[i\eta(\theta,\phi)]$
(and droplet functions) are plotted 
throughout this work by mapping their spherical coordinates $\theta$ and $\phi$ to the radial 
part $r(\theta,\phi)$ and phase $\eta(\theta,\phi)$.}
\newcounter{spherical}
\setcounter{spherical}{\value{footnote}}

We summarize the approach of Ref.~\onlinecite{Garon15} (see also Refs.~\onlinecite{davidtomo,LG18})
to visualize operators of coupled spin systems using multiple droplet functions which are chosen
according to a suitable symmetry-adapted decomposition of the tensor-operator space. 
This allows us to also fix the setting and notation for this work.
The general idea relies on mapping\cite{Silver76,CH98} 
components $T_{jm}^{(\ell)}$ of irreducible tensor operators\cite{Wigner31,Wigner59,Racah42,BL81}
$T_{j}^{(\ell)}$ to 
spherical harmonics\cite{Jac99,Note\thespherical} $Y_{jm}=Y_{jm}(\theta,\phi)$.
An arbitrary operator $A$ in a coupled spin system can be expanded into linear combinations
\begin{equation}
\label{A}
A = \sum_{\ell} A^{(\ell)} 
=\sum_{\ell}\sum_{j \in \mathcal{J}(\ell)}\sum_{m=-j}^{j} c_{jm}^{(\ell)} T_{jm}^{(\ell)}
\end{equation}
of tensor components $T_{jm}^{(\ell)}$
according to rank $j$ and order $m$ with $-j\leq m \leq j$
and suitably chosen labels (or quantum numbers) 
$\ell$, such that the set $\mathcal{J}(\ell)$ of ranks $j$ occurring for each label 
$\ell$ does not contain any rank twice. Depending on the chosen labels, certain
properties and symmetries of the spin system are emphasized.
Each component $A^{(\ell)}$ is now bijectively mapped to a droplet function
$f^{(\ell)}=f^{(\ell)}(\theta,\phi)$, which can be 
decomposed into 
\begin{equation}
\label{spherical_A}
f^{(\ell)}  
= \sum_{j \in \mathcal{J}(\ell)} \sum_{m=-j}^{j} c_{jm}^{(\ell)} Y_{jm},
\end{equation}
where the coefficients $c_{jm}^{(\ell)}$ in Eqs.~\eqref{A} and \eqref{spherical_A} are identical.
This approach enables us to represent each operator component 
$A^{(\ell)}$ by
a droplet function
$f^{(\ell)}$, which is given by its expansion into spherical harmonics, refer to the example on
the r.h.s of Table~\ref{tab:labels_partition}.
The droplet functions
$f^{(\ell)}$ are denoted as {\it droplets} and the set of all droplets form the full DROPS representation of an arbitrary operator $A$.

\begin{table}
\caption{Overview of how irreducible tensor operators $T_{j}^{(\ell)}$ 
with components $T_{jm}^{(\ell)}$
are partitioned
in the LISA basis
according to their label $\ell$ and rank $j$ for the prototypical case of six spins $1/2$ (left).
For a generic operator with randomly chosen complex matrix elements,
the droplet functions $f^{(\ell)}$
are illustrated
separately for each label $\ell$ (right). For all droplet functions, the maximum radii are normalized to one for better visibility.
Each label $\ell$ consists of a number of sublabels:
the cardinality $g$ of the set of involved spins (i.e.\ the $g$-linearity) and the explicit set $G$, 
the symmetry type given by a standard Young tableau $\tau_i^{[g]}$ of size $g$
and, possibly, an {\it ad hoc} label given by a roman numeral.
{\it Ad hoc}  sublabels are necessary for $g=6$ as otherwise one could not distinguish, for instance,
between the doubly occurring rank $2$
(in bold) for the symmetry type $\tau_7^{[6]}$.
The structure of the partitioning is illustrated  on the right for the zero-linear term (Id) and selected linear, 
bilinear, trilinear, and 6-linear components.
Plots for all possible droplet functions are shown in Table~\ref{tab:4spin_1d2} for a system consisting of four spins and in 
 Figs.~\ref{fig:6spin_1d2} and \ref{fig:6spin_1d2_B}
for six spins. \label{tab:labels_partition}}
\includegraphics{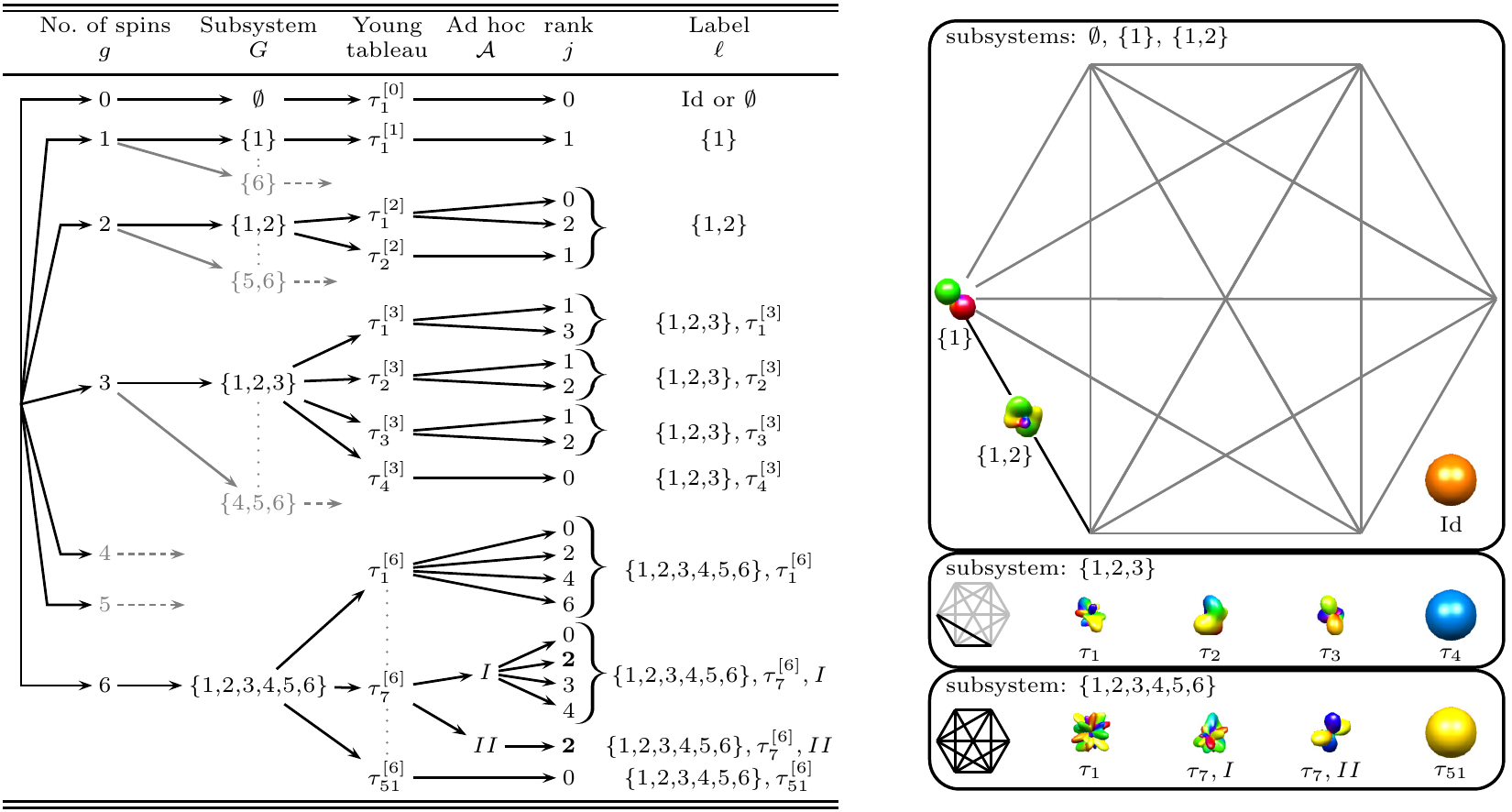}
\end{table}

The task to find suitable labels $\ell$ that allow for a complete 
decomposition of the tensor-operator space according to Eq.~\eqref{A} has been widely studied\cite{JMPW74,Sharp75,IL95,RW10} and is related
 to the search for a
complete set of mutually commuting operators or good quantum numbers.\cite{Merzbacher98}
Different possibilities have been discussed in Ref.~\onlinecite{Garon15}, but here we will focus on the LISA basis,\cite{Garon15} whose labeling scheme is outlined in
Tab.~\ref{tab:labels_partition}. 
First, tensor basis operators are subdivided with respect 
to the cardinality $g \in \{0,1,\dots,N\}$ of the set of involved spins (i.e.\ their $g$-linearity), where
$N$ denotes the total number of spins. Second, tensor operators with identical $g$-linearity are 
further partitioned according to the explicit set $G \in \binom{\{1,2,\dots,N\}}{g}$ of involved spins,  
where $\binom{\{1,2,\dots,N\}}{g}$ denotes the set of all subsets of $\{1,2,\dots,N\}$ with cardinality $\abs{G}=g$.
For example for $g=2$ and $N=4$, we obtain $G \in \{\{1,\!2\},\{1,\!3\},\{1,\!4\},\{2,\!3\},\{2,\!4\},\{3,\!4\}\}$.
Third, we further partition with respect to the symmetry type 
given by a standard Young tableau\cite{Boerner67,Hamermesh62,Pauncz95,Sagan01}  $\tau_i^{[g]}$ of size $g$
(and with at most $(2J{+}1)^2{-}1=4J(J{+}1)$ rows, depending on the spin number $J$),
which results in a decomposition according to symmetries under permutations of the
set $G$. For reference, all potentially occurring symmetry types $\tau_i^{[g]}$ for $g \in \{1,2,3,4,5,6\}$
are uniquely enumerated and specified according to their index $i$ in Tables~\ref{tab:4spin_1d2} and \ref{tab:tableaux}.
For $g=3$ and $G=\{1,\!2,\!3\}$ 
we have the symmetry types
\begin{equation}\label{taus_3}
\tau_{1}^{[3]}=\TTab{1 2 3}\, ,\; \tau_{2}^{[3]}=\TTab{1 2, 3}\, , \; \tau_{3}^{[3]}=
\TTab{1 3, 2}\, ,\; \tau_{4}^{[3]}=\TTab{1, 2, 3},
\end{equation}
and equivalent symmetry types arise for all the other sets $G \in \binom{\{1,2,\dots,N\}}{3}$ of involved spins with $\abs{G}=g=3$.
Fourth, an {\it ad hoc}  sublabel $\mathcal{A}$ given by a roman numeral is used to distinguish between cases 
if the same rank occurs more than one.\cite{FP37,JvW51} For $g=6$ and the symmetry type 
$\tau_{7}^{[6]}$, the rank of $j=2$ (as shown in bold 
on the l.h.s. of Table~\ref{tab:labels_partition})  
would occur twice if these cases would have not been distinguished by the {\it ad hoc}  sublabels $I$ and $II$.
In summary, our labeling scheme for the LISA basis is given by
$\ell:=(G,\tau^{[g]},\mathcal{A}$). We often suppress 
redundant sublabels.
As discussed in more detail in Sec.~\ref{sec:arbitrary},
for systems containing spins with spin numbers larger than $1/2$, 
the decomposition structure is considerably simplified by additional parent sublabels $\mathcal{P}$.

\section{Summary of the computational techniques used to construct the LISA basis\label{sec:sum_method}}
\begin{figure}
\includegraphics{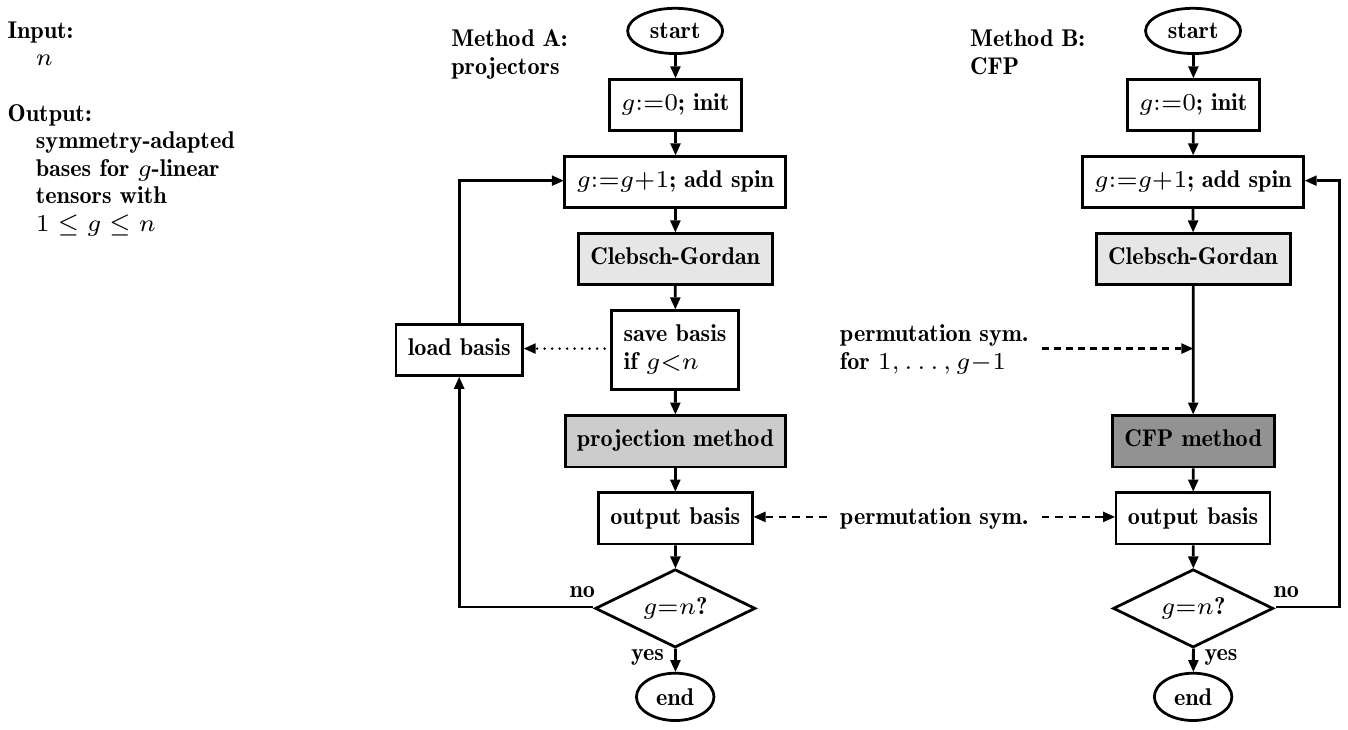}
\caption{Flow charts for methods A and B used to iteratively construct $g$-linear tensors for $g\in\{1,\ldots,n\}$.
Both methods rely first on a Clebsch-Gordan decomposition
to symmetrize tensors according to $\SU(2)$ symmetries
after adding 
an additional spin. In a second step, Method A applies projection operators
to symmetrize the tensors with respect to permutations. Method B uses 
a basis change according to fractional parentage coefficients (CFP),
in order to completely permutation symmetrize the tensors,  which 
already have been partially symmetrized with respect to
the first $g{-}1$ spins in the previous iteration. \label{fig:flow}}
\end{figure}

In this section, we provide an overview how to explicitly construct
the LISA basis, which has been introduced in Sec.~\ref{sec:Vis_summary}.
We focus on spin systems where
each spin has the same spin number $J \in \{1/2,1,3/2,\dots\}$.
The LISA basis is a symmetry-adapted basis according to 
symmetries under simultaneous $\SU(2)$ rotations of spins
as well as under spin permutations. As discussed in Sec.~\ref{sec:Vis_summary},
these symmetries of a tensor operator $T_{jm}^{(G,\tau^{[g]},\mathcal{A})}$ are specified by the rank $j$ and order $m$
as well as the symmetry type $\tau^{[g]}$. We start by discussing the simple cases of zero and one spins and
explain how to use the Clebsch-Gordan decomposition\cite{Wigner59,BL81,Zare88,PDG12} 
to symmetrize tensors according to $\SU(2)$ symmetries when a new spin is added to a spin system. 
This is the first step of the iterative construction, which is schematically illustrated in Fig.~\ref{fig:flow}. 
Depending on the spin system,  in the second step two alternative
methods (denoted A and B) are used for the symmetrizing with respect to 
spin permutations.
Method A relies on explicit projection
operators\cite{Boerner67,Hamermesh62,SancTemMNMRXIII,tung1985group,Sagan01}
and symmetrizes all $g$-linear tensors in one step. 
Method B uses a basis change according to fractional parentage 
coefficients\cite{Racah65,EL57,Kaplan75,Silver76,Chisholm76,KJS81,JvW51}
(CFP) and iteratively symmetrizes  $g$-linear tensors with respect to spin permutations, which 
in the previous iteration have already been partially symmetrized with respect to
the first $g{-}1$ spins. We close this section by discussing sign conventions
and how to embed $g$-linear tensors into larger spin systems.
Further details are deferred to Sec.~\ref{sec:construciondetails}.

For zero-linear tensors (i.e.\ $g=0$), we have the tensor operator
$T^{[0]}_{0}$ with the single component $T^{[0]}_{00}$. 
We use the notation $T^{[g]}_{j}$ for general $g$-linear tensors of rank $j$, but we will often drop the index $[g]$.
For the spin number $J=1/2$,
we in particular obtain\cite{EBW87} 
\begin{equation}
\label{eq:T00}
T^{[0]}_{00} = T_{00} = 
\frac{1}{\sqrt{2}}\begin{pmatrix}
1 & 0 \\
0 & 1
\end{pmatrix}.
\end{equation}
For linear tensors and spin number $J=1/2$ (i.e.\ qubits), we have the three components\cite{EBW87}
\begin{equation}\label{tensor_one}
T^{[1]}_{1,-1}   = T_{1,-1} = \begin{pmatrix}
0 & 0 \\
1 & 0
\end{pmatrix}, \;
T^{[1]}_{10}   = T_{10} = \frac{1}{\sqrt{2}}\begin{pmatrix}
1 & 0 \\
0 & -1
\end{pmatrix}, \;
T^{[1]}_{11}   = T_{11} = \begin{pmatrix}
0 & -1 \\
0 & 0
\end{pmatrix}
\end{equation}
of the tensor operator $T^{[1]}_{1}$.
For a general spin number $J$ (i.e.\ qudits), 
all tensor operators ${}^JT^{[1]}_{j}={}^JT_j$ with $j\in\{1,\ldots,2J\}$ are present.
Their tensor operator components ${}^JT_{jm}$ with $m\in\{-j,\ldots,j\}$
are given as (see, e.g., Refs.~\onlinecite{Fano53,BL81,Brif98})
\begin{equation}\label{general_tensor}
\left[ {}^JT_{jm} \right]_{m_1m_2}
=\sqrt{\tfrac{2j{+}1}{2J{+}1}} C^{Jm_1}_{Jm_2jm}
=(-1)^{J-m_2} C^{jm}_{Jm_1J,-m_2}
\end{equation}
in terms of Clebsch-Gordan coefficients\cite{Wigner59,BL81,Zare88,PDG12}
where $m_1,m_2 \in \{J,\ldots,-J\}$.
Clebsch-Gordan coefficients 
are the expansion coefficients of a (coupled) total angular momentum eigenbasis
in an (uncoupled) tensor product basis.
We note that the Clebsch-Gordan coefficients
in Eq.~\eqref{general_tensor} describe the (tensor-product) combination
of pure states into a density matrix $\ket{\psi_1}\bra{\psi_2} =
\ket{\psi_1}\otimes \bra{\psi_2}$ of a single spin.
Tables for the Clebsch-Gordan coefficients can 
be found in literature\cite{PDG12} and there also exist several methods for their computation including
recursion relations and explicit formulas.\cite{Wigner59,BL81,MesII:1962,Landau_Lifshitz:1977}

After adding an additional spin, a basis change according to Clebsch-Gordan coefficients\cite{Wigner59,BL81,Zare88,PDG12} 
is applied in both methods A and B (see Fig.~\ref{fig:flow}).
The Clebsch-Gordon decomposition\cite{Wigner59,BL81,Zare88,PDG12} describes how
a tensor product of two irreducible representations is expanded
into a direct sum of irreducible representations: the tensor product of two tensor operators
$T_{j_1}$ and $T_{j_2}$ with ranks $j_1$ and $j_2$ are split up according to
\begin{align}
\label{eq:CGdecomp}
T_{j_1} \otimes T_{j_2}
=\bigoplus_{j=|j_l-j_2|}^{j_l+j_2} T_j. 
\end{align}
The $2j{+}1$ tensor components $T_{jm}$ with 
$m \in \{-j,\dots,j\}$ of each tensor $T_j$ on the r.h.s.\ of Eq.~\eqref{eq:CGdecomp} are given by
\begin{equation}
\label{eq:CGcoeffs}
T_{jm} = \sum_{j=|j_l-j_2|}^{j_l+j_2} \sum_{m = m_1 + m_2} C^{jm}_{j_1m_1j_2m_2}\,  T_{j_1m_1}  \otimes T_{j_2m_2} 
\end{equation}
via the Clebsch-Gordon 
coefficients\cite{Wigner59,BL81,Zare88,PDG12} $C_{j_1m_1j_2m_2}^{jm}$. 
Here, the Clebsch-Gordan coefficients in
Eq.~\eqref{eq:CGcoeffs} describe how tensor operators for $g{-}1$ spins are combined with 
the ones for a single spin into tensor operators for $g$ spins.
In the case of spins $1/2$, tensor operators $T_{j_1}$ obtained from the last iteration 
are combined with the tensor operator $T_{j_2}=T_{1}$ (see Eqs.~\eqref{eq:CGdecomp} and \eqref{eq:CGcoeffs}).
For higher spin numbers $J$,
the tensor operator $T_{j_2}$ is substituted by the direct sum
$\oplus_{q=1}^{2J} T_q$. More concretely, a $(g{-}1)$-spin system is joined
with a single spin $J$, which results in a $g$-spin system such that 
a $(g{-}1)$-linear tensor $T_{j_1}$ 
generates a set of $g$-linear tensors $T_{j}$:
\begin{align}\label{tensor_dec}
T_{j_1}  \otimes
\left( \bigoplus_{q=1}^{2J}\,
 T_q \right)
=\bigoplus_{q=1}^{2J} \bigoplus_{j=|j_1-q|}^{j_1+q} T_{j}.
\end{align}
The corresponding $g$-linear tensor components 
$T_{jm}$ 
with $m= m_1 + k $ and $k \in \{-q,\ldots,q\}$ are determined from the 
tensor components $T_{j_1m_1}$ and $T_{qk}$
via Clebsch-Gordan coefficients as detailed in 
Eq.~\eqref{eq:CGcoeffs}. After the Clebsch-Gordan basis change, either method A or B is used for 
the symmetrization with respect to spin permutations.
Details are treated in Sec.~\ref{sec:construciondetails}.

The discussed $g$-linear tensor operator components $T_{jm}$ of a rank $j$ and degree $m$
are only defined
up to a phase. We employ  
the Condon-Shortley phase convention\cite{CondonShortley,Wigner59,BL81}
 $T_{jm} = (-1)^m T_{jm}^\dagger$ 
that restricts the phase freedom to a freedom of choosing an arbitrary 
sign
for each rank $j$. In order to uniquely specify the tensor operators,
we fix these sign factors as detailed in Sec.~\ref{sec:sign_details}.
Finally, the $g$-linear tensor operators are embedded into various 
$N$-spin systems via $N{-}g$ tensor products with suitably positioned 
tensor operators
${}^{J}T^{\emptyset}_{00}$, which are proportional to identity matrices.
For each $N$-spin system, the $g$-linear tensor operators are
embedded according to the $\binom{N}{g}$ available subsets
$G\in\binom{\{1,\ldots,N\}}{g}$. For example, we denote by 
$T^{\emptyset}_{00}$ the embedded variant of the zero-linear tensor operator component
$T^{[0]}_{00}$ and the linear tensor operators $T^{[1]}_{j}$ result in 
the embedded tensor operator $T^{G}_{j}$ for each single-element set 
$G \in \{\{1\},\{2\},\dots,\{N\}\}$ of involved spins.

\section{Examples and applications for multiple spins $1/2$\label{sec:mulitple_1_2}}
In this section, we present examples and applications
for multiple spins $1/2$ and thereby illustrate and motivate 
our visualization approach. We focus on four and more spins $1/2$,
as examples for the case of up to three spins $1/2$
have already been discussed in Ref.~\onlinecite{Garon15}.
Building on the general outline given in Sec.~\ref{sec:Vis_summary},
we start by discussing the labels and their structure for four spins $1/2$.

\begin{table}[t]
\caption{Decomposition structure of four coupled spins $1/2$ into linearity $g$,
tableau $\tau_i^{[g]}$ (or simply $\tau_i$ for fixed $g$), and ranks $j$ (left).
For each subsystem $G\in\binom{\{1,..,4\}}{2}$, the 
{\it bilinear} tensors corresponding to different  tableaux 
are assembled in a single droplet function and hence the label $\ell$ of bilinear droplets does not contain a sublabel corresponding to a specific tableau $\tau_{i}^{[4]}$.
The permutation symmetry corresponding to tableau $\tau_{10}^{[4]}$ does not appear in the four-spin-$1/2$ system, i.e. no rank $j$ exists, which is indicated by ``-'' at the bottom of the last column.
On the right side, all droplet functions visualize together a complex random matrix. 
For each linearity $g$ multiple subsystems $G\in\binom{\{1,..,4\}}{g}$ occur.  \label{tab:4spin_1d2}}
\includegraphics{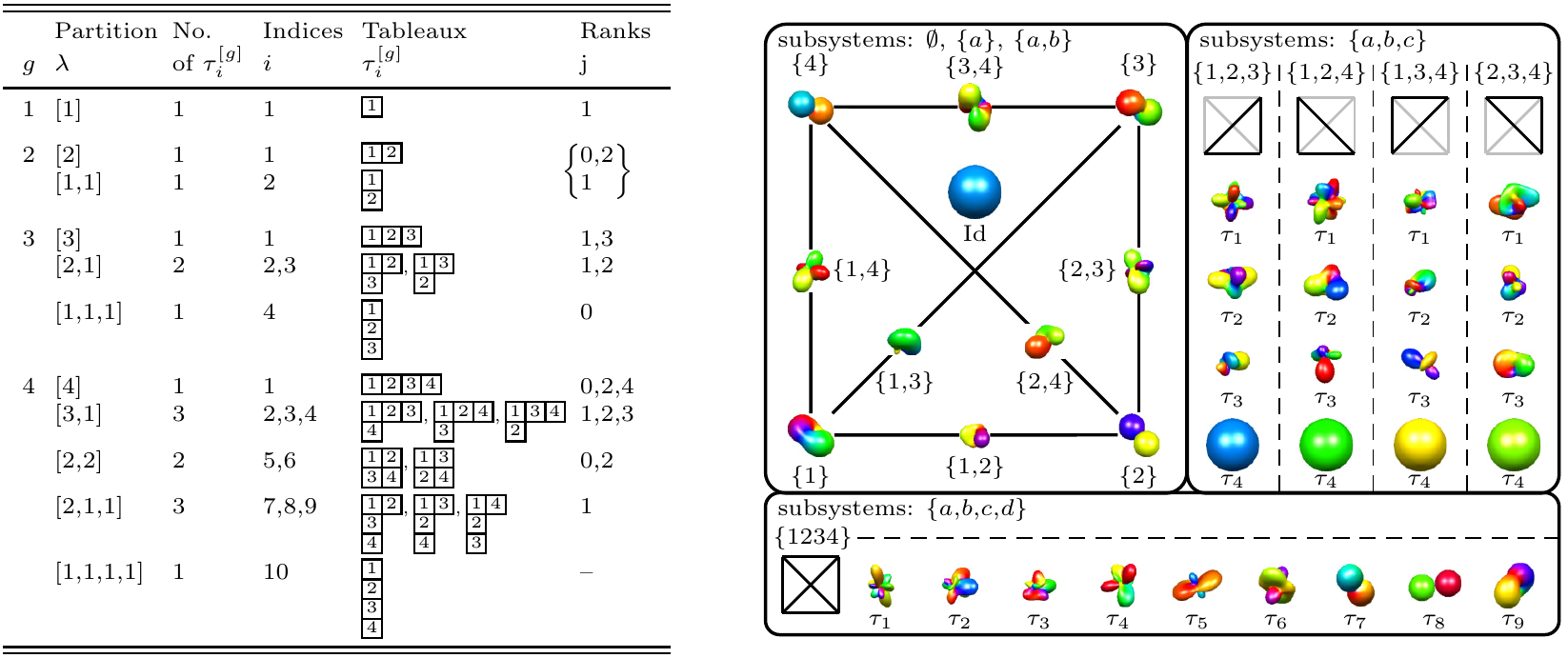}
\end{table}
\begin{figure*}
\includegraphics{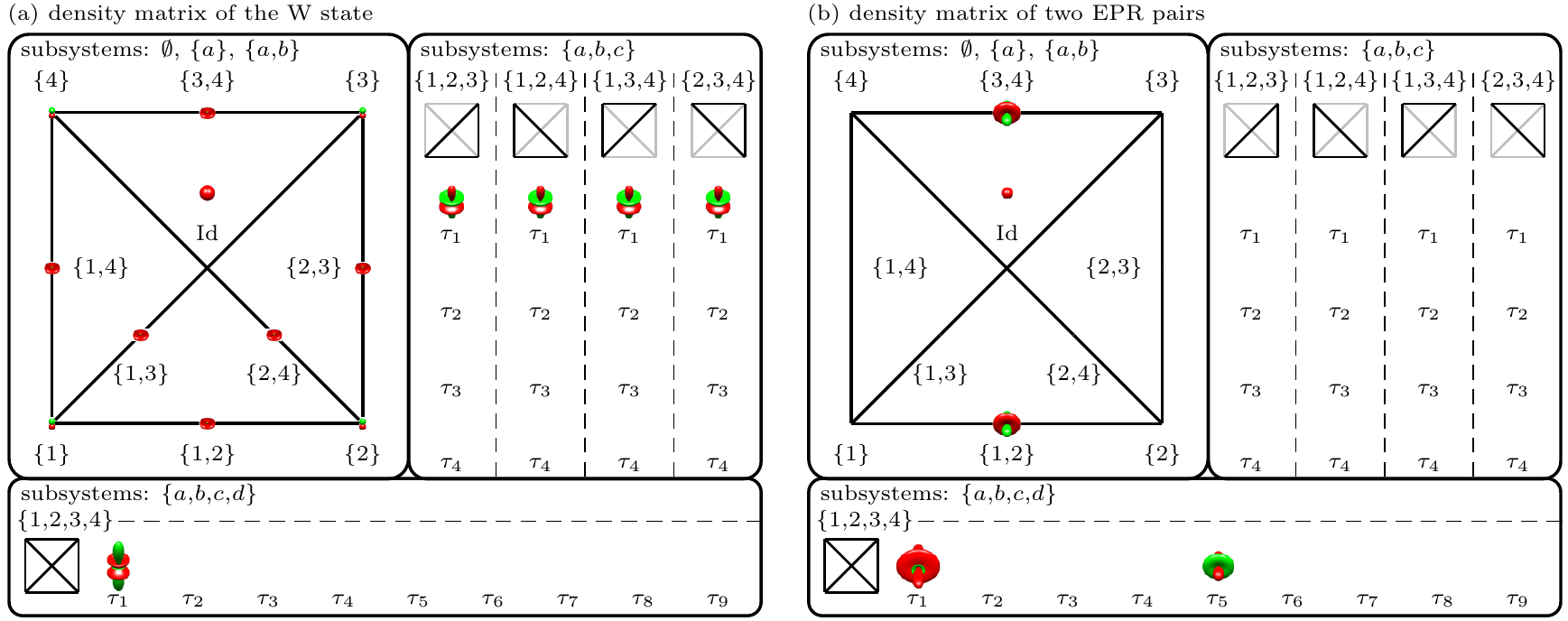}
\caption{(a) Visualizations of the density matrix 
$|W\rangle\langle W|$ of the four-qubit W state $|W\rangle$,
the droplet function for the subsystem $\{1,\!2,\!3,\!4\}$ 
is scaled to 2/3 of its original size. (b) Visualizations of the 
density matrix $| \psi \rangle\langle \psi |$ of two EPR pairs
$| \psi \rangle = (|0000\rangle + |1111\rangle + |0011\rangle + |1100\rangle)/2$
(between spins 1 and 2 as well as 3 and 4). \label{fig:ent_s4}}
\end{figure*}

 The left part of 
Table~\ref{tab:4spin_1d2} describes the decomposition of the tensor space.
For each subsystem size $g$, we list the potentially occurring 
partitions\cite{Boerner67,Hamermesh62,Pauncz95,Sagan01} $\lambda$
and the associated tableaux $\tau_i^{[g]}$, which 
are given together with their quantity and index. 
Also, for each $\lambda$ we 
state the appearing tensor ranks $j$. The bilinear tensors for a fixed subsystem
$G\in\binom{\{1,..,4\}}{2}$ are combined into a single droplet function, which is possible
as the relevant ranks $0$, $2$, and $1$ do not contain any repetition.
Note that for $g=4$, the partition [1,1,1,1] 
and its tableau $\tau_{10}^{[6]}$ do not 
correspond to any rank $j$ 
(indicated by ``-" at the bottom of the last column of the table at the left side of Tab.~\ref{tab:4spin_1d2}).
For each of the possible subsystems $G \in\binom{\{1,..,4\}}{g}$,
we have in total one label for the zero-linear tensor, one label for linear tensors, 
one label for bilinear tensors, four labels for trilinear tensors, and nine labels for four-linear tensors.
This labeling structure
for a system of four spins $1/2$
is reflected on the right of Table~\ref{tab:4spin_1d2}, where a 
$16 \times 16$ complex
random matrix
is visualized using multiple droplet functions.
The upper left panel on the right of Table~\ref{tab:4spin_1d2}
highlights the topology of the spin system, where 
nodes represent single spins $1/2$ 
and edges correspond to bilinear tensors.
Each droplet $f^{(\ell)}$ is arranged according to
its label $\ell$.
The visualization of the zero-linear tensor is labeled by $\ell=\mathrm{Id}$,
linear tensors by their subsystem $\ell=\{a\}$ for $a\in\{1,..,4\}$,
and bilinear tensors also by their subsystem $ \ell=\{a,\!b\}$ for $a,b \in\{1,..,4\}$ with $a<b$,
i.e., $\ell \in \{\{1,\!2\},\{1,\!3\},\{1,\!4\},\{2,\!3\},\{2,\!4\},\{3,\!4\}\}$.
We use $\ell=(G,\tau_i^{[g]})$ 
for $3 \leq g \leq 4$,
which explicitly specifies the tableau $\tau_i^{[g]}$.
On the right of Table~\ref{tab:4spin_1d2}, we also see the labels given by the four 
tableaux $\tau_i^{[3]}$ for each of the trilinear subsystems
$G\in\{\{1,\!2,\!3\},\{1,\!2,\!4\},\{1,\!3,\!4\},\{2,\!3,\!4\}\}$, where
the edges between involved spins are indicated by bold black lines 
whereas edges to non-involved spins are grayed out.
In the four-linear subsystem $\{1,\!2,\!3,\!4\}$, non-zero droplet functions can only occur for the nine
tableaux $\tau_1^{[4]}$-$\tau_9^{[4]}$.
Hence  in a system consisting of four spins $1/2$,
the information contained in an arbitrary operator (consisting of  $(2^4)^2=256$ complex matrix elements) 
is represented by 36 droplet functions, which have the correct transformation properties under non-selective rotations and which are organized according to the subset $G$ of involved spins and the type of permutation symmetry specified by a Young tableau $\tau_i^{[g]}$.

The cases of $g=5$ and $g=6$ are detailed in Table~\ref{tab:tableaux}
and visualizations of a
complex random matrix for systems consisting of five and six spins $1/2$ are shown in 
Figs.~\ref{fig:5spin_1d2} and \ref{fig:6spin_1d2} of Appendix~\ref{sec:app_randomMat}, respectively.
For subsystem sizes $g\geq 6$, in addition to 
the set $G$ of involved spins and the Young tabeleau $\tau_i^{[g]}$, the label
$\ell$ for a given droplet function may also include an additional 
{\it ad hoc} sublabel $\mathcal{A}$, resulting in $\ell=(G,\tau_i^{[g]},\mathcal{A})$.

Next, two examples illustrate how inherent symmetries of density matrices
are made apparent in our visualization approach. 
We consider two entangled pure states\cite{DVC00,BR01,PhysRevA.65.052112} in a four-qubit system 
(i.e.\ a system consisting of four spins $1/2$), where the corresponding density matrices
are highlighted
in Fig.~\ref{fig:ent_s4} 
following exactly the prototype in Table~\ref{tab:4spin_1d2}.
The first example is shown in Fig.~\ref{fig:ent_s4}(a), which represents
the density matrix 
$|W\rangle\langle W|$ of the
four-qubit
W state\cite{DVC00,BR01} $|W\rangle = (|0001\rangle + |0010\rangle + |0100\rangle + |1000\rangle)/2$, 
which is also known as a Dicke state.\cite{Dicke1954,stockton2003}
The highly symmetric structure of 
$|W\rangle\langle W|$
is clearly 
visible in Fig.~\ref{fig:ent_s4}(a).
All droplet functions for different subsystems $G$ of a given linearity $g$ have an 
identical shape. Also, only the fully permutation symmetric tensors 
corresponding to the tableaux $\tau_1^{[2]}$, $\tau_1^{[3]}$, and $\tau_1^{[4]}$ appear.
In total, only 16 droplet functions 
are nonzero. This is reflected by the tensor decomposition
\begin{align*}
|W\rangle\langle W| = &  \; T_{00}^{\text{Id}} - 
\tfrac{1}{2}\sum_{k=1}^4 T_{10}^{\{k\}} 
+ \left ( \sum_{\{k,l\} \in G_2} \tfrac{1}{\sqrt{3}}T_{00}^{\{k,l\}} -\tfrac{1}{\sqrt{6}}T_{20}^{\{k,l\}} \right ) \\ 
&+ \left( \sum_{\{k,l,m\} \in G_3} -\tfrac{3}{\sqrt{60}}T_{10}^{(\{k,l,m\},\tau_1^{[3]})} 
+\tfrac{4}{\sqrt{10}}T_{30}^{(\{k,l,m\},\tau_1^{[3]})} \right) 
+ \left( \tfrac{2}{\sqrt{20}} T_{00}^{\tau_1^{[4]}} 
- \tfrac{1}{\sqrt{7}} T_{20}^{\tau_1^{[4]}} 
- \tfrac{16}{\sqrt{70}} T_{40}^{\tau_1^{[4]}}\right)
\end{align*}
with $G_2 = \{\{1,\!2\},\{1,\!3\},\{1,\!4\},\{2,\!3\},\{2,\!4\},\{3,\!4\}\}$ and $G_3 = \{\{1,\!2,\!3\},\{1,\!2,\!4\},\{1,\!3,\!4\},\{2,\!3,\!4\}\}$.

The second example is given by 
the density matrix
$| \psi \rangle\langle \psi |$
of two EPR pairs\cite{PhysRevA.65.052112} 
$| \psi \rangle = (|0000\rangle + |1111\rangle + |0011\rangle + |1100\rangle)/2$ 
and is illustrated in
Fig.~\ref{fig:ent_s4}(b). Again, the symmetry structure of $| \psi \rangle\langle \psi |$
is readily visible. In this case, linear and trilinear droplet functions are completely absent. 
For the bilinear droplet functions, only the ones corresponding to 
the subsystems $\{1,\!2\}$ and $\{3,\!4\}$  are nonzero as 
the qubits 1 and 2 as well as 3 and 4 form the EPR pairs.
In the second example, we obtain the tensor decomposition
\begin{align*}
|\psi\rangle\langle\psi| = & \;  T_{00}^{\text{Id}} + \left ( \tfrac{1}{\sqrt{3}} T_{00}^{\{1,2\}} +  T_{2,-2}^{\{1,2\}} 
+ \tfrac{2}{\sqrt{6}} T_{20}^{\{1,2\}} +  T_{2,2}^{\{1,2\}} \right ) 
+ \left( \tfrac{1}{\sqrt{3}} T_{00}^{\{3,4\}} +  T_{2,-2}^{\{3,4\}} + \tfrac{2}{\sqrt{6}} T_{20}^{\{3,4\}} +  T_{2,2}^{\{3,4\}} \right ) \\
&+ \left[ \tfrac{7}{\sqrt{45}} T_{00}^{\tau_1^{[4]}} 
{+} \tfrac{2}{\sqrt{63}} T_{20}^{\tau_1^{[4]}} 
{+} \tfrac{6}{\sqrt{70}} T_{40}^{\tau_1^{[4]}} 
{+} \tfrac{2}{\sqrt{42}} \left( \sqrt{6} T_{4,-2}^{\tau_1^{[4]}} 
{+} T_{2,-2}^{\tau_1^{[4]}} {+} T_{22}^{\tau_1^{[4]}} 
{+} \sqrt{6} T_{4,2}^{\tau_1^{[4]}} \right)
+   T_{4,-4}^{\tau_1^{[4]}} 
+  T_{44}^{\tau_1^{[4]}} \right]  \\
&- \left[ \tfrac{2}{3} T_{00}^{\tau_5^{[4]}} 
+ \tfrac{4}{\sqrt{18}} T_{20}^{\tau_5^{[4]}} 
+ \tfrac{2}{\sqrt{3}} \left( T_{2,-2}^{\tau_5^{[4]}} 
+ T_{22}^{\tau_5^{[4]}}\right) \right],
\end{align*}
which explains the occurrence of four-linear components in Fig.~\ref{fig:ent_s4}(b)
even though the state $|\psi\rangle$ is a product state and has no 
four-particle contributions as a \emph{pure} state.
This emphasizes the fact that the DROPS visualization does not (directly) depict the symmetries of 
a pure state $|\psi\rangle$ but of the corresponding
density-matrix $| \psi \rangle\langle \psi |$.

\addtocounter{footnote}{1}
\footnotetext[\value{footnote}]{Hermitian operators
lead to positive and negative values, which 
are shown in red (dark gray) and green (light gray).}
\newcounter{colors}
\setcounter{colors}{\value{footnote}} 
\begin{figure}
\includegraphics{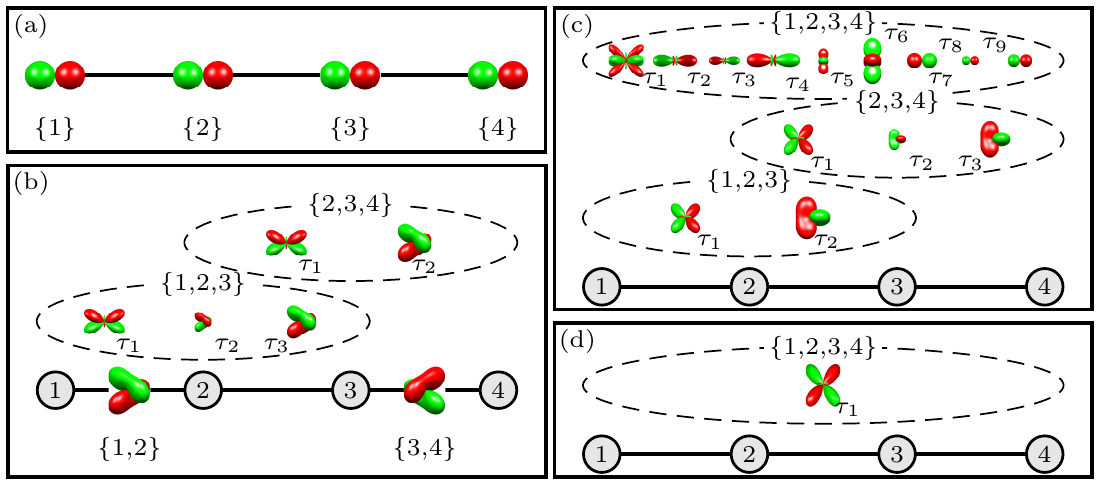}
\caption{Generation of a completely symmetric four-linear state in 
a chain of four spins $1/2$ following Table~S2 in Ref.~\onlinecite{maxq}. 
Starting from
$\rho_0 = \sum_{k=1}^4 I_{kz}$ a 
$[\pi/2]_y$ pulse on each spin results in (a),
the
evolution under coupling with time $t=1/(2\mathrm{J})$ followed by a $[\pi/2]_y$ 
pulse on each spin 
is repeating three times and visualized at various stages (b)-(d). The droplet function in (d) is scaled to 1/3 of its original size.
Linear and bilinear droplet functions are plotted on the nodes (i.e.\ spins)
and edges (i.e.\ couplings), respectively.
General $g$-linear components are indicated by dashed ellipses.
\cite{Note\thecolors}
\label{fig:maxq_4s}}
\end{figure}

The last example in this section illustrates the value of the DROPS visualization for analyzing 
the dynamics of controlled quantum systems.\cite{Roadmap2015}
This enables us to analyze the effect of control schemes
by illustrating the droplets and their symmetries appearing during the time evolution.
A free simulation package\cite{ipad_app,app} is
available, which can be used to simulate systems consisting of up to three spins $1/2$.
In the context of nuclear magnetic resonance spectroscopy, we
 consider the creation of maximum-quantum coherence 
in an Ising chain of four spins $1/2$ 
(see Fig.~\ref{fig:maxq_4s}),
which is based on a $\pi/2$ excitation pulse
followed by a
series of delays and $\pi/2$ pulses.\cite{maxq}
An operator $A_p$ has a defined coherence order\cite{EBW87} $p$ if a rotation around the $z$ axis
by any angle $\alpha$ generates the same operator $A_p$ up to a phase factor $\exp(-ip\alpha)$, i.e., 
$\exp (-i\alpha \sum_{k=1}^N I_{kz}) A_p \exp (i\alpha \sum_{k=1}^N I_{kz}) = A_p \exp(-ip\alpha)$.
Recall that Cartesian operators for single spins are 
$I_x:=\sigma_x/2$, $I_y:=\sigma_y/2$, and $I_z:=\sigma_z/2$, where
the Pauli matrices are
$
\sigma_x=\left(
\begin{smallmatrix}
0 & 1\\
1 & 0
\end{smallmatrix}
\right)
$,
$
\sigma_y=\left(
\begin{smallmatrix}
0 & -i\\
i & 0
\end{smallmatrix}
\right)
$, and
$
\sigma_z=\left(
\begin{smallmatrix}
1 & 0\\
0 & -1
\end{smallmatrix}
\right)
$. For $n$ spins, one has the operators
$I_{k \eta} := \bigotimes_{s=1}^{n} I_{a_{s}}$ 
where $a_{s}$ is equal to $\eta$ for $s{=}k$ 
and is zero otherwise; note $I_{0}:=
\left(\begin{smallmatrix}
1 & 0\\
0 & 1
\end{smallmatrix}\right)$.
All tensor-operator components $T_{jm}$ have the unique coherence order $p = m$.
The Cartesian product operator $I_{kx}$, which corresponds to observable transverse magnetization, contains coherence order $p=\pm 1$
and a triple-quantum coherence state is a linear combination of
tensor operators with rank $j \geq 3$ and order $m = \pm 3$.
The maximal coherence order is limited by the number of spins and thus by the maximal rank $j$ of tensors.
Note that a droplet $f^{(\ell)}$ representing an operator $A_p$ with coherence order $p$ exhibits the
same rotation properties as $A_p$. That is 
$f^{(\ell)}$ is reproduced up to a phase factor $\exp(-ip\alpha)$ if $f^{(\ell)}$
is rotated around the $z$ axis by $\alpha$, refer also to Fig.~\ref{tab:ops}.
The experiment considered in Ref.~\onlinecite{maxq}
generates maximal quantum coherence states starting from the initial state 
$\rho_0 = \sum_{k=1}^4 I_{kz}$,
which is specified using the Cartesian product operators 
$I_{kz}$. All coupling constants in the drift (or system) Hamiltonian
are assumed to be equal, i.e., $\mathrm{J}=\mathrm{J}_{12}=\mathrm{J}_{23}=\mathrm{J}_{34}$.
In a first step, a 
$[\pi/2]_y$ pulse is applied on each spin. Then, 
a transfer block consisting of an evolution under the
coupling with coupling period $t=1/(2\mathrm{J})$ followed by a $[\pi/2]_y$ pulse on 
each spin is repeated three times. The panels in Fig.~\ref{fig:maxq_4s} show the state of the spin 
system for different points in time: 
Panel (a) represents the initial state $\rho_0 = \sum_{k=1}^4 I_{kz}$ 
after a $\pi/2$ pulse with phase $y$ is applied to each spin. 
Panels (b), (c), and (d) depict the state after one, two, and three repetitions of the transfer block, respectively. 
In panel (d), the initial state has been fully transferred to a single 4-linear droplet function corresponding
to fully permutation-symmetric tensors (as denoted by $\tau_{1}^{[4]}$), which also contains the desired maximum-coherence orders\cite{maxq} $p=\pm4$.
 A similar example
for an Ising chain consisting of five spins $1/2$ is shown in Appendix~\ref{sec:app_exp}
[refer to Fig.~\ref{fig:exps_spins1d2}(a1)-(a5)]. 

\begin{figure}
\includegraphics{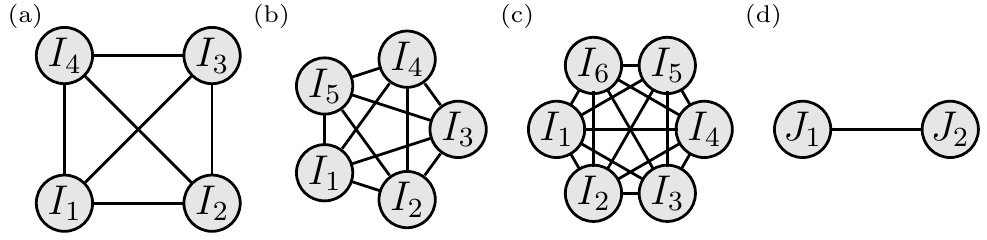}
\caption{Interaction structure of visualized spin systems (nodes represent spins):
(a)-(c) systems with $N\in\{4,5,6\}$ spins $1/2$
(see Sec.~\ref{sec:mulitple_1_2}),
(d) two spins with arbitrary spin numbers $J_1$ and $J_2$
as discussed in Sec.~\ref{sec:arbitrary}.
\label{fig:systems}}
\end{figure}

Additional examples and applications of the DROPS visualization are illustrated in Appendix~\ref{sec:further_viz}.
In 
Fig.~\ref{fig:systems}(a)-(c), general systems consisting of four to six spins $1/2$ are schematically represented as complete graphs.
In the following, we discuss the generalization of the DROPS representation to 
systems consisting of two, see Fig.~\ref{fig:systems}(d), or more
spins with arbitrary spin numbers.


\section{Representation of systems consisting of spins with arbitrary spin numbers\label{sec:arbitrary}}

Building on our description in Sec.~\ref{sec:Vis_summary}, we now consider the case of 
two coupled spins with arbitrary spin numbers. Even though spins $1/2$ (which are also known
as qubits) constitute the most important case, spins with higher spin number $J>1/2$
are highly relevant and widely studied as exemplified by 
bosonic systems, such as photons and gluons, composite particles as deuterium or helium-4, and
quasiparticles such as Cooper pairs or phonons. We start in Sec.~\ref{sec:equal_J}
with the case of two coupled spins with arbitrary but identical spin number $J$.
We extend this case to two coupled spins with different spin numbers
$J_1 \neq J_2$ in Sec.~\ref{sec:unequal}, which
also discusses
examples and illustrations
for the concrete spin numbers $J_1=1/2$ and $J_2=1$.
Generalizations of our approach to an arbitrary number of coupled spins 
with arbitrary spin numbers
are discussed in Sec.~\ref{sec:generalization}.

\begin{table}
\caption{Multiplicities of ranks $j$ occurring for 
bilinear tensors
in two-spin systems 
with equal spin numbers $J_1=J_2=J$ (left) and different spin numbers $J_1\ne J_2$ (right).
\label{tab:tensors_arb}}
\begin{tabular}{@{\hspace{2mm}} c@{\hspace{3mm}} c @{\hspace{2mm}}}
\\[-2mm]
\hline\hline
\\[-2mm]
\begin{tabular}[t]{@{\hspace{2mm}} c @{\hspace{2mm}} l @{\hspace{1.5mm}} l @{\hspace{1mm}} r
@{\hspace{1.5mm}} r @{\hspace{1.5mm}} r @{\hspace{1.5mm}} r @{\hspace{1.5mm}} r 
@{\hspace{1.5mm}} r @{\hspace{1.5mm}} r @{\hspace{1.5mm}} r @{\hspace{1.5mm}} r 
@{\hspace{1.5mm}} r @{\hspace{1.5mm}} r @{\hspace{1.5mm}} r @{\hspace{1.5mm}} r 
@{\hspace{1.5mm}} r @{\hspace{1.5mm}} r @{\hspace{1.5mm}}} 
$J$ & $\lambda$ & $j=$ & $0$ & $1$ & $2$ & $3$ & $4$ & $5$ & $6$ 
& $7$ & $8$ & $9$ & $10$ & $11$ & $12$ & $13$ & $14$
\\[1mm]  \hline \\[-2mm]
$1/2$ & [2] && 1 && 1\\
          & [1,1] && & 1\\
$1$ & [2] && 2 & 1 & 3 & 1 & 1\\
       & [1,1] && & 3 & 1 & 2\\
$3/2$ & [2] && 3 & 2 & 6 & 3 & 4 & 1 & 1\\
          & [1,1] && & 5 & 3 & 5 & 2 & 2\\
$2$ & [2] && 4 & 3 & 9 & 6 & 8 & 4 & 4 & 1 & 1\\
       & [1,1] && & 7 & 5 & 9 & 5 & 6 & 2 & 2\\
$5/2$ & [2] && 5 & 4 & 12 & 9 & 13 & 8 & 9 & 4 & 4 & 1 & 1\\
          & [1,1] && & 9 & 7 & 13 & 9 & 11 & 6 & 6 & 2 & 2\\
$3$ & [2] && 6 & 5 & 15 & 12 & 18 & 13 & 15 & 9 & 9 & 4 & 4 & 1 & 1\\
       & [1,1] && & 11 & 9 & 17 & 13 & 17 & 11 & 12 & 6 & 6 & 2 & 2\\
$7/2$ & [2] && 7 & 6 & 18 & 15 & 23 & 18 & 22 & 15 & 16 & 9 & 9 & 4 & 4 & 1 & 1\\
          & [1,1] && & 13 & 11 & 21 & 17 & 23 & 17 & 19 & 12 & 12 & 6 & 6 & 2 & 2\\[1mm]              
\end{tabular} 
&
\begin{tabular}[t]{@{\hspace{2mm}} c @{\hspace{2mm}} c @{\hspace{3mm}} l  @{\hspace{1mm}} r
@{\hspace{1.5mm}} r @{\hspace{1.5mm}} r @{\hspace{1.5mm}} r @{\hspace{1.5mm}} r 
@{\hspace{1.5mm}} r @{\hspace{1.5mm}} r @{\hspace{1.5mm}} r @{\hspace{1.5mm}} r 
@{\hspace{1.5mm}} r @{\hspace{1.5mm}} r @{\hspace{1.5mm}} r @{\hspace{1.5mm}} r 
@{\hspace{1.5mm}} r @{\hspace{1.5mm}} r @{\hspace{1.5mm}}} 
$J_1$ & $J_2$ & $j=$ & $0$ & $1$ & $2$ & $3$ & $4$ & $5$ & $6$ & $7$ & $8$ & $9$ & $10$ & $11$
\\[1mm]  \hline \\[-2mm]
$1/2$ & $1$ && 1 & 2 & 2 & 1\\
$1/2$ & $3/2$ && 1 & 2 & 3 & 2 & 1\\
$1/2$ & $2$ && 1 & 2 & 3 & 3 & 2 & 1\\
$1/2$ & $5/2$ && 1 & 2 & 3 & 3 & 3 & 2 & 1\\
$1/2$ & $3$ && 1 & 2 & 3 & 3 & 3 & 3 & 2 & 1\\
$1/2$ & $7/2$ && 1 &  2 & 3 & 3 & 3 & 3 & 3 & 2 & 1\\
$1$ & $3/2$ && 2 & 5 & 6 & 5 & 3 & 1\\
$1$ & $2$ && 2 & 5 & 7 & 7 & 5 & 3 & 1\\
$1$ & $5/2$ && 2 & 5 & 7 & 8 & 7 & 5 & 3 & 1\\
$1$ & $3$ && 2 & 5 & 7 & 8 & 8 & 7 & 5 & 3 & 1\\
$1$ & $7/2$ && 2 & 5 & 7 & 8 & 8 & 8 & 7 & 5 & 3  & 1\\
$3/2$ & $2$ && 3 & 8 & 11 & 11 & 9 & 6 & 3 & 1\\
$3/2$ & $5/2$ && 3 & 8 & 12 & 13 & 12 & 9 & 6 & 3 & 1\\
$3/2$ & $3$ && 3 & 8 & 12 & 14 & 14 & 12 & 9 & 6 & 3 & 1\\
$2$ & $5/2$ && 4 & 11 & 16 & 18 & 17 & 14 & 10 & 6 & 3 & 1\\
$2$ & $3$ && 4 & 11 & 17 & 20 & 20 & 18 & 14 & 10 & 6 & 3 & 1\\[1mm]
\end{tabular}
\\[1mm]  \hline \hline  \\[-2mm]
\end{tabular} 
\end{table}

\begin{table*}
\caption{Labeling scheme for bilinear tensors of two coupled spins. 
For $J_1=J_2=J$ (left), parent sublabels $\mathcal{P}$ and Young tableaux sublabels $\tau_i^{[g]}$ are used.
For $J_1\neq J_2$ (right), Young tableaux are replaced by {\it ad hoc}  sublabels.
Both cases result in $4J_1 J_2$ droplet functions.
 \label{tab:2s_labels}}
\begin{tabular}{@{\hspace{2mm}} c @{\hspace{4mm}} c @{\hspace{2mm}}} 
\\[-2mm]
\hline\hline
\\[-2mm]
\begin{tabular}[t]{@{\hspace{1mm}} l @{\hspace{3mm}} l @{\hspace{3mm}} l @{\hspace{3mm}} l @{\hspace{3mm}} l @{\hspace{1mm}}} 
 $\mathcal{P}$ & $\tau_i^{[g]}$ & $j$ &  $\ell$ \quad ($J_1=J_2=J$) 
\\[1mm]  \hline \\[-2mm]
 $1$,$1$  & $\tau_1$ & $0$,$2$\rdelim\}{1.8}{11pt} &  \multirow{2}{*}{$\{1,\!2\},1,1$} \\
 & $\tau_2$ & $1$ \\[2mm] 
 $2$,$2$ & $\tau_1$ & $0$,$2$,$4$\rdelim\}{1.8}{11pt} & \multirow{2}{*}{$\{1,\!2\},2,2$} \\
 & $\tau_2$ & $1$,$3$ \\[-1mm]
  $\vdots$ & & $\vdots$ &  $\vdots$ \\   
 $2J$,$2J$ & $\tau_1$ & $0$,$2$,$4$,..,$2J$\hspace{1.5mm}\rdelim\}{1.8}{11pt} &  \multirow{2}{*}{$\{1,\!2\},2J,2J$} \\
 &$\tau_2$ & $1$,$3$,..,$2J{-}1$ \\[2mm] 
 $1$,$2$ & $\tau_1$ & $1$,$2$,$3$ &  $\{1,\!2\},1,2,\tau_1$ \\[1mm] 
 & $\tau_2$ & $1$,$2$,$3$ &  $\{1,\!2\},1,2,\tau_2$  \\[-1mm]
  $\vdots$ & & $\vdots$ &  $\vdots$ \\   
 $k$,$l$ \; $(l\neq k)$ & $\tau_1$ & $|k{-}l|$,..,$k{+}l$ & $\{1,\!2\},k,l,\tau_1$ \\[1mm] 
 & $\tau_2$ & $|k{-}l|$,..,$k{+}l$ &  $\{1,\!2\},k,l,\tau_2$  \\[-1mm]
  $\vdots$ & & $\vdots$ & $\vdots$ \\ 
 $2J{-}1$,$2J$ & $\tau_1$ & $1$,..,$2J{-}1$ &  $\{1,\!2\},2J{-}1,2J,\tau_1$ \\[1mm] 
 & $\tau_2$ & $1$,..,$2J{-}1$ &  $\{1,\!2\},2J{-}1,2J,\tau_2$\\[1mm]
\end{tabular}
&
\begin{tabular}[t]{@{\hspace{1mm}} l @{\hspace{3mm}} l @{\hspace{3mm}} l @{\hspace{3mm}} l @{\hspace{3mm}} l @{\hspace{1mm}}} 
 $\mathcal{P}$ & $\mathcal{A}$ & $j$ & $\ell$ \quad ($J_1\neq J_2$)
\\[1mm]  \hline \\[-2mm]
 $1,1$ & & $0,1,2$ &  $\{1,\!2\},1,1$ \\
 $2,2$ & & $0,1,2,3,4$ &  $\{1,\!2\},2,2$ \\[-1mm]
 $\vdots$ & &  $\vdots$ \\
 $2J_1,2J_1$ & & $0$,..,$4J_1$ &  $\{1,\!2\},2J_1,2J_1$ \\ 
 $2,3$ & $I$ & $1$,..,$5$ & $\{1,\!2\},2,3,I$ \\
  & $II$ & $1$,..,$5$ &  $\{1,\!2\},2,3,II$ \\[-1mm]  
 $\vdots$ & & $\vdots$ \\
 $k,l$ \quad $(l>k)$ & $I$ & $l{-}k$,..,$l{+}k$ &  $\{1,\!2\},k,l,I$ \\ 
 & $II$ & $l{-}k$,..,$l{+}k$ & $\{1,\!2\},k,l,II$ \\[-1mm] 
 $\vdots$ & & $\vdots$  \\  
 $2J_1{-}1,2J_1$ & $I$ & $1$,..,$2J_1{-}1$ &  $\{1,\!2\},2J_1{-}1,2J_1,I$ \\ 
  & $II$ & $1$,..,$2J_1{-}1$ &  $\{1,\!2\},2J_1{-}1,2J_1,II$ \\   
 $2J_1,2J_1{+}1$ & & $1$,..,$2J_1+1$ &  $\{1,\!2\},2J_1,2J_1{+}1$ \\[-1mm]   
 $\vdots$ & & $\vdots$ \\
 $2J_1,2J_2$ & & $2J_1{-}2J_2$,..,$2J_1{+}2J_2$ &  $\{1,\!2\},2J_1,2J_2$\\[1mm]
\end{tabular}
\\[1mm]  \hline \hline  \\[-2mm]
\end{tabular}
\end{table*}

\subsection{Two coupled spins with equal spin numbers \label{sec:equal_J}}

Recall from Sec.~\ref{sec:Vis_summary} that the state of
a single spin $J$ can be described by $2J{+}1$ tensor operators $T_j$ with ranks
$j \in \{0, \dots, 2J \}$ where each tensor operator $T_j$
has $2j{+}1$ tensor-operator components 
$T_{jm} \in \mathbb{C}^{(2J+1) \times (2J+1)}$ with $m \in \{-j, \dots, j\}$.
The rank $j=0$ corresponds to a zero-linear tensor operator
and the ranks $1\leq j \leq 2J$ correspond to linear tensor operators.
Compared to the case of spins $1/2$, 
the number and multiplicity of the occurring ranks $j$ 
in tensor decompositions for multiple spins 
grow even more rapidly for general spin numbers. This is
already appreciable for bilinear tensors of two spins as detailed 
for different values of $J_1=J_2=J$
on the left of Table~\ref{tab:tensors_arb},
where multiplicities of the occurring ranks are listed
separately for the permutation symmetries
corresponding to the
partitions [2] and [1,1]. 
Additional  sublabels are required to distinguish
between 
multiply appearing ranks $j$ in order to
maintain the bijectivity of the mapping
from tensor operators to spherical harmonics following Sec.~\ref{sec:Vis_summary}.

For two coupled spins, there are zero-linear, linear, and bilinear tensors as given by the different numbers 
$g \in \{0,1,2\}$ of involved spins. The treatment of the cases with $g \in \{0,1\}$ follows 
Sec.~\ref{sec:Vis_summary}. For $g=0$, the set $G = \emptyset$ of involved spins is
empty. The corresponding single zero-linear tensor operator of rank $j=0$ requires no further partitioning
and is given the label $\ell = \text{Id}$. The linear tensors are partitioned according to
the set $G \in \{\{1\},\{2\}\}$ of  involved spins, which contains either the first or the second spin.
For both cases, $2J$ linear tensor operators with ranks $j \in \{1,2,\dots, 2J\}$ are present
and no rank appears twice. This ensures that no additional sublabels are necessary and
 the labels $\ell = \{1\}$ and $\ell = \{2\}$ can be used to uniquely specify the linear tensor operators.
So far, the tensor operators corresponding to the labels $\ell \in \{\text{Id},\{1\},\{2\}\}$ 
result jointly in three droplet functions.

\begin{figure}
\includegraphics{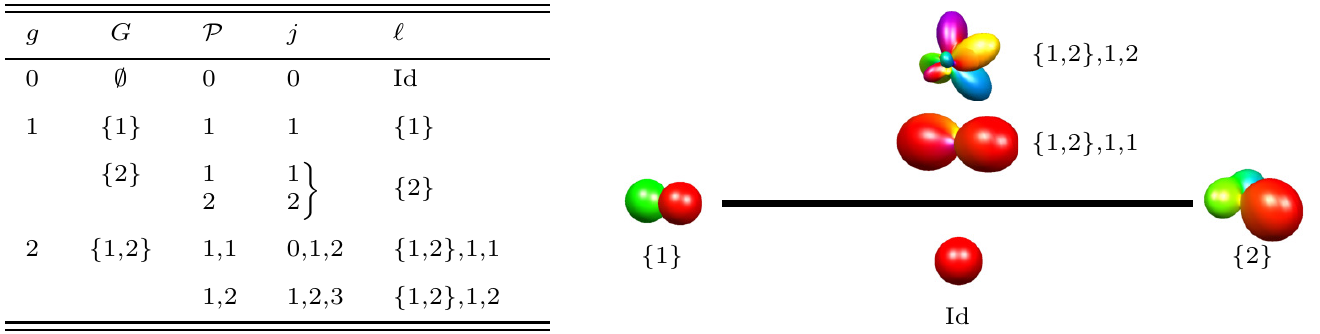}
\caption{The labeling scheme for two spins with 
spin numbers $J_1=1/2$ and $J_2=1$ results in five groups of tensors
(left). The right panel visualizes the corresponding droplet functions for
a $6\times 6$-dimensional complex random  matrix. \label{fig:spin_1d2_1}}
\end{figure}

\begin{figure}
\includegraphics{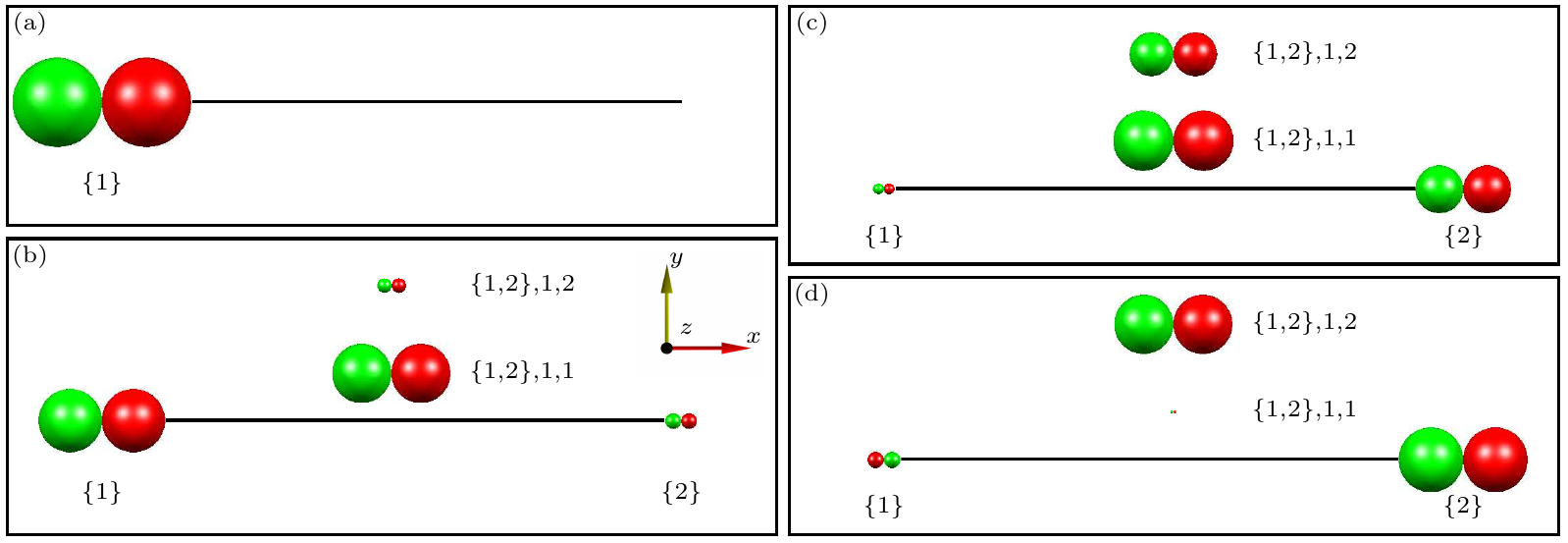}
\caption{Visualization of a
negative polarization transfer under isotropic mixing conditions
in a two-spin system consisting of
a spin $1/2$ 
and a spin $1$ (see Ref.~\onlinecite{Luy2000}):  
(a) 0 ms, (b) 10 ms, (c) 20 ms, and (d) 30 ms. \label{fig:exp_spin_1d2_1}}
\end{figure}

For bilinear tensors, the occurring ranks $j$ and their multiplicity are detailed on the left
of Table~\ref{tab:tensors_arb} separately for the partitions [2] and [1,1].
Additional sublabels are necessary for $J>1/2$ to uniquely distinguish the appearing tensor operators.
This is also true after the sublabels for permutation symmetries
given by the partitions [2] and [1,1] (or the related Young tableaux $\tau_i$)
have been applied.
{\it Ad hoc}  sublabels could be used, but they usually do not correlate with
any physical properties of the quantum system. Instead, here we 
employ so-called parent sublabels (or parents), which are 
motivated by classical 
methods.\cite{CondonShortley,Racah42,Racah43,Racah49,EL57,
Racah65,Kaplan75,Chisholm76,KJS81,RW10}
Recall that a bilinear tensor operator $T_j$ of rank $j$ is obtained 
in the Clebsch-Gordon decomposition [see Eq.~\eqref{eq:CGdecomp}]
from the tensor product of the two linear tensor operators $T_{j_1}$ and $T_{j_2}$.
The ranks $j_1$ and $j_2$ (with $j_1\leq j_2$) form the parent sublabel $\mathcal{P}=(j_1,\!j_2)$ of $T_j$.
For example, the bilinear tensor operator $T_1$ appears in
the decomposition of $T_1\otimes T_2$. This results in the parent sublabel (or parents)
$\mathcal{P} = (1,\!2)$ for this bilinear tensor operator 
$T_1$, representing the ranks $j_1=1$  and 
$j_2=2$ of the linear tensor operators $T_1$ and $T_2$.
One significant advantage of using parent sublabels is that they naturally arise in 
the construction of tensor operators. 
All parents that appear for bilinear tensors of two coupled spins
with arbitrary but equal spin number 
are detailed on the left-hand side of
Table~\ref{tab:2s_labels}. The bilinear tensors
are grouped according to  
their parents and their Young tableaux $\tau_i$,
which specify permutation symmetries as discussed above.
This scheme results in $(2J)^2$ droplet functions representing 
bilinear tensors. In total, $(2J)^2 + 3$ droplet functions are needed
to completely specify the quantum state of two coupled spins with identical spin number $J$.
Recall that for two coupled spins $1/2$ (i.e. with $J_1=J_2=J=1/2$), 
bilinear tensors can be uniquely represented by
only one ($(2J)^2=1$) droplet function,
which is fully specified by the label $\ell=G=\{1,2\}$, which indicates that it contains operators acting on the first and second spin.
However, for two coupled spins with $J_1=J_2=J^\prime=1$, four ($(2J^\prime)^2=4$) droplet functions  are necessary to represent all bilinear tensors, which obviously are not uniquely specified 
by the set  $G=\{1,2\}$ of involved spins. Of these four bilinear droplet functions, 
two function have identical parent ranks ($j_1=j_2$) and are fully characterized by a label of the form
$\ell=(G, \mathcal{P})$: the complete label 
for $j_1=j_2=1$
is $(\{1,\!2\},1,1)$ and $(\{1,\!2\},2,2)$
for $j_1=j_2=2$. The two remaining bilinear droplet functions have parent ranks $j_1=1$ and $j_2=2$
but different Young tableaux $\tau_i$. They are fully specified by the 
labels $(\{1,\!2\},1,2,\tau_1)$ and $(\{1,\!2\},1,2,\tau_2)$, respectively (c.f. fourth column in Table~\ref{tab:2s_labels}).

\subsection{Two coupled spins with different spin numbers\label{sec:unequal}}

Building on the methodology introduced in Sec.~\ref{sec:equal_J}, we address 
in this section the case of two coupled spins with different spin numbers $J_1\neq J_2$.
As before, the appearing bilinear tensor ranks $j$ 
and their multiplicity grows rapidly as shown on the right of 
Table~\ref{tab:tensors_arb}. Zero-linear and linear tensors can be---as before---represented using three droplet functions.
In contrast to the case of equal spin numbers,
we can no longer rely on permutation symmetries to label
bilinear droplet functions, because permuting spins with different spin numbers
does not preserve the global structure of the quantum system.
This forces us to combine 
parent sublabels with
{\it ad hoc}  sublabels in order to completely subdivide all
bilinear tensors. The resulting 
labeling scheme for bilinear tensors is summarized 
on the right of Table~\ref{tab:2s_labels}. Overall, $4 J_1 J_2$ different droplet functions exist for bilinear tensors
and arbitrary operators are represented by $4 J_1 J_2 + 3$ droplet functions.

A concrete example is given in Fig.~\ref{fig:spin_1d2_1}
for the case of two coupled spins with the spin numbers $J_1=1/2$ and $J_2=1$.
The labeling scheme is detailed on left of Fig.~\ref{fig:spin_1d2_1}.
One observes
the tensor rank of zero for the zero-linear tensors,
the linear tensor rank of one for the spin $1/2$, and
the linear tensor ranks of one and two for the spin $1$.
The bilinear tensor ranks are given by zero, one, and two
for the parent sublabel $P=(1,\!1)$
as well as one, two, and three for the parent sublabel $P=(1,\!2)$.
The right panel of Fig.~\ref{fig:spin_1d2_1} shows the corresponding
droplet functions, which are arranged according
to their labels.

For the same case of one spin $1/2$ and one spin $1$, 
we visualize in Fig.~\ref{fig:exp_spin_1d2_1}
the dynamics of quantum states during 
an isotropic mixing polarization transfer experiment.
In this experiment,
$x$ polarization
of the first spin (represented by droplet $\{1\}$), which corresponds to the initial density operator
$S_{1x}$,
is transferred 
via bilinear operators [represented by the droplets $(\{1,\!2\},1,1)$ and $(\{1,\!2\},1,1)$]
to $x$ polarization of the second spin (represented by droplet $\{2\}$) under 
the effective isotropic mixing (Heisenberg) coupling Hamiltonian
$H_{\text{iso}}=2 \pi \mathrm{J}_{\text{iso}} (S_{1x}S_{2x}+S_{1y}S_{2y}+S_{1z}S_{2z})$.\cite{Luy2000}
The operators in this case are defined by
$S_{1\eta_1} := I_{\eta_1} \otimes \text{id}_3$
and $S_{2\eta_2} := \text{id}_2 \otimes S_{\eta_2}$, where 
\begin{equation*}
S_{x}= \tfrac{1}{\sqrt{2}} \left(
\begin{smallmatrix}
0 & 1 & 0\\
1 & 0 & 1\\
0 & 1 & 0
\end{smallmatrix}
\right),
\quad
S_{y}= \tfrac{1}{\sqrt{2} i} \left(\!\!
\begin{smallmatrix}
\phantom{-}0 & \phantom{-}1 & \phantom{-}0\\
-1 & \phantom{-}0 & \phantom{-}1\\
\phantom{-}0 & -1 & \phantom{-}0
\end{smallmatrix}
\right), 
\quad
\text{and}
\quad
S_{z}=\left(
\begin{smallmatrix}
1 & \phantom{-}0 & \phantom{-}0\\
0 & \phantom{-}0 & \phantom{-}0\\
0 & \phantom{-}0 & -1
\end{smallmatrix}
\right)
\end{equation*}
are the spin-$1$ matrices and $\text{id}_n$ denotes the $n \times n$ identity matrix.
For a coupling constant $\mathrm{J}_{\text{iso}}$ of $11$~Hz,
the four panels in Fig.~\ref{fig:exp_spin_1d2_1} show DROPS representations of
the density matrix after (a) 0 ms, (b) 10 ms, (c) 20 ms, and (d) 30 ms, respectively.
The time-dependent $x$ polarization of the first spin is given by the function
$T_{1x} (t) =\{11+16\cos(3\pi \mathrm{J}_{\text{iso}}\,t)\}/18$, which is
negative for $t$=30 ms.
This is visible in panel (d), where the sign of the linear droplet corresponding to the first spin (labeled $\{1\}$) 
is inverted compared to panels (a) to (c): Whereas initially, the positive (red) lobe of the droplet $\{1\}$ points 
in the positive $x$ direction, after 30 ms the positive (red) lobe of the droplet $\{1\}$ points in the negative $x$ 
direction. The occurrence of polarization with inverted sign in such a simple two-spin system (consisting of a 
spin $1/2$ and a spin $1$) is of interest\cite{Luy2000} because at least five spins are necessary to 
achieve negative polarization in isotropic mixing experiments in systems consisting exclusively of spins $1/2$.

\subsection{Generalization to an arbitrary number of spins with arbitrary spin numbers \label{sec:generalization}}

We discuss now how parent sublabels can be also applied to more than two spins.
The most general spin system is composed of an arbitrary number of coupled spins
with arbitrary spin numbers $J_k$. The zero-linear and linear tensors
can be described as before. In particular, one has $2J_k$ linear tensors
with rank $j \in \{1, \dots, 2J_k \}$. Bilinear and general $g$-linear tensors
can be initially divided with respect to the set $G$ of involved spins.
A $g$-linear tensor operator $T_j$ is obtained via repeated Clebsch-Gordan
decompositions from $g$ linear tensor operators $T_{j_k}$ of rank $j_k$ 
with $1\leq k \leq g$. And
the parent sublabel $\mathcal{P} = (j_1,\!j_2,\ldots,\!j_g)$ of $T_j$ is given 
by the sequence of ranks. For example, the trilinear tensor operator $T_2$ is contained
in the Clebsch-Gordon decomposition of the tensor product
of the three linear tensor operators $T_1$, $T_1$, and $T_2$ and its parent sublabel
is given by $\mathcal{P} = (1,\!1,\!2)$. Young tableaux  specifying permutation symmetries
could be at least applied to subsystems with equal spin numbers.
Theoretically, {\it ad hoc}  sublabels can always be used to discern between any remaining tensor operators
with equal rank.
However, the practicability of this approach, which is related to the scaling of the number of necessary {\it ad hoc}  sublabels, has to be investigated in future work
together with the option of subgroup labels.\cite{CondonShortley,Racah42,Racah43,Racah49,EL57,
Racah65,Kaplan75,Chisholm76,KJS81,RW10}

\section{Explicit construction of the symmetry-adapted bases\label{sec:construciondetails}}

Here, we present the details for constructing symmetry-adapted bases
as outlined in Sec.~\ref{sec:sum_method}. Each tensor operator has to be
uniquely identified by a set of sublabels (or quantum numbers).
After the space of all tensors has been divided according to their $g$-linearity and 
the subsystem $G$ of involved spins, the tensors can be further subdivided 
with respect to their parents $\mathcal{P}$ (as introduced in  Sec.~\ref{sec:arbitrary}),
their permutation symmetries as given by a Young tableau $\tau^{[g]}$ of size $g$,
and/or necessary {\it ad hoc}  sublabels $\mathcal{A}$ that together with 
the rank $j$ and order $m\in\{-j,\ldots,j\}$
finally identify a one-dimensional
tensor subspace. Some of this information might be redundant or inapplicable
in certain cases (as permutation symmetries in the scenario of Sec.~\ref{sec:unequal}), and
we also do not utilize parent sublabels in spin-$1/2$ systems.
Our explanations start below with the initial construction of 
zero-linear and linear tensors. 
In
Sec.~\ref{sec:proj_comp_detail} and Sec.~\ref{sec:CFP_details},
we then separately describe the iterative construction of $g$-linear tensor operators (for $g\geq 2$)
based on  the projection method (denoted as Method A in Sec.~\ref{sec:sum_method})
and on the CFP method relying on fractional parentage coefficients
(denoted as Method B in Sec.~\ref{sec:sum_method}).
We conclude by explaining the chosen phase convention for DROPS basis tensor operators (see  Sec.~\ref{sec:sign_details})
and how tensors are embedded into a full $N$-spin system (see Sec.~\ref{sec:embeddingdetails}).

Let us first recall the  tensor-operator notation
$T_{jm}^{(G,\mathcal{P}, \tau, \mathcal{A})}$,
which uses the rank $j$ and order $m$ together with
all possible sublabels given by the set $G$ of involved spins, the parent sublabel $\mathcal{P}$,
the permutation symmetry $\tau$, and the {\it ad hoc}  sublabel $\mathcal{A}$.
Below, a superscript $[g]$ is used for each sublabel to indicate a specific linearity $g$.
Before accounting for the embedding in Sec.~\ref{sec:embeddingdetails},
the label $G^{[g]}$, 
is dropped. By default, we assume
that for a $g$-linear term the set of spins consists of the first $g$ spins of the system, i.e.
$G^{[g]}=\{1,\ldots,g\}$  for a linearity $g\geq 1$ and  $G^{[0]}=\emptyset$.

In the zero-linear case ($g=0$),
the parent sublabel is
an empty list $\mathcal{P}^{[0]}=()$, the tableau sublabel is empty ($\tau^{[0]} =\varnothing$), and the {\it ad hoc}  label is canonically initialized to $\mathcal{A}^{[0]}=I$; also $j^{[0]}=0$ and $m^{[0]}=0$.
We use the abbreviations $T_{0}$ and $T_{00}$  for the tensor operator and 
its component in the zero-linear case, while emphasizing that
their explicit form depends on 
the spin number $J$ as detailed in Eqs.~\eqref{eq:T00}
and \eqref{general_tensor}.

For the case of linear tensors,  $\mathcal{P}^{[1]}=(j^{[1]})$
for the rank $j^{[1]}$,
$\tau^{[1]} =\TTab{1}$, and $\mathcal{A}^{[1]}=I$. The linear tensor operators and their components 
can be uniquely identified using the simplified notations $T_{j}$ and $T_{jm}$ with $j=j^{[1]}\neq 0$.
Their explicit form depends again on the spin number $J$, see Eqs.~\eqref{tensor_one} and \eqref{general_tensor}.
After addressing these notational issues and default initializations,
we discuss the iterative construction process.

\subsection{Projection method\label{sec:proj_comp_detail}}

In the first phase of the projection method, 
the tensor decomposition from Eq.~\eqref{tensor_dec} is iteratively applied
in order to construct $g$-linear tensors from $(g{-}1)$-linear ones as outlined
in Sec.~\ref{sec:sum_method} and Fig.~\ref{fig:flow}. The explicit 
form of the corresponding tensor components can be computed with the help of 
Eq.~\eqref{eq:CGcoeffs} and the knowledge of Clebsch-Gordan coefficients.
During this iteration the Young-tableau sublabels are ignored since 
permutation symmetries are only accounted for in the second and third phase of the projection method.
{\it Ad hoc}  sublabels can be suppressed during this phase.
The parent sublabels are updated in each iteration by extending the list of parents
with that rank $q\in\{1,\ldots,2J\}$ from the added spin $J$ in Eq.~\eqref{tensor_dec} that resulted 
in the tensor operator under consideration.
When Eq.~\eqref{tensor_dec} 
has been repeated sufficiently many times such that
the desired linearity $g$ is attained, the first phase of the projection method is completed.

In the second phase of the projection method, we explicitly determine projection operators,
which will allow us to project tensor operators (and their components) onto subspaces of well-determined 
permutation symmetry. We follow the
account of Ref.~\onlinecite{Garon15} and start by recalling some basic ideas and 
notations.\cite{Boerner67,Hamermesh62,Pauncz95,Sagan01}
A permutation $\sigma \in S_g$ contained in the symmetric group $S_g$ maps  elements $i \in G
= \{1,\dots,g \}$ to elements $\sigma(i) \in G$ such that $\sigma(i_1) \neq \sigma(i_2)$ for 
$i_1 \neq i_2$. The multiplication of two elements $\sigma_2,\sigma_1 \in S_g$ is defined
by the composition $(\sigma_2 \sigma_1)(i):=(\sigma_2 \circ \sigma_1)(i)=\sigma_2[\sigma_1(i)]$ for $i \in G$.
For example, we have $(1,\!2)(1,\!3)=(1,\!3,\!2)$ using the cycle notation for elements of $S_3$.
Young tableaux are combinatorial objects built from a set
of boxes, arranged in left-orientated rows, with the row lengths in non-increasing order. 
The boxes are filled with the numbers $\{1,2,\dots,g\}$ but without repeating any number. A Young tableau 
is called {\it standard} if the entries in each row and each column are increasing. The number of 
boxes $a_j$ in each row $j$ determines 
a partition $\lambda=[a_1,a_2,\dots]$, which characterizes
the shape of a Young tableau. We use a superscript $[g]$
in a Young tableau $\tau^{[g]}$ in order to clarify the number $g$ of involved spins.
The  standard Young tableaux for $g \in \{1,2,3,4\}$ are presented in Fig.~\ref{fig:ent_s4} and
$g=5$ and $g=6$ are summarized in Tables~\ref{tab:tableaux} and~\ref{tab_cfp_a}.
The set of row-wise permutations $R(\tau)$ of a Young tableau $\tau$ is given by
all permutations of entries of $\tau$ that leave the set of elements in each row of $\tau$ fixed.
The set of column-wise permutations $C(\tau)$ can be defined similarly.
The Young symmetrizer $e_{\tau}$ is an element of the group ring $\R[S_g]$ of 
$S_g$ and can then be written for each Young tableau $\tau$ as the product 
\begin{equation}\label{eq:young}
e_{\tau} := f_{\lambda(\tau)}\, H_\tau V_\tau,
\end{equation}
where $H_\tau = \sum_{\sigma \in R(\tau)} \sigma$, 
$V_\tau = \sum_{\sigma \in C(\tau)} (-1)^{\abs{\sigma}} \sigma$
and  $\abs{\sigma}$ denotes the minimal number of transpositions
necessary to write $\sigma$ as a product thereof.
The rational factor
$f_{\lambda(\tau)} \in \R$ is equal to the number of
standard Young tableaux with the same shape $\lambda(\tau)$ as $\tau$ divided by $g!$ and 
ensures the correct normalization such that 
$e_{\tau} e_{\tau} = e_{\tau}$; note that $f_\lambda := f_{\lambda(\tau)}$ is fixed
by the shape $\lambda(\tau)$ of $\tau$.
Next, we determine the projection operators $P_p$, which 
are orthogonalized versions of the Young symmetrizers $e_{\tau_i}$. 
Let us consider the ordered sequence $\tau_r , \dots , \tau_s$
of all standard Young tableaux of fixed shape, where $r$ denotes the first index in the list and 
$s$ the last one. The projection operators $P_p$ are defined as
\begin{equation}
\label{eg:proj}
P_{p}=\begin{cases}
  e_{\tau_p}  & \text{if } p=r,\\
  f[d\,(a,\!b)+\idperm]P_{t} & \text{if } p>r.
\end{cases}
\end{equation}
For $r < p \leq s$, the index $t$ and the two boxes
$\TTab{a}$ and $\TTab{b}$ (with $b:=a+1$)
can be found as follows: There exists $t \in \{r, \dots , p{-}1\}$ such that
the tableau $\tau_t$
differs from $\tau_p$ only by the position of two boxes $\TTab{a}$
and $\TTab{b}$.
The signed axial distance $d \in Z$ from the box $\TTab{a}$ to $\TTab{b}$ 
in $e_{\tau_t}$ is the number of steps from $\TTab{a}$ to $\TTab{b}$ while counting
steps down or to the left positively and steps up or to the
right negatively. The transposition $(a,\!b)$ permutes $a$ and $b$, while 
$\idperm$ denotes the identity permutation. The normalization 
factor $f \in \R$ is chosen such that $P_{p} P_{p} =P_{p}$. 
We also refer to the example computations in 
Ref.~\onlinecite{Garon15}. Note that challenges related to 
applicability of this orthogonalization procedure (and under which conditions 
the projection property $P_{p} P_{p} =P_{p}$ holds) are discussed 
in Sec.~\ref{sec:discussion}. This completes the second phase and the projection operator $P_{p}$
can be used in the third phase.

In the third phase, each projection operator $P_{p}$  
corresponding to a standard Young tableau $\tau_p$
is applied to the space of 
tensor operators. Tensor operators (and their components) are 
projected onto the tensor subspace, the permutation symmetry of which is defined by $\tau_p$ and $P_{p}$.
In many cases, the tensor components $T_{jm}^{(\mathcal{P},\tau_p)}$ will be uniquely determined by the
image of the projection operator $P_{p}$, the rank $j$, the order $m$, and, possibly, the parent sublabel  $\mathcal{P}$.
But additional {\it ad hoc}  sublabels $\mathcal{A}\in \{I,II,\dots\}$ 
and an {\it ad hoc}  procedure to partition the space of all possible $T_{jm}^{(\mathcal{P},\tau_p)}$
into one-dimensional subspaces identified by $\mathcal{A}$ are necessary 
in the most general case. 
It is critical to coordinate the choice of these one-dimensional
subspaces for (at least) all projection operators $P_{p}$ corresponding to Young tableaux $\tau_p$
that have the same shape. Therefore, this  procedure corresponding to the {\it ad hoc}  sublabels could be applied
even before the projection operators.
An example where {\it ad hoc}  sublabels are necessary is given by six coupled spins $1/2$
(where we do not use parent sublabels)
as detailed in Table~\ref{tab:tableaux} in Appendix~\ref{sec:app_randomMat}.

\subsection{CFP method\label{sec:CFP_details}}

We describe in the following how to construct symmetry-adapted bases using a method
 based on fractional parentage coefficients (CFP).\cite{Racah65,EL57,Kaplan75,Silver76,Chisholm76,KJS81,JvW51}
We limit our presentation to multiple coupled spins $1/2$ and we do not consider any parent sublabels.
As explained in Sec.~\ref{sec:sum_method} and Eq.~\eqref{tensor_dec}, tensors of linearity $g$ are constructed
iteratively from the ones with linearity $g{-}1$ in two steps.
These two steps can be repeated until the desired linearity has been achieved.
In the first step, the 
Clebsch-Gordan decomposition in Eq.~\eqref{tensor_dec} 
is used to construct $g$-linear tensors operators
$T^{(\tau^{[g-1]},\mathcal{A}^{[g-1]},j^{[g-1]})}_{j^{[g]}}$
from $(g{-}1)$-linear ones
$T^{(\tau^{[g-1]},\mathcal{A}^{[g-1]})}_{j^{[g-1]}}$, where the explicit tensor-operator components 
are again determined using Clebsch-Gordan coefficients and Eq.~\eqref{eq:CGcoeffs}.
While executing the Clebsch-Gordan decomposition of Eq.~\eqref{tensor_dec}, we 
temporarily record 
$\tau^{[g-1]}$ and $\mathcal{A}^{[g-1]}$ from the 
previous generation together with the old rank $j^{[g-1]}$ in the labels of the provisional tensor operators
$T^{(\tau^{[g-1]},\mathcal{A}^{[g-1]},j^{[g-1]})}_{j^{[g]}}$.
This information is used in the second step below to recombine the provisional tensor operators into their 
final form and to specify this final form using updated labels.
But first, the fractional parentage coefficients and their structure
are explained which will finally lead to a characterization
of how this second step can be accomplished.

Fractional parentage coefficients can be interpreted as a block-diagonal
transformation matrix $CFP^g$ acting on the space of $g$-linear tensors. This 
transformation results in $g$-linear
tensor operators that are fully permutation symmetrized assuming that the input tensor
operators are permutation symmetrized with respect to the first $g{-}1$ spins.
The transformation matrix 
\begin{equation}
\label{eq:cfp_mtx}
CFP^{g} = \bigoplus_{j^{[g]}} CFP_{j^{[g]}}^{g}
= \bigoplus_{j^{[g]}}
\bigoplus_{\tau^{[g-1]}} CFP_{j^{[g]},\tau^{[g-1]}}^{g}
\end{equation}
can be block-diagonally decomposed according to the rank $j^{[g]}$ of
the target tensor operator and the permutation symmetry $\tau^{[g-1]}$ of 
the initial $(g{-}1)$-linear tensor operator. In the example of $g=4$ and $j^{[g]}=1$,
one obtains
\begingroup
\allowdisplaybreaks
\begin{subequations}
\label{cfp_blocks}
\begin{align}
CFP_{1}^{4} 
&=
CFP^{4}_{1,\text{\tiny{\TTab{123}}}} \oplus  CFP^{4}_{1,\text{\tiny{\TTab{12,3}}}} 
\oplus  CFP^{4}_{1,\text{\tiny{\TTab{13,2}}}} \oplus  CFP^{4}_{1,\text{\tiny{\TTab{1,2,3}}}}
\label{cfp_blocks_a}
\\
\label{cfp_blocks_b}
&= 
\left [
\begin{tabular}{ l | c }
\text{\tiny{\TTab{123}}} &  1 \\[1mm] \hline & \\[-3mm]
\text{\tiny{\TTab{123,4}}} &  1
\end{tabular}
\right ]
\oplus
\left [
\begin{tabular}{ l | c c}
\text{\tiny{\TTab{12,3}}} &  1 &  2 \\[1mm] \hline & \\[-3mm]
\text{\tiny{\TTab{124,3}}} &  $-\sqrt{\tfrac{5}{8}}$ &  $\sqrt{\tfrac{3}{8}}$ \\[1mm]
\text{\tiny{\TTab{12,3,4}}} &  $\sqrt{\tfrac{3}{8}}$ &  $\sqrt{\tfrac{5}{8}}$
\end{tabular}
\right ]
\oplus
\left [
\begin{tabular}{ l | c c}
\text{\tiny{\TTab{13,2}}} &  1 &  2 \\[1mm] \hline & \\[-3mm]
\text{\tiny{\TTab{134,2}}} &  $-\sqrt{\tfrac{5}{8}}$ &  $\sqrt{\tfrac{3}{8}}$ \\[1mm]
\text{\tiny{\TTab{13,2,4}}} &  $\sqrt{\tfrac{3}{8}}$ &  $\sqrt{\tfrac{5}{8}}$
\end{tabular}
\right ]
\oplus
\left [
\begin{tabular}{ l | c }
\text{\tiny{\TTab{1,2,3}}} &  0 \\[2mm] \hline & \\[-3mm]
\text{\tiny{\TTab{14,2,3}}} &  1
\end{tabular}
\right ] \\[1mm]
\label{cfp_blocks_c}
&= 
\left[
\begin{tabular}{ l  l @{\hspace{4mm}} l |c c c c c c}
 & &  $\tau^{[3]}$ & \text{\tiny{\TTab{123}}} & \text{\tiny{\TTab{12,3}}} & \text{\tiny{\TTab{12,3}}} & 
 \text{\tiny{\TTab{13,2}}} & \text{\tiny{\TTab{13,2}}} & \text{\tiny{\TTab{1,2,3}}}\\[1mm]
$j^{[4]}$  & $\tau^{[4]}$ &  $j^{[3]}$ & 1 & 1 & 2 & 1 & 2 & 0\\ \hline
& & \\[-2.5mm]
1 & \text{\tiny{\TTab{123,4}}} && 1 & & & & &\\[1mm] 
1 & \text{\tiny{\TTab{124,3}}} && & $-\sqrt{5/8}$ & $\sqrt{3/8}$ & & & \\[1mm]
1 & \text{\tiny{\TTab{12,3,4}}} && & $\sqrt{3/8}$ & $\sqrt{5/8}$ & & & \\[2mm] 
1 & \text{\tiny{\TTab{134,2}}} && & & & $-\sqrt{5/8}$ & $\sqrt{3/8}$ &\\[1mm]
1 & \text{\tiny{\TTab{13,2,4}}} && & & & $\sqrt{3/8}$ & $\sqrt{5/8}$ &\\[2mm]
1 & \text{\tiny{\TTab{14,2,3}}} && & & & & & 1 
\end{tabular}
\right]
\end{align}
\end{subequations}
\endgroup
for the transformation matrix resulting in tensor operators of fixed rank $j^{[g]}=1$ but with
varying permutation symmetry $\tau^{[g]}$. We have supplemented the formal decomposition
in Eq.~\eqref{cfp_blocks_a} with an explicit description of the column basis 
for the provisional tensor operators 
as well as the row basis for the final tensor operators in Eqs.~\eqref{cfp_blocks_b}-\eqref{cfp_blocks_c}.
For each block in Eq.~\eqref{cfp_blocks_b}, the upper-left corner contains $\tau^{[g-1]}$,
the left column enumerates the row basis specified by $\tau^{[g]}$, and the row on the upper right
lists the column basis determined by the ranks $j^{[g-1]}$. The associated transformation matrix 
is located in the lower-right quadrant. Equation~\eqref{cfp_blocks_c} provides essentially the same information.
Consequently, one block $CFP^{g}_{j^{[g]},\tau^{[g-1]}}$ of the transformation matrix $CFP^{g}$
can be interpreted as the matrix $[CFP^{g}_{j^{[g]},\tau^{[g-1]}}]_{\tau^{[g]},j^{[g-1]}}$
with row and column indices given by $\tau^{[g]}$ and $j^{[g-1]}$, respectively.
A tensor operator 
\begin{equation}\label{provisional_comb}
T^{(\tau^{[g]})}_{j^{[g]}} = \sum_{j^{[g-1]}} [CFP^{g}_{j^{[g]},\tau^{[g-1]}}]_{\tau^{[g]},j^{[g-1]}}\;
T^{(\tau^{[g-1]},j^{[g-1]})}_{j^{[g]}}
\end{equation}
of fixed 
rank $j^{[g]}$ and permutation symmetry $\tau^{[g]}$
is now linearly combined from certain provisional tensor operators $T^{(\tau^{[g-1]},j^{[g-1]})}_{j^{[g]}}$.
Note that the value of $\tau^{[g-1]}$ is implicitly determined by $\tau^{[g]}$ (refer also to the next paragraph).
In general, Eq.~\eqref{provisional_comb} has to be extended to account for potential {\it ad hoc}  sublabels 
$\mathcal{A}$ by
substituting permutation symmetries $\tau$ with combinations $(\tau,\mathcal{A})$
of permutation symmetries and {\it ad hoc}  sublabels (and possibly summing over
multiple values of $\mathcal{A}^{[g-1]}$). 
Note that the tensor-operator components
$T^{(\tau^{[g]})}_{j^{[g]},m^{[g]}}$
have compared to the tensor operators $T^{(\tau^{[g]})}_{j^{[g]}}$
an additional dimension given by the order $m^{[g]}\in \{-j^{[g]},\ldots,j^{[g]}\}$.
The tensor operator components 
can be directly computed 
by extending the transformation matrix $CFP_{j^{[g]}}^{g}$
to
$CFP_{j^{[g]}}^{g} \otimes \mathrm{id}_{2j^{[g]}+1}$
(where $\mathrm{id}_{2j^{[g]}+1}$ is the identity matrix of dimension
${2j^{[g]}{+}1}$)
since the fractional parentage coefficients do not depend on the value of 
the order $m^{[g]}$. In summary, our description of the fractional parentage coefficients
provides with Eq.~\eqref{provisional_comb} an explicit formula to perform 
the second step to linearly recombine the provisional tensor operators into their 
final form.

We close this subsection by further exploring the structure of fractional parentage coefficients.
For example, note that one block is repeated
in Eq.~\eqref{cfp_blocks}, even though the corresponding row and column bases differ
with respect to the appearing permutation symmetries $\tau^{[g]}$ and $\tau^{[g-1]}$
for $CFP^{4}_{1,\text{\tiny{\TTab{12,3}}}}$
and  $CFP^{4}_{1,\text{\tiny{\TTab{13,2}}}}$.  The structure of the transformations
$CFP^{4}_{j^{[g]},\tau^{[g-1]}}$ is still completely determined 
when we substitute the occurring standard Young tableaux $\tau$ 
with partitions $\lambda(\tau)$ given by the shape of $\tau$. The fractional parentage
coefficients do \emph{not} explicitly depend on the standard Young tableaux, but only on their shape.
For example, the information in Eq.~\eqref{cfp_blocks} is equivalent to
\begin{align}
\label{eq:CFP_0}
CFP^{4}_{1,[3]} \oplus  CFP^{4}_{1,[2,1]} \oplus  CFP^{4}_{1,[1,1,1]} 
= 
\left [
\begin{tabular}{ l |@{\hspace{2mm}} c }
[3]&  1 \\[0.5mm] \hline & \\[-2.5mm]
[3,1] &  1
\end{tabular}
\right ]
\oplus
\left [
\begin{tabular}{ l |@{\hspace{2mm}} c c}
[2,1] &  1 &  2 \\[0.5mm] \hline & \\[-2.5mm]
[3,1] &  $-\sqrt{\tfrac{5}{8}}$ &  $\sqrt{\tfrac{3}{8}}$ \\[1mm]
[2,1,1] &  $\sqrt{\tfrac{3}{8}}$ &  $\sqrt{\tfrac{5}{8}}$
\end{tabular}
\right ]
\oplus
\left [
\begin{tabular}{ l | @{\hspace{2mm}} c }
[1,1,1] &  0 \\[0.5mm] \hline & \\[-2.5mm]
[2,1,1] &  1
\end{tabular}
\right ].
\end{align}
One can recover $CFP^{4}_{1,\text{\tiny{\TTab{12,3}}}}$ together
with the standard Young tableaux in its row basis from $CFP^{4}_{1,[2,1]}$.
Note that $\tau^{[g]}$ is completely determined by $\tau^{[g-1]}$ and the shape
$\lambda(\tau^{[g]})$ of $\tau^{[g]}$. For example, $\tau^{[g]}=\text{\tiny{\TTab{1 2 4,3}}}$ for
$\tau^{[g-1]}=\text{\tiny{\TTab{1 2,3}}}$ and $\lambda(\tau^{[g]})=[3,\!1]$ as there is only one possibility 
to add the box $\TTab{4}$ while observing $\lambda(\tau^{[g]})=[3,\!1]$. This argument holds in general.
The repeated block in Eq.~\eqref{cfp_blocks} is a consequence of the two 
possible standard Young tableaux for the partition $[2,\!1]$.
One might wonder why no standard Young tableaux of shape
$[2,\!2]$ or $[1,\!1,\!1,\!1]$ appear for the rank $j^{[4]}=1$ in Eq.~\eqref{cfp_blocks}.
But these cases are ruled out by a priori arguments\cite{Garon15} leading to left part of Fig.~\ref{fig:ent_s4},
and similar restrictions significantly reduce the appearing cases in general.
In this regard, note that $0\leq j^{[g]} \leq g$. The full dimension of the transformation matrix
$CFP^{g}$ is given by the number of occurring tensor operators.
For the examples of systems consisting of three, four, five, and six spins $1/2$, 
the matrices $CFP^{g}$ have the dimension
$7 \times 7$, $19 \times 19$, $51 \times 51$, and $141 \times 141$, respectively.
The explicit form of the fractional parentage coefficients
for up to six spins $1/2$ has been extracted from tables in Ref.~\onlinecite{JvW51}
and is given in Appendix~\ref{CFP_app}.

\subsection{Phase and sign convention\label{sec:sign_details}}

The phase and sign of tensor operator components are not uniquely
determined by the methods for constructing symmetry-adapted bases
and they can be chosen arbitrarily.
We follow the convention of Condon and Shortley\cite{CondonShortley},
which fixes the phase up to a sign.
We have developed in Ref.~\onlinecite{Garon15}
criteria to select this sign factor such that droplet functions
reflect the properties of the depicted operators:
First, droplet functions of Hermitian operators should only feature
the colors red and green (for the phases zero and $\pi$).
Second, droplet functions of identity operators have a positive
value that is shown in red.
Third, droplet functions of a linear Cartesian operator $I_{n\eta}$ with $\eta \in \{x,y,z \}$ acting on the $n$th spin
are oriented according to its Bloch vector representation.
Fourth, the droplet function of a fully permutation-symmetric
Cartesian operator $\bigotimes_n I_{n \eta} $ with $\eta 
\in \{x,y,z \}$, has an elongated shape, and its positive lobe points in 
the direction of $\eta$.
Fifth,
raising and lowering operators are visualized by donut-shaped and 
rainbow-colored droplet functions. The number of rainbows  directly reflects the coherence 
order and the color transition of the raising operator is inverted when compared to the one of the lowering operator.
Finally, droplet functions of coupling Hamiltonians
$2I_{1x}I_{2y}+2I_{1y}I_{2x}$ exhibit a planar shape.
This motivates the sign adjustments in Table~\ref{tab:sign_spin1d2} for $g \leq 3$, which are multiplied
to $g$-linear tensors of spins $1/2$ that have been obtained
using the fractional-parentage approach in Sec.~\ref{sec:CFP_details}.
This convention is consistent with the one used for three spins $1/2$
in Ref.~\onlinecite{Garon15}. The phase factors for tensors with $g > 3$ and rank $j$ can be obtained 
via the formula $\exp[i \pi (g{-}j)/2]$.
In the following, we assume that the phase factors of tensors have been adjusted according to these rules.

\begin{table}[t]
\caption{Sign adjustments, which are multiplied to the $g$-linear tensors of
spins $1/2$ that have been obtained using  Sec.~\ref{sec:CFP_details} for up to $g \leq 3$.
\label{tab:sign_spin1d2}}
\begin{tabular}{@{\hspace{2mm}} c @{\hspace{6mm}} c @{\hspace{2mm}} c 
@{\hspace{2mm}} c @{\hspace{6mm}} c @{\hspace{2mm}} c @{\hspace{2mm}} c 
@{\hspace{6mm}} c @{\hspace{2mm}} c @{\hspace{2mm}}  c @{\hspace{6mm}}  c @{\hspace{2mm}}} 
\\[-2mm]
\hline\hline
\\[-2mm]
$g$ & $j\hphantom{^{[g]}}=$ & \multicolumn{1}{r}{0} & & & 1 & & & 2 & & 3 \\
    & $\tau_i^{[g]}=$ & $\tau_1$ & $\tau_4$ & $\tau_1$ & $\tau_2$ & $\tau_3$ & $\tau_1$ & $\tau_2$ & $\tau_3$ & $\tau_1$
\\[1mm]  \hline \\[-2mm]
0 && \hphantom{-}1 &  &  &   \\
1 && & & \hphantom{-}1 &  &   \\
2 && -1 & & & -i & & \hphantom{-}1 &  \\
3 && & \hphantom{-}i & -1 & \hphantom{-}1 & \hphantom{-}1 & & \hphantom{-}i  & \hphantom{-}i & \hphantom{-}1  
\\[1mm]  \hline \hline  \\[-2mm]
\end{tabular}
\end{table}

\subsection{Embedding tensors into the full $N$-spin system\label{sec:embeddingdetails}}

Let us finally explain how to embed $g$-linear tensors into a full $N$-spin system.
We consider $g$-linear tensor-operator components $T_{jm}^{[g]}$
where additional sublabels such as parent sublabels $\mathcal{P}$,
permutation symmetries $\tau$, and  sublabels $\mathcal{A}$ have been suppressed
for simplicity. We also assume that the $n$th spin has spin number $J_n$.
For $g=0$,
the zero-linear tensor component $T_{00}^{[0]}$ is
mapped to the embedded tensor operator
component $T_{00}^{\emptyset}:=\otimes_{n=1}^{N} {}^{J_n}T_{00}$.
For $g>0$, we assume that the set of involved spins is given by
$G=\{b_1,\ldots,b_g\}$  where $b_p < b_q$ for $p<q$. This enables us to
define the permutation $\zeta:=(1,\!b_1)\cdots(g,\!b_g)$ while
adopting the convention that $(p,\!p):=\idperm$ denotes the identity permutation.
The $g$-linear tensor-operator components
$T_{jm}^{[g]}$ are transformed into their embedded counterparts
$T_{jm}^{G}$ relative to the set $G$
of involved spins using the definition
\begin{equation}
T_{jm}^{G} := \zeta \cdot \left[T_{jm}^{[g]} \otimes \left(\bigotimes_{n=g+1}^{N}   {}^{J_{\zeta(n)}}T_{00} \right)    \right],
\end{equation}
where $\zeta$ acts by permuting the tensor factors. We assume that $T_{jm}^{[g]}$ fits to
the spins and their spin number into which it is embedded.
For $N=3$ and $G=\{2,3\}$,
one obtains the example of $\zeta=(1,\!2)(2,\!3)=(1,\!2,\!3)$ and
\begin{equation}
T_{jm}^{\{2,3\}} =
\zeta \cdot \left( T_{jm}^{[2]} \otimes {}^{J_1}T_{00}   \right).
\end{equation}

\section{Discussion and open problems 
related to the projection method for more than four spins $1/2$\label{sec:discussion}}

In this section, we discuss challenges related to the projection method which appear for
more than five spins $1/2$. 
For up to six spins $1/2$, we have verified that the projectors $P_{\tau_i} = P_i$
that have been computed using the method explained in Sec.~\ref{sec:proj_comp_detail}
are in \emph{almost} all cases
compatible with the tensor-operator basis that has been obtained using the method based 
on the fractional parentage coefficients as detailed in Sec.~\ref{sec:CFP_details}.
Everything is fine for up to four spins $1/2$.
But for five and six spins, a few projectors which are given 
as elements of the group ring of the symmetric group
are corrupted as they
do not even observe the projection property $P_{\tau_i} P_{\tau_i} = P_{\tau_i}$
(or more precisely, they cannot be normalized such that they are projections):
For five spins, 
the single projector corresponding to the Young tableau $$\tau_{16}= \text{{\tiny$\TTab{145,2,3}$}}$$
is corrupted.
For six spins,
the four projectors corresponding to the Young tableaux
$$\tau_{15}=\text{{\tiny$\TTab{1356,24}$}},\; \tau_{21}=\text{{\tiny$\TTab{1256,3,4}$}},\; 
\tau_{24}=\text{{\tiny$\TTab{1356,2,4}$}}, \; \text{ and }\;
\tau_{25}=\text{{\tiny$\TTab{1456,2,3}$}}$$
are corrupted. This very limited failure of the projection method as explained in Sec.~\ref{sec:proj_comp_detail}
is puzzling. In the following, we explain the corresponding mathematical structure in further detail
and discuss potential reasons for this limited failure. But from an applications point of view,
the second method based on the fractional parentage coefficients (see Sec.~\ref{sec:CFP_details})
works without any problems and we have used it as a substitute in order to determine the 
symmetry-adapted decomposition of tensor operators for up to six spins $1/2$.

In order to clarify the subsequent discussion, we shortly recall  
how an element of the symmetric group $S_g$ acts on the tensor space,
but we limit ourselves to the case of spins $1/2$ (i.e.\ qubits).
Given $\sigma \in S_g$, one has $\sigma(A_1 \otimes \cdots \otimes A_g)
:=A_{\sigma^{-1}(1)} \otimes \cdots \otimes A_{\sigma^{-1}(g)}$ for $A_i \in \C^{2\times 2}$.
The action on the full tensor space is then obtained by linearity. The symmetric group $S_g$ is generated
by the transpositions $(i,\!i{+}1)$ with $i \in \{1,\ldots,g{-}1\}$ and the action of $S_g$ on the
tensor space can consequently be made even more explicit if we identify the action of the transpositions
$(i,\!i{+}1)$. In particular, the action of  $(1,\!2) \in S_2$
can be described using the commutation (or swap) matrix\cite{HJ2,HS81} $K$
as follows
\begin{equation}\label{eq:swap}
(1,\!2) (A_1 \otimes A_2) = K\, (A_1 \otimes A_2)\, K = A_2 \otimes A_1
\; \text{ with }\;
K
=
\left(
\begin{smallmatrix}
1 & 0 & 0 & 0\\
0 & 0 & 1 & 0\\
0 & 1 & 0 & 0\\
0 & 0 & 0 & 1
\end{smallmatrix}\right).
\end{equation}
Equation~\eqref{eq:swap} can be vectorized using the formula\cite{HJ2,HS81}
$\vec{ABC}=(C^T \otimes A)\, \vec{B}$, where $\vec{B}$ 
denotes the vector of stacked columns 
of a matrix $B$. One obtains 
$(K^T \otimes K)\, \vec{A_1 \otimes A_2} = \vec{A_2 \otimes A_1}$ and (e.g.)
$[(K \otimes I_0 \otimes I_0)^T \otimes (K \otimes I_0 \otimes I_0)]\,
\vec{A_1 \otimes A_2 \otimes A_3 \otimes A_4} =
\vec{A_2 \otimes A_1 \otimes A_3 \otimes A_4}$, where $I_0$ is the $2\times 2$ identity matrix. 
This approach allows us to explicitly specify 
the action of elements $\sigma$ of the symmetric group or its group ring on the tensor space using
(albeit large) matrices $\Upsilon(\sigma)$ that operate linearly (by multiplication) on vectorized tensor-operator components.
Note that $\Upsilon(\sigma)$ acts implicitly on all tensor-operator components
and not only the $g$-linear ones (assuming that $g$ is equal to the number of spins).
Also, the transformation based on fractional parentage coefficients 
(i.e.\ the second step in Sec.~\ref{sec:CFP_details})
operates directly on tensor-operator components and can be therefore interpreted
as a matrix transformation on the same space as  $\Upsilon(\sigma)$ but restricted
to $g$-linear tensor operators. The explicit form of the action of the symmetric group ring
on tensors given by $\Upsilon$ will facilitate our further analysis.
As $\Upsilon$ is a linear representation of the group ring of $S_g$,
projection operators $P \in \R[S_g]$ with $P^2=P$ are mapped by $\Upsilon$ to projection operators
$\Upsilon(P)$ with $\Upsilon(P)^2=\Upsilon(P)\Upsilon(P)=\Upsilon(P^2)=\Upsilon(P)$.
The representation $\Upsilon$ of the group ring is faithful (i.e.\ the map $\sigma\in \R[S_g] \mapsto \Upsilon(\sigma)$ 
is injective) for $g\leq 4$, but it has a one-dimensional kernel for $g=5$
and a 26-dimensional kernel for $g=6$.
The existence of a kernel unfortunately complicates the analysis of the corrupt projectors $P_{\tau_i}$.
We, however, do not believe that this is the cause for the corruption.

We continue by summarizing important, general properties of projection operators. If a projector $\mathcal{P}$
is given as a matrix [as is, e.g., $\Upsilon(P_{\tau_i})$], then it has only the eigenvalues zero and one,
which will usually appear with multiplicity. The 
eigenvalue-zero
eigenspace  is equal to the kernel of $\mathcal{P}$,
and the image of $\mathcal{P}$ (i.e.\ the invariant subspace under the projection $\mathcal{P}$)
is equal to the eigenvalue-one eigenspace, the dimension of which
is given by the trace $\tr(\mathcal{P})$. In the following, it will be important to distinguish
two notions of orthogonality: First, we have introduced in Sec.~\ref{sec:proj_comp_detail}
the projectors $P_{\tau_i}$ as \emph{orthogonalized} versions of the Young symmetrizers
$e_{\tau_i}$ with the intention that the eigenvalue-one eigenspaces of $\Upsilon(P_{\tau_i})$
are \emph{orthogonal} for different Young tableaux $\tau_i$.
Second, two projectors $P_1$ and $P_2$ [as, e.g., $e_{\tau_i}$ or $P_{\tau_i}$, or even $\Upsilon(e_{\tau_i})$
or $\Upsilon(P_{\tau_i})$]
are denoted as \emph{orthogonal} if $P_1 P_2 = P_2 P_1 =0$, i.e., if their
sequential application maps everything to zero. These two notions of orthogonality
are not necessarily related. For example, one has for $g=3$ the
Young symmetrizers 
\begin{equation}
e_{\tau_2}=e_\text{{\tiny$\TTab{12,3}$}}=[\idperm + (1,\!2) -  (1,\!3) - (1,\!3,\!2)]/3\; \text{ and }\;
e_{\tau_3}=e_\text{{\tiny$\TTab{13,2}$}}=[\idperm - (1,\!2) +  (1,\!3) - (1,\!2,\!3)]/3
\end{equation}
and the projection operators
\begin{equation}\label{eq:P}
P_{\tau_2}=e_{\tau_2}\; \text{ and }\;
P_{\tau_3}=[\idperm - (1,\!2)  + 2\, (2,\!3) - (1,\!3) - 2\, (1,\!2,\!3) + (1,\!3,\!2) ]/3.
\end{equation}
One obtains that $e_{\tau_2}$ and $e_{\tau_3}$ are orthogonal (i.e.\  $e_{\tau_2} e_{\tau_3} = e_{\tau_3} e_{\tau_2} =0$)
while $P_{\tau_2}$ and $P_{\tau_3}$ are not. But the eigenvalue-one eigenspaces of $\Upsilon(e_{\tau_2})$
and $\Upsilon(e_{\tau_3})$ are not orthogonal, while the ones
of $\Upsilon(P_{\tau_2})$ and $\Upsilon(P_{\tau_3})$ are.  
Orthogonal projections are particularly convenient
and, in general, for a given
direct-sum decomposition $V= V_1 \oplus \cdots \oplus V_v$ of a vector space $V$, one can always \emph{choose}
$v$ projections $\mathcal{P}_i$ such that
(i) all projections $\mathcal{P}_i$ are mutually orthogonal,
(ii) $\mathcal{P}_{1} + \cdots + \mathcal{P}_{v} = \mathcal{I}$ (where $\mathcal{I}$ is the identity projection onto $V$), and
(iii) the image of $\mathcal{P}_{i}$ is equal to $V_i$ (see, e.g., Theorem 4.50 on p.~92 of Ref.~\onlinecite{Fuhrmann12}).
Also, the properties (i) and (ii) are closely related as a sum of several projections is again a projection if and only
if all projections are mutually orthogonal (see, e.g., Ref.~\onlinecite{MO115067}).

After these preparations, we can study certain peculiarities of the Young symmetrizers
$e_{\tau}$ as defined in Eq.~\eqref{eq:young} for $g\geq 5$. We will not necessarily assume that 
the Young tableau $\tau$ is a standard Young tableaux, i.e., the boxes
of $\tau$ are allowed to be arbitrarily filled with the numbers $\{1,\ldots,g\}$ but without repeating any number.
It is well known\cite{Boerner67,Simon1995} that Young symmetrizers are not necessarily orthogonal,
even if one only considers standard Young tableaux.
In particular, one has $e_{\tau'}e_{\tau}=0$ for the Young symmetrizers $e_{\tau}$ and $e_{\tau'}$
if there exist two integers $i,j \in \{1,\dots,g\}$ such that
$i$ and $j$ are in the same row of $\tau$ and the same column of $\tau'$
(see, e.g., Proposition VI.3.2 in Ref.~\onlinecite{Simon1995}).
For example, we have for $g=5$ only two pairs $(\tau',\tau)$
of (non-equal) standard Young tableaux such that $e_{\tau'}e_{\tau}\neq 0$, i.e.\
$(\tau',\tau) \in \{ (\tau_6, \tau_{10}), (\tau_{17}, \tau_{21}) \}$. The corresponding shapes
are $[3,\!2]$ and $[2,\!2,\!1]$.
Similarly, one has 13 such pairs for $g=6$ and  in particular
the pairs $(\tau_{8},\tau_{15})$ and $(\tau_{9},\tau_{15})$.
The shapes of all the occurring standard Young tableaux (for $g=6$) are 
$[4,\!2]$, $[3,\!3]$, $[3,\!2,\!1]$, and $[2,\!2,\!2]$. This non-orthogonality has also been studied
in Ref.~\onlinecite{Keppeler2014,Alcock2017b,Alcock2017c} together
with the question of how to find orthogonal sets of projectors.
Also, Stembridge\cite{Stembridge11} notes that all Young symmetrizers for standard Young tableaux of fixed
shape $\lambda$ are mutually orthogonal if and only if $\lambda=[2,\!2]$, $\lambda=[m]$, or $\lambda=[m,\!1,\ldots,\!1]$
for some positive integer $m$. This observed non-orthogonality may, however,
not have any implications for the corruption of the projection operators
$P_{\tau_i}$: 
Both symptoms appear for $g=6$ and $\tau_{15}$, 
but this is the only case were 
both symptoms occur simultaneously 
for standard Young tableaux of 
the same shape and  $g\in\{5,6\}$. In addition, the projection operators $P_{\tau_i}$ are not even orthogonal for 
$g=3$ [as discussed below Eq.~\eqref{eq:P}].
The non-orthogonality of Young symmetrizers of standard Young tableaux is therefore
most likely not the cause (or at least not the only one) for the 
corruption of the projection operators $P_{\tau_i}$.

In a final step, we restrict our focus to Young tableaux $\tau$ of fixed shape as 
the construction in Sec.~\ref{sec:proj_comp_detail} essentially operates only
on Young tableaux of fixed shape and the corresponding Young symmetrizers $e_\tau$.
For a given partition $\lambda$,
let us define  the projector $e_\lambda:= 
f_\lambda \sum_\tau  e_{\tau} = f_{\lambda}^2 \sum_\tau H_\tau V_\tau$ where
the sums go over all 
(not necessarily standard)
Young tableaux $\tau$ of shape $\lambda$
[cf.\ Eq.~\eqref{eq:young}]. The projector $e_\lambda$ is contained in the center of
 the group ring $\R[S_g]$, i.e., it commutes with $\R[S_g]$
(see Cor.~VI.3.7 in Ref.~\onlinecite{Simon1995}).
All projectors $e_\lambda$ are mutually orthogonal and 
one obtains the identity by summing the  $e_\lambda$ for arbitrary partitions $\lambda$.
In addition, $e_\lambda$ projects onto the left ideal of $\R[S_g]$ spanned by 
the Young symmetrizers $e_{\tau}$ for standard Young tableaux $\tau$ of shape
$\lambda$ and this left ideal describes an irreducible representation of $S_g$.\cite{Boerner67,Simon1995}
Our orthogonalization construction for the projection operators $P_{\tau_i}$ (see Sec.~\ref{sec:proj_comp_detail})
aims at splitting the eigenvalue-one eigenspace of $\Upsilon(e_\lambda)$  into 
the orthogonal eigenvalue-one eigenspaces of $\Upsilon(P_{\tau_i})$. This, however,
fails for (e.g.) $\tau_{16}$ and $g=5$, even though 
an extension of the relevant eigenspaces of the projections $\Upsilon(P_{\tau_{11}}), \ldots, \Upsilon(P_{\tau_{15}})$
to the one of $\Upsilon(e_{[3,1,1]})$ is possible. An analysis along these lines might give further insight
into how the projection method
of Sec.~\ref{sec:proj_comp_detail} is connected to the method based on fractional parentage coefficients
(see Sec.~\ref{sec:CFP_details}) and why the corruption of the projection operators $P_{\tau_i}$ arises.
But the high-dimensionality of the corresponding matrices
significantly complicates the analysis. In summary, we are currently not able to explain the corruption in the
projection method and leave this as an open question.
However, the method based on fractional parentage coefficients provides
a suitable substitute for practical purposes.

\section{Conclusion\label{sec:conclusion}}
We have extended the DROPS representation of Ref.~\onlinecite{Garon15} to visualize finite-dimensional 
quantum systems for up to six spins $1/2$ and two spins of arbitrary spin number.
A general multi-spin operator can be completely characterized and visualized using multiple
spherical plots that are each assembled  
from linear combinations of spherical harmonics $Y(\theta,\phi)$. 
The DROPS representation relies on decomposing spin operators into a
symmetry-adapted tensor basis and subsequently mapping it to 
linear combinations of spherical harmonics.
The construction algorithm in its original form 
for up to three spins
relies on explicit projection operators.\cite{Garon15} 
Due to the challenges discussed in Sec.~\ref{sec:discussion},
the projection method is only directly applicable for up to four coupled spins $1/2$.
By applying a methodology
based on fractional parentage coefficients, we have circumvented
these challenges. This methodology
relies on consecutive transformations from partially to fully permutation-symmetrized tensors.
With this technique, tensors of systems consisting of arbitrary numbers of spins $1/2$ can be 
identified by the sublabels $g$, $G$, $\tau^{[g]}$, and, for larger systems 
with six particles and more, additionally by $\mathcal{A}$, as well as the rank $j$ and order $m$.  
These tensors and their mapping to generalized Wigner functions were calculated explicitly 
for various examples for up to six spins $1/2$.
Note that the necessity of {\it ad hoc}  sublabels for six and more spins had been already anticipated
in Ref.~\onlinecite{Garon15}.  

We further extended the projection method to spins with arbitrary spin numbers. 
In particular, we discuss the cases of two coupled spins with $J_1 = J_2$ and $J_1 \neq J_2$. 
Since the number of appearing tensors is rapidly increasing with the spin number, 
the partitioning of the tensors according to
physical features of the system and inherent properties of tensors characterized by $g$, $G$, and $\tau^{[g]}$
do not suffice to obtain groups in which every tensor rank $j$ appears only once.
Although {\it ad hoc}  sublabels, analogously introduced as in the case of spins $1/2$, could resolve this problem,
 they suffer from a lack of systematics and connections related to tensor properties.
For larger spin numbers, the number of occurring tensors is substantially 
larger compared to systems consisting of spins $1/2$ and
a large set of $\mathcal{A}$ would be required even for two spins.
This inconvenience can be circumvented by relying on parent sublabels 
which are in particular suitable for larger spin numbers. Parent sublabels
can be more methodically and consistently applied and are better connected to tensor properties.
Tensors of a system consisting of two spins with $J_1 = J_2$ can be conveniently grouped according to 
the sublabels
$g$, $G$, $\mathcal{P}$, and $\tau^{[g]}$. In systems with $J_1 \neq J_2$, where permutation symmetries 
are not meaningful, tensors are organized with
respect to the sublabels
$g$, $G$, $\mathcal{P}$, and $\mathcal{A}$. 
We also discuss the extension to a larger number of spins  (with arbitrary spin numbers),
but an explicit treatment is beyond the scope of the current work.

Illustrative examples for up to six spins $1/2$ and a spin $1/2$ coupled to a spin $1$ are 
provided. This also includes entangled quantum states.
Quantum systems are frequently described by abstract operators or matrices
and our methodology is in this regard particularly useful in visualizing quantum concepts 
and systems by conveniently partitioning the inherent information.
The DROPS representation has the favorable property to naturally 
reflect transformations under non-selective spin rotations as well as spin permutations.
This approach is also convenient for highlighting the time evolution of experiments 
as animations. A free software package\cite{ipad_app,app} 
for the interactive exploration of coupled spin dynamics based on the DROPS visualization
in real time
is already available for up to three coupled spins $1/2$.
Potential applications of the DROPS visualization for larger spin systems and for particles with spin 
number larger than $1/2$ range from electron and nuclear magnetic
resonance applications in physics, chemistry, biology, and medicine
to theoretical
and experimental quantum information theory\cite{Acin2018} in which
quantum information is stored for example
by electron or nuclear spins, trapped ions, quantum dots, and superconducting circuits or 
(quasi-)particles of arbitrary spin numbers. 

\section{Acknowledgment}
We acknowledge preliminary work towards this project by Ariane Garon.
This work was supported in part by the Excellence Network of Bavaria (ENB) through ExQM. R.Z. 
and  S.J.G.  acknowledge  support  from  the  Deutsche Forschungsgemeinschaft (DFG) through
Grant No.\ Gl 203/7-2. We have
relied
on the computer algebra systems {\sc Magma},\cite{MAGMA} {\sc Matlab},\cite{matlab} and {\sc Sage}\cite{sagemath}  
for explicit computations.

\appendix

\section{Further visualizations for systems consisting of four, five or six spins $1/2$\label{sec:further_viz}}

In this appendix, we provide additional examples to 
further illustrate experimental spin 
operators using the DROPS representation. In Appendix~\ref{sec:app_ops}, we analyze the Wigner representation
of fully symmetric operators [see Eq.~\eqref{eq_fully_sym}], raising operators, and anti-phase operators typically arising in 
NMR spectroscopy for up to six coupled spins $1/2$. In Appendix~\ref{sec:app_exp}, we 
visualize multiple experiments: First, we show the evolution of droplet 
functions in the generation of multiple-quantum coherence in a five-spin system, followed by an efficient state-transfer 
experiment in a spin chain consisting of six spins $1/2$. Finally, we present snapshots of droplet functions 
during an isotropic mixing experiment in a system consisting of four spins $1/2$. In Appendix~\ref{sec:app_randomMat}, we present 
the DROPS representations for (complex) random matrices for systems consisting of five and six spins $1/2$.

\subsection{Wigner representations of prominent spin operators\label{sec:app_ops}}

\begin{figure}[b]
\includegraphics{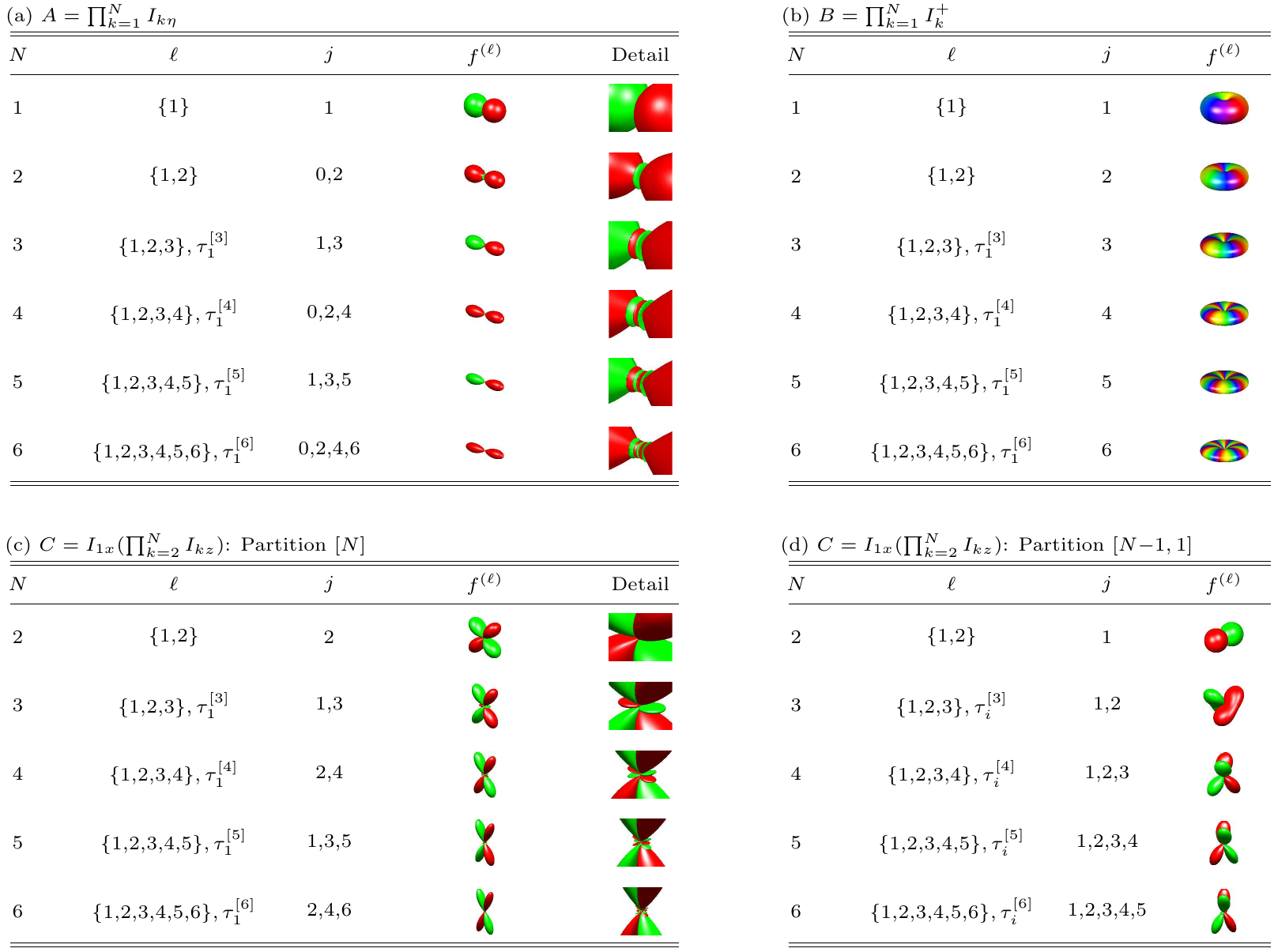}
\caption{The droplet functions $f^{(\ell)}$ visualizing various operators for up to six coupled spins $1/2$ ($N=6$). 
The full labels are given in 
the second column and the
occurring tensors ranks $j$ are shown in the third column of each table.  The detail 
column shows the corresponding magnified centers of $f^{(\ell)}$.
In Table (a), the spherical functions $f^{(\ell)}$  of the fully symmetrical 
operator $A=\prod_{k=1}^{N} I_{kx}$ are shown. 
Table (b) depicts the representations of the non-Hermitian operators $B=\prod_{k=1}^{p} I^{+}_{k}$.  
In Table (c) and (d), the visualizations of the 
antiphase operators $C=I_{1x}(\prod_{k=2}^{N} I_{kz})$ are depicted. 
There is only one symmetry type $\tau_1^{[N]}$ for the partition $[N]$ and the related droplet is shown in
Table (c) . For the partition $[N{-}1,1]$,
we find $N{-}1$ different standard Young tableaux $\tau_i^{[N]}$ with $i \in \{2,\dots,N\}$ and thus, have $N{-}1$ droplets  $f^{(\ell)}$.
They all have identical shapes but different sizes and one representative spherical
function for this case is illustrated in Table (d).
In total $N$ droplets visualize the antiphase operator
from Eq.~\eqref{eq:antiphase}.
In contrast to our usual strategy,
droplets for two spins (i.e.\ $N=2$) are plotted here separately for the fully permutation-symmetric
part (i.e.\ for the partition $[N]$) in (c)
and the remaining part in (d).
\label{tab:ops}}
\end{figure}

We show the visualization for some
prominent operators in NMR spectroscopy. In Table~\ref{tab:ops}(a), 
the droplet functions representing the fully symmetric operator
\begin{equation}\label{eq_fully_sym}
A=\prod_{k=1}^{N} I_{k\eta}
\end{equation}
with $\eta = x$ for different systems consisting of up to $N=6$ spin $1/2$. The only non-vanishing 
tensor components have permutation symmetries $\tau_1^{[N]}$ and hence, we find only one droplet 
function labeled by $G$ or $(G,\tau_1^{[N]})$. 
The elongated shape with $N{-}1$ rings, having alternating phases in the center 
of droplet, is characteristic for the DROPS representation of these operators. The operators
for $\eta \in \{y,z\}$ (not shown) exhibit the same shape but are orientated along the $y$ and $z$ axis, respectively.

\begin{figure}[b]
\includegraphics{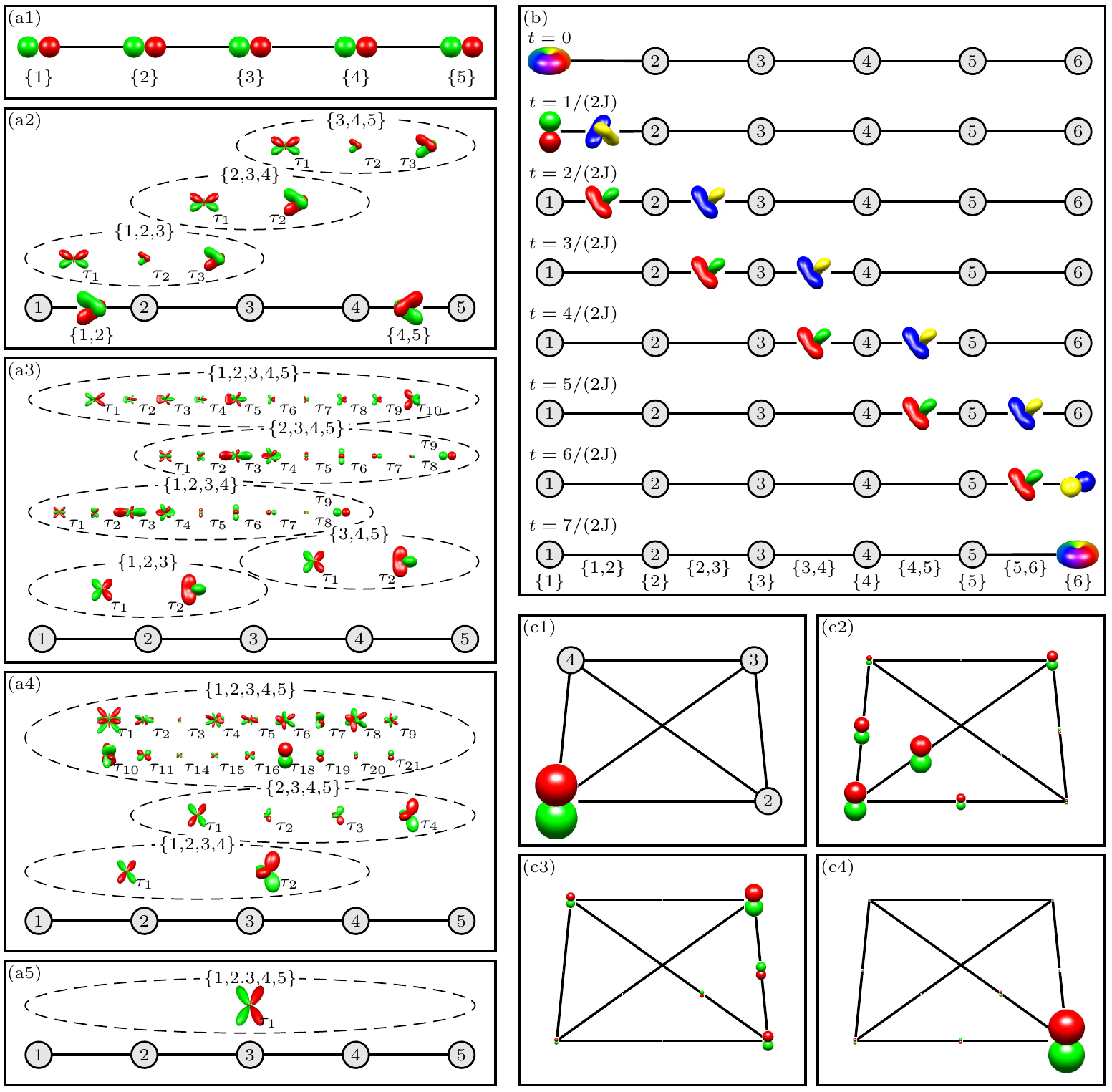}
\caption{Visualization of experiments: In panel (a1)-(a5), the
generation of five-quantum coherence in a chain consisting of five spins $1/2$ is shown for a sequence of 
$\pi/2$ pulses and optimized delays.
The initial state is 
$\rho_0 = \sum_{k=1}^5 I_{kz}$. Panel (a1) depicts the visualization of 
the density matrix after an $[\pi/2]_y$ pulse on each spin. Panels (a2)-(a5) illustrate 
the state after repeated  evolutions under coupling with 
time $t=1/(2\mathrm{J})$ followed by $[\pi/2]_y$ pulses on each spin,
see also  Fig.~\ref{fig:maxq_4s}.
The droplet in panel (a5) is scaled to 1/3 of its original size. 
Panel (b) shows the coherent transfer of the initial state $\rho_0 = I_{1x} + iI_{1y}$ to 
the target state $\rho_N = I_{6x} + iI_{6y}$ by unitary transformations in a six-spin-$1/2$ chain:
bilinear
encoded states are created,
which can be efficiently transferred, 
and they are eventually decoded at the end of the spin chain (see Ref.~\onlinecite{KG02}).
In panels (c1)-(c4), the
polarization transfer of a four-spins-$1/2$ system under isotropic 
mixing conditions is shown for the times (c1) 0 ms, (c2) 20 ms, (c3) 40 ms, 
and (c4) 133 ms, while 
only the linear and bilinear terms are explicitly displayed. \label{fig:exps_spins1d2}}
\end{figure}

Table~\ref{tab:ops}(b) shows the droplet functions representing the $p$-quantum operators
\begin{equation}
B=\prod_{k=1}^{p} I^{+}_{k}
\end{equation}
with the single-spin raising operators defined as $I^{+}_{k} = I_{kx} + i I_{ky}$.
In the DROPS representation, $p$-quantum operators
are represented by rainbow-colored donut shapes with $p$ rainbows coding for $p$ 
phase transitions from 0 to $2 \pi$ when the operator is rotated by 360$^\circ$ around the $z$ axis. The color transition for an operator 
$I^{-}_{k} = I_{kx} - i I_{ky}$ is inverted (not shown). Again, only the  coefficients of
tensors with symmetry $\tau^{[N]}_{1}$ are non-zero for both $I^{+}_{k}$ and $I^{-}_{k}$ and thus, only one droplet is found.

In Tables~\ref{tab:ops}(c) and (d), the droplet functions representing the antiphase operators 
\begin{equation}\label{eq:antiphase}
C=I_{1x} \left(\prod_{k=2}^{N} I_{kz} \right)
\end{equation}
for different sizes of spin-$1/2$ systems with number of particles $N \in \{2,3,4,5,6\}$ 
are shown. 
Only coefficients of tensors with symmetries given by the partitions $\lambda(\tau_i^{[g]})=[N]$ and $\lambda(\tau_i^{[g]})=[N{-}1,1]$ are non-vanishing.
There is only one symmetry type $\tau_1^{[N]}$ for the partition $[N]$ and the related droplet is shown in
Table \ref{tab:ops}(c). The typical features of the droplet functions $f^{(G,\tau_1^{[N]})}$ are four arms with $N{-}2$ plates with alternating phases separating the two pairs of arms.  For the partition $[N{-}1,1]$,
we find $N{-}1$ different occurring symmetries $\tau_i^{[N]}$ with $i \in \{2,\dots,N\}$ and thus, have $N{-}1$ droplets  $f^{(\ell)}$.
They all have identical shapes but different sizes and one representative droplet
function for this case is illustrated in Table \ref{tab:ops}(d).
In total $N$ droplets visualize the antiphase operator from Eq.~\eqref{eq:antiphase}.
In addition,  for each tensor only orders 
with $|m| = \pm 1$ occur.

\subsection{Visualization of experiments\label{sec:app_exp}}

We use our approach to represent and visualize 
experiments with up to six spins $1/2$. First we show maximum quantum 
coherence generation \cite{maxq} in a chain of five spins $1/2$ using 
$\pi/2$ hard pulses and delays [see Table S2 in Ref.~\onlinecite{maxq}]. 
This is the five-spin analog to the experiment visualized given in Fig.~\ref{fig:maxq_4s} of Sec.~\ref{sec:mulitple_1_2} for four spins.
The initial state is $\rho_0 = \sum_{k=1}^5 I_{kz}$ and the coupling is given
by an Ising Hamiltonian.
All coupling constants in the drift Hamiltonian
are assumed to be equal, i.e., $\mathrm{J}=\mathrm{J}_{12}=\mathrm{J}_{23}=\mathrm{J}_{34}=\mathrm{J}_{45}=\mathrm{J}$.
Fig.~\ref{fig:exps_spins1d2}(a1)-(a5) shows the droplet functions for different points in time. 
Panel (a1) shows the droplet functions after $\pi/2$ pulses with phases $y$ on each spin. 
A coupling evolution of duration $t=1/(2\mathrm{J})$ followed again by $\pi/2$ pulses with phases $y$
on all spins is repeated four times.
Panels (a2)-(a5) depict the droplets 
representing the state after each of this sequence block.
In the course of the experiment, higher orders of coherence are created,
which is reflected by the occurrence of droplets of larger $g$. Although many different tensors in 
various subsystems and symmetries appears, the information can still be partitioned in a clear scheme. Eventually,
after the experiment in panel (e), the state is fully described by a single 5-linear droplet 
(representing $G=\{1,\!2,\!3,\!4,\!5\}$ with the Young tableau sublabel
$\tau_{1}^{[5]}$),
which also contains the desired maximum-quantum coherence.

As an additional illustrative example, we present an efficient transfer of an initial state 
$\rho_0 = I_{1x} + iI_{1y}$ to the target state $\rho_t = I_{6x} + iI_{6y}$ by unitary transformations. 
We consider a linear chain of six
coupled spins $1/2$
and only assume Ising couplings (with identical coupling constant $\mathrm{J}$) between next neighbors and the free evolution
Hamiltonian is given by $H=2\pi\sum_{k=2}^6 \mathrm{J} \ I_{(k-1)z}I_{kz}$.
The approach in Ref.~\onlinecite{KG02} first encodes the initial linear operators into bilinear operators, 
which can then be efficiently propagated through the spin chain.
Fig.~\ref{fig:exps_spins1d2}(b) shows the state visualized by droplet functions for different points in time.
The nodes represent the particles and the edges the couplings between the spins.
The first row shows the visualization of $\rho_0$.
This initial state is then encoded by applying a $\pi/2$ pulse with phase $-x$  
followed by $\pi/2$ pulse with phase $y$ on the first spin, which then evolves under the the coupling Hamiltonian $H$
for a duration $1/(2\mathrm{J})$ resulting in the state shown in the second row of 
Fig.~\ref{fig:exps_spins1d2}(b). Subsequently, a sequence of a $\pi/2$ pulse with phase $x$ 
on the first spin, a $\pi/2$ pulse with phase $y$ on the second spin and a free evolution period under 
the coupling Hamiltonian $H$ with duration  $1/(2\mathrm{J})$ generates the 
encoded state, which is shown in
the third row of  Fig.~\ref{fig:exps_spins1d2}(b). This encoded state, which consists only of bilinear 
operators can then be efficiently propagated along the 
spin chain by applying an effective soliton sequence composed of a $\pi/2$ pulse with phase $y$ on all spins 
followed by a free evolution under coupling with duration  $1/(2\mathrm{J})$, which results in the propagation of the encoded state by 
one spin position. This is repeated three times and the resulting states are depicted in row four to six. 
The state is then decoded first by repeating the soliton sequence one more time (row seven) and 
then by a sequence consisting of a $\pi/2$ with phase $-y$ on the fifth spin, a $\pi/2$ with phase $x$ 
on the sixth spin, a free evolution with duration $1/(2\mathrm{J})$, and a $\pi/2$ with phase $x$ on the sixth spin 
is applied. This finally generates the desired state $\rho_t = I_{6x} + iI_{6y}$ depicted in 
row eight. Neglecting the durations of the hard pulses, the total transfer time is $7/(2\mathrm{J})$.
For comparison,\cite{KG02} the same transfer could be achieved by a sequence of five next-neighbor 
SWAP operations (each with a duration of $3/(2\mathrm{J})$) which would require a total transfer time of $15/(2\mathrm{J})$.

\begin{figure*}
\includegraphics{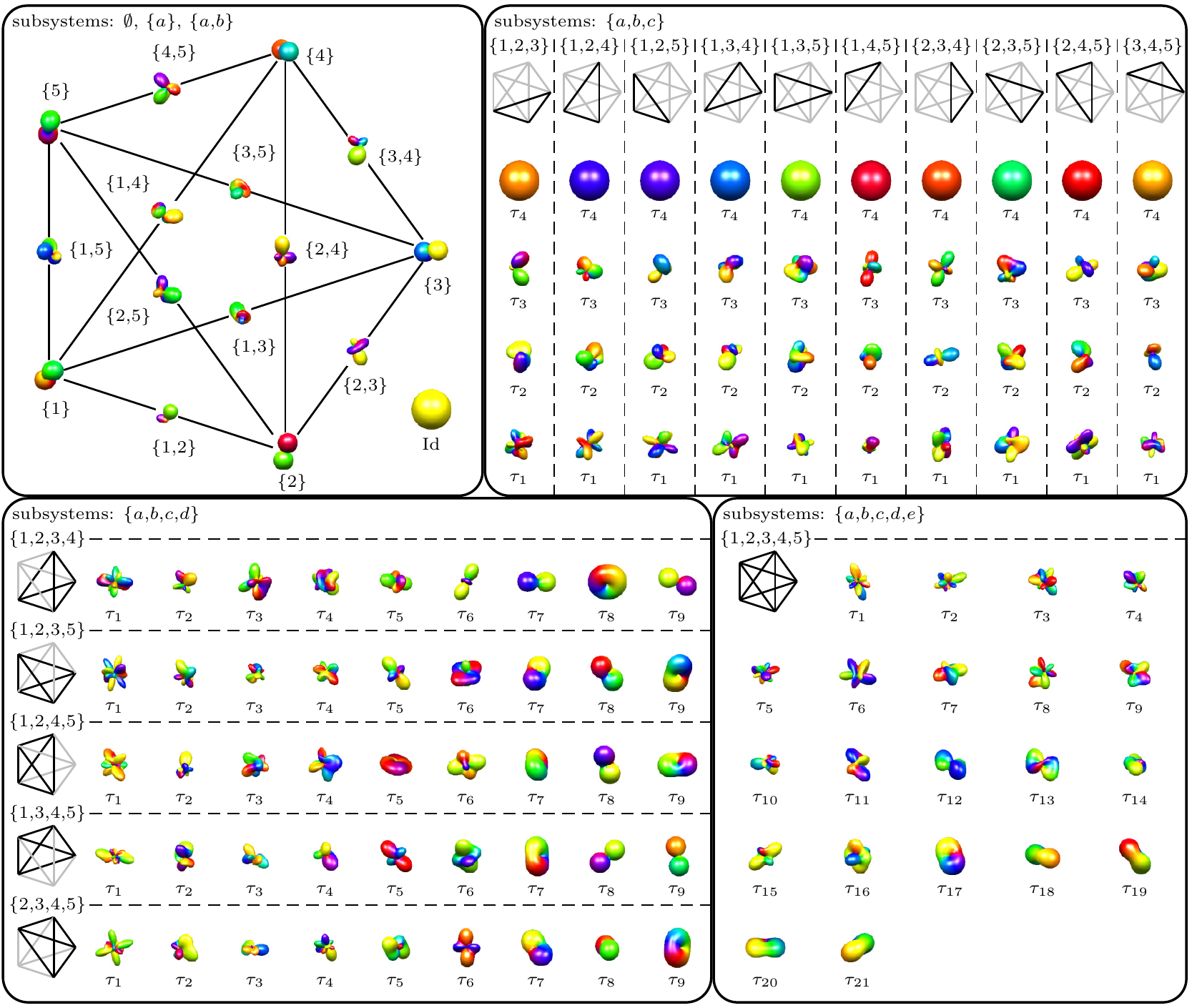}
\caption{Visualization of a complex random matrix for five spins $1/2$.
The droplet functions are arranged according to their $g$-linearity.
The top left panel shows the topology of the system with nodes representing the
spins and edges their couplings. Here, $f^{\text{Id}}$ is placed beneath the diagram,
the droplets corresponding to $g=1$ are plotted on the nodes and the
bilinear droplets for $g=2$ are placed on the edges.
The top right panel illustrates $f^{(G,\tau_i^{[3]})}$ for all possible subsystems with
$g=3$. The bottom left and  right panels depict the droplet functions
for all occurring subsystems with $g=4$ and $g=5$, respectively.
The topologies of each $G$ are visualized by diagrams located at each subpanel.  
Droplet functions are normalized for better visibility. \label{fig:5spin_1d2}}
\end{figure*}

\afterpage{%
\begin{figure}
\includegraphics{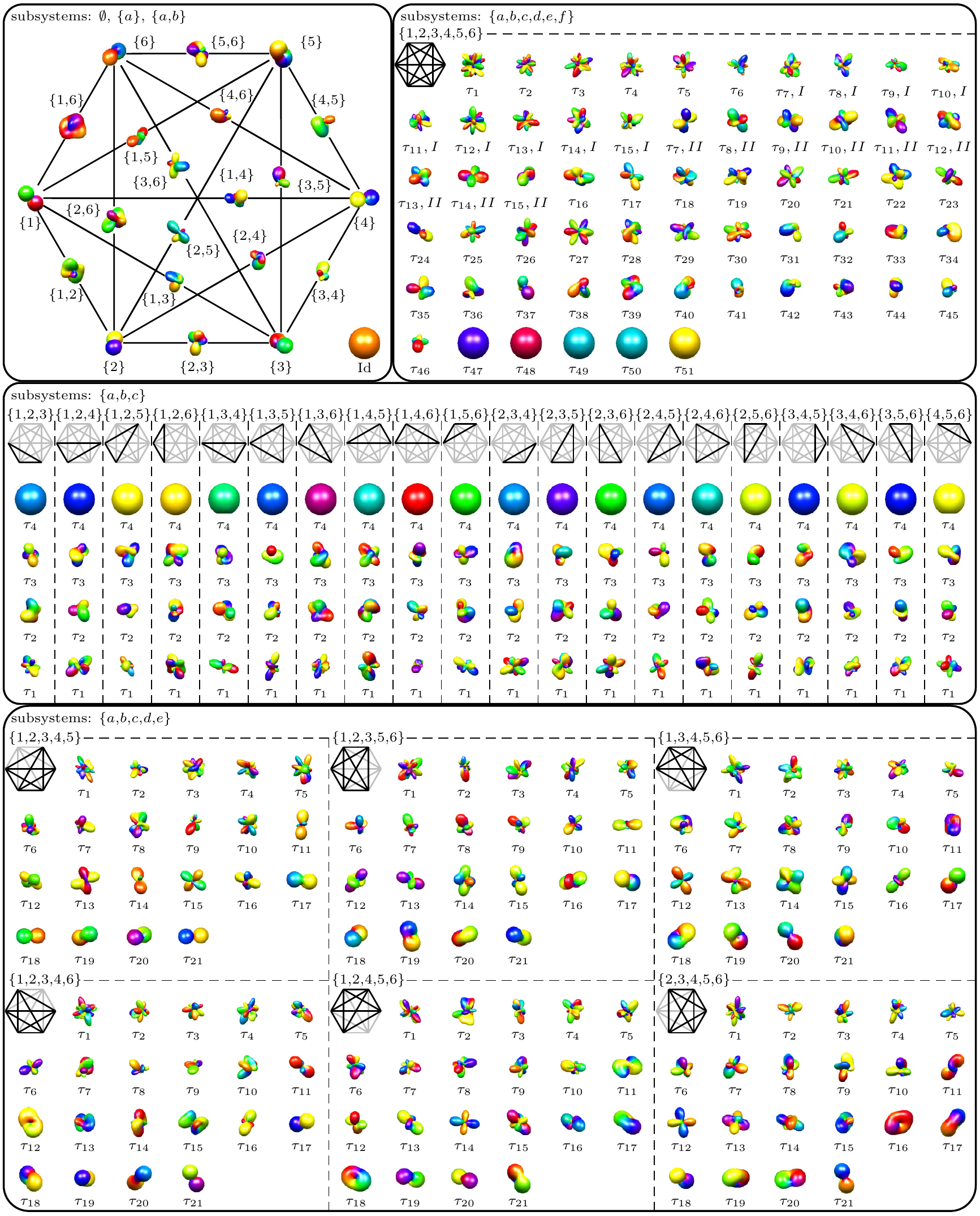}
\caption{
Visualization of a complex random matrix for six spins $1/2$ analog to
Fig.~\ref{fig:5spin_1d2}.
Droplet functions for $g\in \{0,1,2\}$ are placed on top left panel.
The top right, the middle, and the bottom panel illustrate all appearing droplet functions for all subsystems for $g=6$, $g=3$, and $g=5$.
Droplet functions are normalized. 
The four-linear contributions are given in Fig.~\ref{fig:6spin_1d2_B}.
\label{fig:6spin_1d2}}
\end{figure}
\clearpage
\begin{figure}
\includegraphics{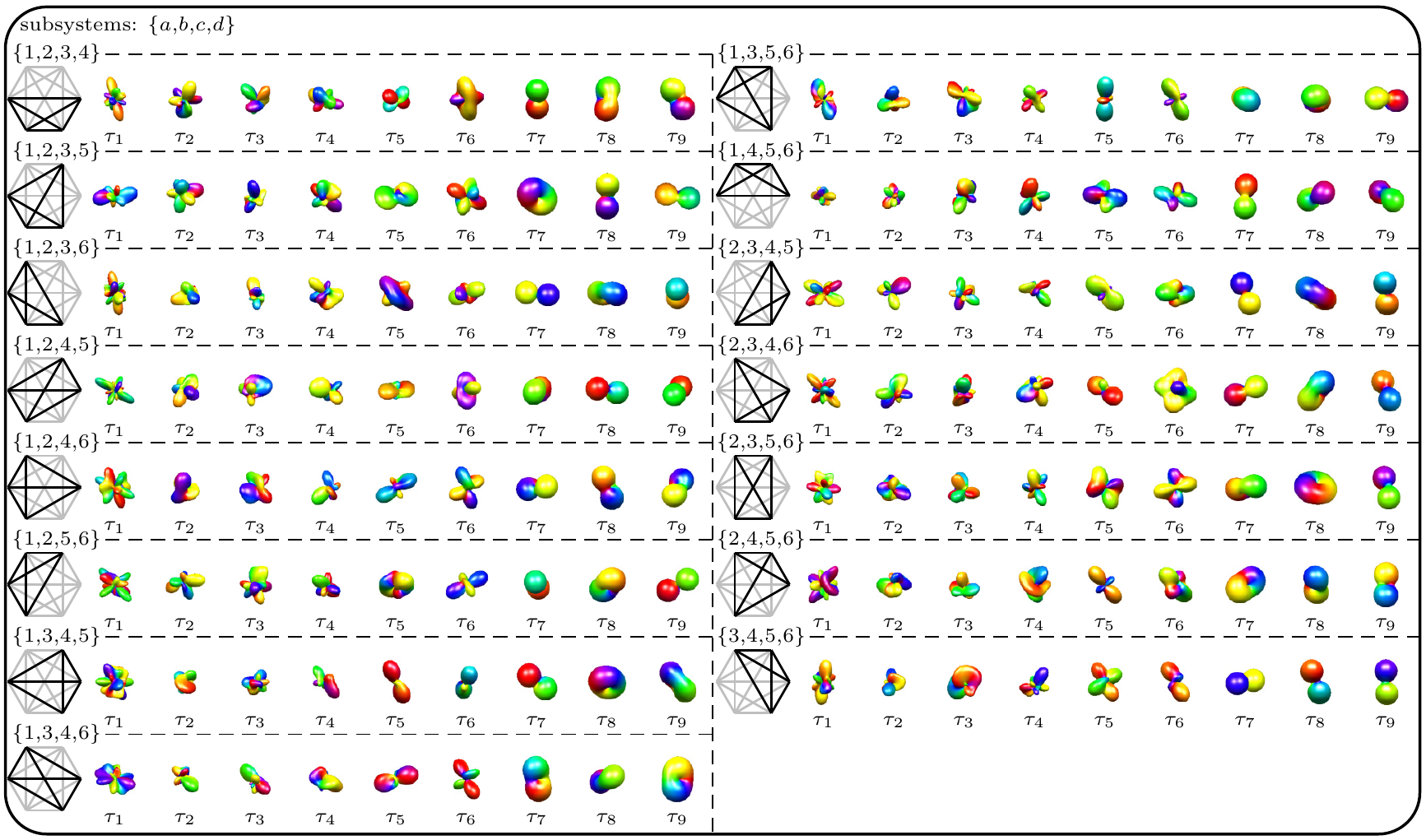}
\caption{Four-linear contributions missing in Fig.~\ref{fig:6spin_1d2}.
Spherical functions are normalized.
 \label{fig:6spin_1d2_B}}
\end{figure}
\begin{table}
\ytableausetup{smalltableaux,aligntableaux=top}
\caption{Standard Young tableaux $\tau^{[g]}$ for $g \in \{5,6\}$ with
the corresponding partitions $\lambda$ ordered with index $i$.
Also the appearing ranks $j$ for each $\tau^{[g]}$ are shown, see also Ref.~\onlinecite{Garon15}. 
{\it Ad hoc}  sublabels are required in the case $g=6$ for $j=2$ and $\lambda = [4,\!2]$.
\label{tab:tableaux}}
\begin{tabular}{@{\hspace{2mm}}c@{\hspace{3mm}}c@{\hspace{2mm}}}
\\[-2mm]
\hline\hline
\\[-2mm]
\begin{tabular}[t]{@{\hspace{0mm}} c @{\hspace{2mm}} l @{\hspace{1.5mm}} l 
@{\hspace{2mm}} l @{\hspace{2mm}} l @{\hspace{2mm}} l @{\hspace{0mm}}} 
& Partition & No.\  & Indices & {\it Ad hoc}  & Ranks
\\
$g$ & $\lambda$ & of $\tau_i^{[g]}$  
& $i$ for $\tau_i^{[g]}$  & $\mathcal{A}$ & $j$
\\[1mm]  \hline \\[-2mm]  
5 & [5] & 1 & 1 && 1,3,5\\
  & [4,1] & 4 & 2,..,5 && 1,2,3,4\\
  & [3,2] & 5 & 6,..,10 && 1,2,3\\  
  & [3,1,1] & 6 & 11,..,16 && 0,2\\
  & [2,2,1] & 5 & 17,..21 && 1\\
  & [2,1,1,1] & 4 & 22,..,25 && --\\
  & [1,1,1,1,1] & 1 & 26 && --\\[1mm]    
6 & [6] & 1 & 1 && 0,2,4,6\\ 
  & [5,1] & 5 & 2,..,6 && 1,2,3,4,5\\  
  & [4,2] & 9 & 7,..,15 & $I$ & 0,2,3,4\\
  & & 9 & 7,..,15 & $II$ & 2\\  
  & [4,1,1] & 10 & 16,..,25 && 1,3\\  
 & [3,3] & 5 & 26,..,30 && 1,3\\ 
 & [3,2,1] & 16 & 31,..,46 && 1,2\\  
 & [2,2,2] & 5 & 47,..,51 && 0\\ 
 & [3,1,1,1] & 10 & 52,..,61 && --\\   
 & [2,2,1,1] & 9 & 62,..,70 && --\\
 & [2,1,1,1,1] & 5 & 71,..,75 && --\\        
 & [1,1,1,1,1,1] & 1 & 76 && --\\[1mm]
\end{tabular}
&
\begin{tabular}[t]{@{\hspace{0mm}} l @{\hspace{0mm}}}
\\[-3mm]
\begin{tabular}{l|l@{\hspace{1mm}}l@{\hspace{1mm}}l@{\hspace{1mm}}l
@{\hspace{1mm}}l@{\hspace{1mm}}l@{\hspace{1mm}}l@{\hspace{1mm}}l
@{\hspace{1mm}}l@{\hspace{1mm}}l@{\hspace{1mm}}l@{\hspace{1mm}}l
@{\hspace{1mm}}l@{\hspace{1mm}}l@{\hspace{1mm}}l@{\hspace{1mm}}l
@{\hspace{1mm}}l@{\hspace{1mm}}l@{\hspace{1mm}}l@{\hspace{1mm}}l@{\hspace{1mm}}l}
$i$  & 1 & 2 & 3 & 4 & 5 & 6 & 7 & 8 & 9 & 10  \\
$\tau_i^{[5]}$ 
& {\tiny$\TTab{12345}$} 
& {\tiny$\TTab{1234,5}$} 
& {\tiny$\TTab{1235,4}$} 
& {\tiny$\TTab{1245,3}$} 
& {\tiny$\TTab{1345,2}$}
& {\tiny$\TTab{123,45}$}
& {\tiny$\TTab{124,35}$}
& {\tiny$\TTab{125,34}$}
& {\tiny$\TTab{134,25}$}
& {\tiny$\TTab{135,24}$}
\\
\end{tabular}
\\[4mm]
\begin{tabular}{l|l@{\hspace{1mm}}l@{\hspace{1mm}}l@{\hspace{1mm}}l
@{\hspace{1mm}}l@{\hspace{1mm}}l@{\hspace{1mm}}l@{\hspace{1mm}}l
@{\hspace{1mm}}l@{\hspace{1mm}}l@{\hspace{1mm}}l@{\hspace{1mm}}l
@{\hspace{1mm}}l@{\hspace{1mm}}l@{\hspace{1mm}}l@{\hspace{1mm}}l
@{\hspace{1mm}}l@{\hspace{1mm}}l@{\hspace{1mm}}l@{\hspace{1mm}}l@{\hspace{1mm}}l}
$i$ & 11 & 12 & 13 & 14 & 15 & 16 & 17 & 18 & 19 & 20 & 21 
\\ 
$\tau_i^{[5]}$ 
& {\tiny$\TTab{123,4,5}$}
& {\tiny$\TTab{124,3,5}$}
& {\tiny$\TTab{125,3,4}$}
& {\tiny$\TTab{134,2,5}$}
& {\tiny$\TTab{135,2,4}$}
& {\tiny$\TTab{145,2,3}$}
& {\tiny$\TTab{12,34,5}$}
& {\tiny$\TTab{12,35,4}$}
& {\tiny$\TTab{13,24,5}$}
& {\tiny$\TTab{13,25,4}$}
& {\tiny$\TTab{14,25,3}$}
\end{tabular}
\\[5mm]
\begin{tabular}{l|l@{\hspace{1mm}}l@{\hspace{1mm}}l@{\hspace{1mm}}l
@{\hspace{1mm}}l@{\hspace{1mm}}l@{\hspace{1mm}}l@{\hspace{1mm}}l
@{\hspace{1mm}}l@{\hspace{1mm}}l@{\hspace{1mm}}l@{\hspace{1mm}}l
@{\hspace{1mm}}l@{\hspace{1mm}}l@{\hspace{1mm}}l@{\hspace{1mm}}l
@{\hspace{1mm}}l@{\hspace{1mm}}l@{\hspace{1mm}}l@{\hspace{1mm}}l@{\hspace{1mm}}l}
$i$ & 1 & 2 & 3 & 4 & 5 & 6  
\\
$\tau_i^{[6]}$ 
& {\tiny$\TTab{123456}$}
& {\tiny$\TTab{12345,6}$}
& {\tiny$\TTab{12346,5}$}
& {\tiny$\TTab{12356,4}$}
& {\tiny$\TTab{12456,3}$}
& {\tiny$\TTab{13456,2}$}
\\
\end{tabular}
\\[4mm]
\begin{tabular}{l|l@{\hspace{1mm}}l@{\hspace{1mm}}l@{\hspace{1mm}}l
@{\hspace{1mm}}l@{\hspace{1mm}}l@{\hspace{1mm}}l@{\hspace{1mm}}l
@{\hspace{1mm}}l@{\hspace{1mm}}l@{\hspace{1mm}}l@{\hspace{1mm}}l
@{\hspace{1mm}}l@{\hspace{1mm}}l@{\hspace{1mm}}l@{\hspace{1mm}}l
@{\hspace{1mm}}l@{\hspace{1mm}}l@{\hspace{1mm}}l@{\hspace{1mm}}l@{\hspace{1mm}}l}
$i$ & 7 & 8 & 9 & 10 & 11 & 12 & 13 & 14 & 15\\
$\tau_i^{[6]}$ 
& {\tiny$\TTab{1234,56}$}
& {\tiny$\TTab{1235,46}$}
& {\tiny$\TTab{1236,45}$}
& {\tiny$\TTab{1245,36}$}
& {\tiny$\TTab{1246,35}$}
& {\tiny$\TTab{1256,34}$}
& {\tiny$\TTab{1345,26}$}
& {\tiny$\TTab{1346,25}$}
& {\tiny$\TTab{1356,24}$}
\end{tabular}
\\[4mm]
\begin{tabular}{l|l@{\hspace{1mm}}l@{\hspace{1mm}}l@{\hspace{1mm}}l
@{\hspace{1mm}}l@{\hspace{1mm}}l@{\hspace{1mm}}l@{\hspace{1mm}}l
@{\hspace{1mm}}l@{\hspace{1mm}}l@{\hspace{1mm}}l@{\hspace{1mm}}l
@{\hspace{1mm}}l@{\hspace{1mm}}l@{\hspace{1mm}}l@{\hspace{1mm}}l
@{\hspace{1mm}}l@{\hspace{1mm}}l@{\hspace{1mm}}l@{\hspace{1mm}}l@{\hspace{1mm}}l}
$i$  & 16 & 17 & 18 & 19 & 20 & 21
& 22 & 23 & 24 & 25 
\\
$\tau_i^{[6]}$ 
& {\tiny$\TTab{1234,5,6}$}
& {\tiny$\TTab{1235,4,6}$}
& {\tiny$\TTab{1236,4,5}$}
& {\tiny$\TTab{1245,3,6}$}
& {\tiny$\TTab{1246,3,5}$}
& {\tiny$\TTab{1256,3,4}$}
& {\tiny$\TTab{1345,2,6}$}
& {\tiny$\TTab{1346,2,5}$}
& {\tiny$\TTab{1356,2,4}$}
& {\tiny$\TTab{1456,2,3}$}
\end{tabular}
\\[5mm]
\begin{tabular}{l|l@{\hspace{1mm}}l@{\hspace{1mm}}l@{\hspace{1mm}}l
@{\hspace{1mm}}l@{\hspace{1mm}}l@{\hspace{1mm}}l@{\hspace{1mm}}l
@{\hspace{1mm}}l@{\hspace{1mm}}l@{\hspace{1mm}}l@{\hspace{1mm}}l
@{\hspace{1mm}}l@{\hspace{1mm}}l@{\hspace{1mm}}l@{\hspace{1mm}}l
@{\hspace{1mm}}l@{\hspace{1mm}}l@{\hspace{1mm}}l@{\hspace{1mm}}l@{\hspace{1mm}}l}
$i$ & 26 & 27 & 28 & 29 & 30 
& 31 & 32 & 33 & 34 & 35 & 36 & 37
\\
$\tau_i^{[6]}$ 
& {\tiny$\TTab{123,456}$}
& {\tiny$\TTab{124,356}$}
& {\tiny$\TTab{125,346}$}
& {\tiny$\TTab{134,256}$}
& {\tiny$\TTab{135,246}$}
& {\tiny$\TTab{123,45,6}$}
& {\tiny$\TTab{123,46,5}$}
& {\tiny$\TTab{124,35,6}$}
& {\tiny$\TTab{124,36,5}$}
& {\tiny$\TTab{125,34,6}$}
& {\tiny$\TTab{125,36,4}$}
& {\tiny$\TTab{126,34,5}$}
\end{tabular}
\\[5mm]
\begin{tabular}{l|l@{\hspace{1mm}}l@{\hspace{1mm}}l@{\hspace{1mm}}l
@{\hspace{1mm}}l@{\hspace{1mm}}l@{\hspace{1mm}}l@{\hspace{1mm}}l
@{\hspace{1mm}}l@{\hspace{1mm}}l@{\hspace{1mm}}l@{\hspace{1mm}}l
@{\hspace{1mm}}l@{\hspace{1mm}}l@{\hspace{1mm}}l@{\hspace{1mm}}l
@{\hspace{1mm}}l@{\hspace{1mm}}l@{\hspace{1mm}}l@{\hspace{1mm}}l@{\hspace{0mm}}l}
$i$  & 38 & 39 & 40
& 41 & 42 & 43 & 44 & 45 & 46 & 47 & 48 & 49 & 50 
& 51
\\
$\tau_i^{[6]}$
& {\tiny$\TTab{126,35,4}$}
& {\tiny$\TTab{134,25,6}$}
& {\tiny$\TTab{134,26,5}$}
& {\tiny$\TTab{135,24,6}$}
& {\tiny$\TTab{135,26,4}$}
& {\tiny$\TTab{136,24,5}$}
& {\tiny$\TTab{136,25,4}$}
& {\tiny$\TTab{145,26,3}$}
& {\tiny$\TTab{146,25,3}$}
& {\tiny$\TTab{12,34,56}$}
& {\tiny$\TTab{12,35,46}$}
& {\tiny$\TTab{13,24,56}$}
& {\tiny$\TTab{13,25,46}$}
& {\tiny$\TTab{14,25,36}$}
\end{tabular}
\end{tabular}
\\[1mm]  \hline \hline  \\[-2mm]
\end{tabular} 
\ytableausetup{smalltableaux,centertableaux}
\end{table}
\clearpage  
\begin{table}
\ytableausetup{smalltableaux,aligntableaux=top}
\caption{Explicit values of fractional parentage coefficients, see also Table~\ref{tab_cfp_b}. 
Empty boxes in the standard Young tableaux $\tau^{[g-1]}$
have to be filled with all possible values as detailed in Tables~\ref{tab:4spin_1d2} and \ref{tab:tableaux}.
\label{tab_cfp_a}}
\begin{tabular}{@{\hspace{2mm}} c @{\hspace{4mm}} c @{\hspace{2mm}}}
\\[-2mm]
\hline\hline
\\[-2mm]
\begin{tabular}[t]{@{\hspace{1mm}} l @{\hspace{2mm}} l @{\hspace{2mm}} l 
@{\hspace{1mm}} l @{\hspace{1mm}} l  @{\hspace{2mm}} l @{\hspace{-16.0cm}} @{\hspace{1mm}}} 
& &   & Transf.\ & Input  & Output 
\\
$g$ & $j^{[g]}$ & $\tau^{[g-1]}$  
& matrix & $j^{[g-1]}$ &  $\tau^{[g]}$ and $\mathcal{A}$ 
\\[1mm]  \hline & & & & & \\[-2mm]  
2 & 0 & \TTab{1} 
& 
\begin{adjustbox}{valign=t}
$
\begin{bmatrix}
1
\end{bmatrix}
$
\end{adjustbox}
&
1
& \TTab{1 2}
\\[1.5mm]
& 1 & \TTab{1}
&
\begin{adjustbox}{valign=t}
$
\begin{bmatrix}
1
\end{bmatrix}
$
\end{adjustbox}
&
1
&
\TTab{1, 2}
\\[2.5mm]
& 2 & \TTab{1}
&
\begin{adjustbox}{valign=t}
$
\begin{bmatrix}
1
\end{bmatrix}
$
\end{adjustbox}
&
1
&
 \TTab{1 2}
\\[3.5mm]  
3 & 0 & \TTab{1, 2}
&
\begin{adjustbox}{valign=t}
$
\begin{bmatrix}
1
\end{bmatrix}
$
\end{adjustbox}
&
1
&
\TTab{1, 2, 3}
\\[5mm]
& 1 & \TTab{1 2}
&
\begin{adjustbox}{valign=t}
$
\begin{bmatrix}
\tfrac{\sqrt{5}}{3} & \tfrac{2}{3}  \\[1mm] 
\tfrac{2}{3} & \tfrac{-\sqrt{5}}{3} 
\end{bmatrix}
$
\end{adjustbox}
&
0,2
&
\multirow{3}{\hsize}{\TTab{1 2 3},\\[0.5mm] \TTab{1 2, 3}}
\\[8mm]
& & \TTab{1, 2}
&
\begin{adjustbox}{valign=t}
$
\begin{bmatrix}
1
\end{bmatrix}
$
\end{adjustbox}
&
1
&
\TTab{1 3, 2}
\\[2.5mm]
& 2 & \TTab{1 2}
&
\begin{adjustbox}{valign=t}
$
\begin{bmatrix}
1
\end{bmatrix}
$
\end{adjustbox}
&
2
&
\TTab{1 2, 3}
\\[2.5mm]
& & \TTab{1, 2}
&
\begin{adjustbox}{valign=t}
$
\begin{bmatrix}
1
\end{bmatrix}
$
\end{adjustbox}
&
1
&
\TTab{1 3, 2}
\\[2.5mm]
& 3 & \TTab{1 2}
&
\begin{adjustbox}{valign=t}
$
\begin{bmatrix}
1
\end{bmatrix}
$
\end{adjustbox}
&
2
&
\TTab{1 2 3}
\\[3.5mm]
4 & 0 & \TTab{1 2 3}
&
\begin{adjustbox}{valign=t}
$
\begin{bmatrix}
1
\end{bmatrix}
$
\end{adjustbox}
&
1
&
\TTab{1 2 3 4}
\\[1.5mm]
& & \TTab{1 \,, \,}
&
\begin{adjustbox}{valign=t}
$
\begin{bmatrix}
1
\end{bmatrix}
$
\end{adjustbox}
&
1
&
\TTab{1 \,, \, 4}
\\[2.5mm]
& 1 & \TTab{1 2 3}
&
\begin{adjustbox}{valign=t}
$
\begin{bmatrix}
1
\end{bmatrix}
$
\end{adjustbox}
&
1
&
\TTab{1 2 3, 4}
\\[2.5mm]
& & \TTab{1 \,, \,}
&
\begin{adjustbox}{valign=t}
$
\begin{bmatrix}
\tfrac{-\sqrt{5}}{\sqrt{8}} &  \tfrac{\sqrt{3}}{\sqrt{8}} \\[1mm]
\tfrac{\sqrt{3}}{\sqrt{8}} &  \tfrac{\sqrt{5}}{\sqrt{8}}
\end{bmatrix}
$
\end{adjustbox}
&
1,2
&
\TTab{1 \, 4, \,},\TTab{1 \,, \,, 4}
\\[8mm]
& & \TTab{1, 2, 3}
&
\begin{adjustbox}{valign=t}
$
\begin{bmatrix}
1
\end{bmatrix}
$
\end{adjustbox}
&
0
&
\TTab{1 4, 2, 3}
\\[5mm]
& 2 & \TTab{1 2 3}
&
\begin{adjustbox}{valign=t}
$
\begin{bmatrix}
\tfrac{\sqrt{7}}{\sqrt{10}} & \tfrac{\sqrt{3}}{\sqrt{10}} \\[1mm]
\tfrac{\sqrt{3}}{\sqrt{10}} & \tfrac{-\sqrt{7}}{\sqrt{10}}
\end{bmatrix}
$
\end{adjustbox}
&
1,3
&
\multirow{3}{\hsize}{\TTab{1 2 3 4},\\[0.5mm] \TTab{1 2 3, 4}}
\\[8mm]
& & \TTab{1 \,, \,}
&
\begin{adjustbox}{valign=t}
$
\begin{bmatrix}
\tfrac{\sqrt{3}}{2} & \tfrac{1}{2} \\[1mm]
\tfrac{-1}{2} & \tfrac{\sqrt{3}}{2}
\end{bmatrix}
$
\end{adjustbox}
&
1,2
&
\TTab{1 \, 4, \,},\TTab{1 \,, \, 4}
\\[8mm]
& 3 & \TTab{1 2 3}
&
\begin{adjustbox}{valign=t}
$
\begin{bmatrix}
1
\end{bmatrix}
$
\end{adjustbox}
&
3
&
\TTab{1 2 3, 4}
\\[2.5mm]
& & \TTab{1 \,, \,}
&
\begin{adjustbox}{valign=t}
$
\begin{bmatrix}
1
\end{bmatrix}
$
\end{adjustbox}
&
2
&
\TTab{1 \, 4, \,}
\\[2.5mm]
& 3 & \TTab{1 2 3}
&
\begin{adjustbox}{valign=t}
$
\begin{bmatrix}
1
\end{bmatrix}
$
\end{adjustbox}
&
3
&
\TTab{1 2 3 4}
\\[3.5mm]
5 & 0 & \TTab{1 \, \,, \,}
&
\begin{adjustbox}{valign=t}
$
\begin{bmatrix}
1
\end{bmatrix}
$
\end{adjustbox}
&
1
&
\TTab{1 \, \,, \,, 5}
\\[5mm]
& & \TTab{1 \,, \,, \,}
&
\begin{adjustbox}{valign=t}
$
\begin{bmatrix}
1
\end{bmatrix}
$
\end{adjustbox}
&
1
&
\TTab{1 \, 5, \,, \,}
\\[5mm]
& 1 & \TTab{1 2 3 4}
&
\begin{adjustbox}{valign=t}
$
\begin{bmatrix}
\tfrac{\sqrt{7}}{\sqrt{15}} & \tfrac{\sqrt{8}}{\sqrt{15}} \\[1mm]
\tfrac{\sqrt{8}}{\sqrt{15}} & \tfrac{-\sqrt{7}}{\sqrt{15}}
\end{bmatrix}
$
\end{adjustbox}
&
0,2
&
\multirow{3}{\hsize}{\TTab{1 2 3 4 5},\\[0.5mm] \TTab{1 2 3 4, 5}}
\\[8mm]
& & \TTab{1 \, \,, \,}
&
\begin{adjustbox}{valign=t}
$
\begin{bmatrix}
\tfrac{\sqrt{2}}{\sqrt{3}} & \tfrac{1}{\sqrt{3}} \\[1mm]
\tfrac{1}{\sqrt{3}} & \tfrac{-\sqrt{2}}{\sqrt{3}}
\end{bmatrix}
$
\end{adjustbox}
&
1,2
&
\TTab{1 \, \, 5, \,},\TTab{1 \, \,, \, 5}
\\[8mm]
& & \TTab{1 \,, \, 4}
&
\begin{adjustbox}{valign=t}
$
\begin{bmatrix}
\tfrac{-\sqrt{5}}{\sqrt{6}} & \tfrac{1}{\sqrt{6}} \\[1mm]
\tfrac{1}{\sqrt{6}} & \tfrac{\sqrt{5}}{\sqrt{6}}
\end{bmatrix}
$
\end{adjustbox}
&
0,2
&
\TTab{1 \, 5, \, 4},\TTab{1 \,, \, 4, 5}
\\[8mm]
& & \TTab{1 \,, \,, \,}
&
\begin{adjustbox}{valign=t}
$
\begin{bmatrix}
1
\end{bmatrix}
$
\end{adjustbox}
&
1
&
\TTab{1 \,, \, 5, \,}
\\[5mm]
\end{tabular}
&
\begin{tabular}[t]{@{\hspace{1mm}} l @{\hspace{2mm}} l @{\hspace{2mm}} l
@{\hspace{1mm}} l @{\hspace{1mm}} l @{\hspace{2mm}} l @{\hspace{-14.8cm}} @{\hspace{1mm}}}
& &   & Transf.\ & Input & Output
\\
$g$ & $j^{[g]}$ & $\tau^{[g-1]}$  
& matrix & $j^{[g-1]}$ &  $\tau^{[g]}$ and $\mathcal{A}$
\\[1mm]  \hline & & & & &  \\[-2mm]  
5 & 2 & \TTab{1 2 3 4}
&
\begin{adjustbox}{valign=t}
$
\begin{bmatrix}
1
\end{bmatrix}
$
\end{adjustbox}
&
2
&
\TTab{1 2 3 4, 5}
\\[2.5mm]
& & \TTab{1 \, \,, \,}
&
\begin{adjustbox}{valign=t}
$
\begin{bmatrix}
\tfrac{\sqrt{126}}{15} &  \tfrac{-\sqrt{35}}{15} &  \tfrac{\sqrt{64}}{15} \\[1mm]
\tfrac{\sqrt{18}}{\sqrt{45}} &  \tfrac{\sqrt{20}}{\sqrt{45}} &  \tfrac{-\sqrt{7}}{\sqrt{45}} \\[1mm] 
\tfrac{-1}{5} &  \tfrac{\sqrt{10}}{5} &  \tfrac{\sqrt{14}}{5}
\end{bmatrix}
$
\end{adjustbox}
&
1,2,3
&
\multirow{4}{\hsize}{\TTab{1 \, \, 5, \,},\TTab{1 \, \,, \, 5},\\[0.5mm] \TTab{1 \, \,, \,, 5}}
\\[12mm]
& & \TTab{1 \,, \, 4}
&
\begin{adjustbox}{valign=t}
$
\begin{bmatrix}
1
\end{bmatrix}
$
\end{adjustbox}
&
2
&
\TTab{1 \, 5, \, 4}
\\[2.5mm]
& & \TTab{1 \,, \,, \,}
&
\begin{adjustbox}{valign=t}
$
\begin{bmatrix}
1
\end{bmatrix}
$
\end{adjustbox}
&
1
&
\TTab{1 \, 5, \,, \,}
\\[5mm]
& 3 & \TTab{1 2 3 4}
&
\begin{adjustbox}{valign=t}
$
\begin{bmatrix}
\tfrac{\sqrt{27}}{\sqrt{35}} & \tfrac{\sqrt{8}}{\sqrt{35}} \\[1mm]
\tfrac{\sqrt{8}}{\sqrt{35}} & \tfrac{-\sqrt{27}}{\sqrt{35}}
\end{bmatrix}
$
\end{adjustbox}
&
2,4
&
\multirow{3}{\hsize}{\TTab{1 2 3 4 5},\\[0.5mm] \TTab{1 2 3 4, 5}}
\\[8mm]
& & \TTab{1 \, \,, \,}
&
\begin{adjustbox}{valign=t}
$
\begin{bmatrix}
\tfrac{\sqrt{8}}{3} & \tfrac{1}{3} \\[1mm]
\tfrac{1}{3} & \tfrac{-\sqrt{8}}{3}
\end{bmatrix}
$
\end{adjustbox}
&
2,3
&
\TTab{1 \, \, 5, \,},\TTab{1 \, \,, \, 5}
\\[8mm]
& & \TTab{1 \,, \, 4}
&
\begin{adjustbox}{valign=t}
$
\begin{bmatrix}
-1
\end{bmatrix}
$
\end{adjustbox}
&
2
&
\TTab{1 \, 5, \, 4}
\\[2.5mm]
& 4 & \TTab{1 2 3 4}
&
\begin{adjustbox}{valign=t}
$
\begin{bmatrix}
1
\end{bmatrix}
$
\end{adjustbox}
&
4
&
\TTab{1 2 3 4, 5}
\\[2.5mm]
& & \TTab{1 \, \,, \,}
&
\begin{adjustbox}{valign=t}
$
\begin{bmatrix}
1
\end{bmatrix}
$
\end{adjustbox}
&
3
&
\TTab{1 \, \, 5, \,}
\\[2.5mm]
& 5 & \TTab{1 2 3 4}
&
\begin{adjustbox}{valign=t}
$
\begin{bmatrix}
1
\end{bmatrix}
$
\end{adjustbox}
&
4
&
\TTab{1 2 3 4 5}
\\[3.5mm]
6 & 0 & \TTab{1 2 3 4 5}
&
\begin{adjustbox}{valign=t}
$
\begin{bmatrix}
1
\end{bmatrix}
$
\end{adjustbox}
&
1
&
\TTab{1 2 3 4 5 6}
\\[1.5mm]
& & \TTab{1 \, \, \,, \,}
&
\begin{adjustbox}{valign=t}
$
\begin{bmatrix}
1
\end{bmatrix}
$
\end{adjustbox}
&
1
&
\TTab{1 \, \, \,, \, 6}
\\[2.5mm]
& & \TTab{1 \, \,, \, \,}
&
\begin{adjustbox}{valign=t}
$
\begin{bmatrix}
-1
\end{bmatrix}
$
\end{adjustbox}
&
1
&
\TTab{1 \, \, 6, \, \,}
\\[2.5mm]
& & \TTab{1 \,, \, \,, \,}
&
\begin{adjustbox}{valign=t}
$
\begin{bmatrix}
1
\end{bmatrix}
$
\end{adjustbox}
&
1
&
\TTab{1 \,, \, \,, \, 6}
\\[5mm]
& 1 & \TTab{1 2 3 4 5}
&
\begin{adjustbox}{valign=t}
$
\begin{bmatrix}
1
\end{bmatrix}
$
\end{adjustbox}
&
1
&
\TTab{1 2 3 4 5, 6}
\\[2.5mm]
& & \TTab{1 \, \, \,, \,}
&
\begin{adjustbox}{valign=t}
$
\begin{bmatrix}
\tfrac{\sqrt{7}}{\sqrt{12}} & \tfrac{-\sqrt{5}}{\sqrt{12}} \\[1mm]
\tfrac{\sqrt{5}}{\sqrt{12}} & \tfrac{\sqrt{7}}{\sqrt{12}} 
\end{bmatrix}
$
\end{adjustbox}
&
1,2
&
\TTab{1 \, \, \, 6, \,},\TTab{1 \, \, \,, \,, 6}
\\[8mm]
& & \TTab{1 \, \,, \, \,}
&
\begin{adjustbox}{valign=t}
$
\begin{bmatrix}
\tfrac{\sqrt{2}}{\sqrt{3}} & \tfrac{-1}{\sqrt{3}} \\[1mm]
\tfrac{1}{\sqrt{3}} & \tfrac{\sqrt{2}}{\sqrt{3}} 
\end{bmatrix}
$
\end{adjustbox}
&
1,2
&
\TTab{1 \, \,, \, \, 6},\TTab{1 \, \,, \, \,, 6}
\\[8mm]
& & \TTab{1 \, \,, \,, \,}
&
\begin{adjustbox}{valign=t}
$
\begin{bmatrix}
\tfrac{\sqrt{5}}{3} & \tfrac{2}{3} \\[1mm]
\tfrac{2}{3} & \tfrac{-\sqrt{5}}{3} 
\end{bmatrix}
$
\end{adjustbox}
&
0,2
&
\TTab{1 \, \, 6, \,, \,},\TTab{1 \, \,, \, 6, \,}
\\[8mm]
& & \TTab{1 \,, \, \,, \,}
&
\begin{adjustbox}{valign=t}
$
\begin{bmatrix}
1
\end{bmatrix}
$
\end{adjustbox}
&
1
&
\TTab{1 \, 6, \, \,, \,}
\\[5mm]
6 & 2 & \TTab{1 2 3 4 5}
&
\begin{adjustbox}{valign=t}
$
\begin{bmatrix}
\tfrac{\sqrt{15}}{5} & \tfrac{\sqrt{10}}{5} \\[1mm]
\tfrac{\sqrt{10}}{5} & \tfrac{-\sqrt{15}}{5} 
\end{bmatrix}
$
\end{adjustbox}
&
1,3
&
\TTab{1 2 3 4 5 6},\TTab{1 2 3 4 5, 6}
\\[8mm]
& & \TTab{1 \, \, \,, \,}
&
\begin{adjustbox}{valign=t}
$
\begin{bmatrix}
\tfrac{\sqrt{63}}{\sqrt{120}} & \tfrac{\sqrt{25}}{\sqrt{120}} & \tfrac{\sqrt{32}}{\sqrt{120}} \\[1mm]
\tfrac{-\sqrt{9}}{\sqrt{240}} & \tfrac{\sqrt{175}}{\sqrt{240}} & \tfrac{-\sqrt{56}}{\sqrt{240}} \\[1mm]
\tfrac{\sqrt{7}}{4} & \tfrac{-1}{4} & \tfrac{-\sqrt{8}}{4} 
\end{bmatrix}
$
\end{adjustbox}
&
1,2,3
&
\multirow{3.1}{\hsize}{\TTab{1 \, \, \, 6, \,},\TTab{1 \, \, \,, \, 6}\, $I$,\\[0.5mm] \TTab{1 \, \, \,, \, 6}\, $II$}
\\[12mm]
& & \TTab{1 \, \,, \, \,}
&
\begin{adjustbox}{valign=t}
$
\begin{bmatrix}
\tfrac{\sqrt{18}}{\sqrt{30}} & \tfrac{-\sqrt{5}}{\sqrt{30}} & \tfrac{-\sqrt{7}}{\sqrt{30}} \\[1mm]
\tfrac{\sqrt{14}}{\sqrt{50}} & \tfrac{\sqrt{35}}{\sqrt{50}} & \tfrac{1}{\sqrt{50}} \\[1mm]
\tfrac{-\sqrt{9}}{\sqrt{75}} & \tfrac{\sqrt{10}}{\sqrt{75}} & \tfrac{-\sqrt{56}}{\sqrt{75}} 
\end{bmatrix}
$
\end{adjustbox}
&
1,2,3
&
\multirow{3.8}{\hsize}{\TTab{1 \, \, 6, \, \,}\, $I$,\TTab{1 \, \, 6, \, \,}\, $II$,\\[0.5mm] \TTab{1 \, \,, \, \,, 6}}
\\[13mm]
\end{tabular}
\\
\hline \hline  \\[-2mm]
\end{tabular}
\ytableausetup{smalltableaux,centertableaux}
\end{table}
\clearpage
\begin{table}
\ytableausetup{smalltableaux,aligntableaux=top}
\caption{Explicit values of fractional parentage coefficients absent in Table~\ref{tab_cfp_a}.
Empty boxes in the standard Young tableaux $\tau^{[g-1]}$
have to be filled with all possible values as detailed in Table~\ref{tab:tableaux}.
\label{tab_cfp_b}}
\begin{tabular}{@{\hspace{2mm}} c @{\hspace{4mm}} c @{\hspace{2mm}}}
\\[-2mm]
\hline\hline
\\[-2mm]
\begin{tabular}[t]{@{\hspace{2mm}} l @{\hspace{2mm}} l @{\hspace{2mm}} l
@{\hspace{1mm}} l @{\hspace{1mm}} l @{\hspace{2mm}} l @{\hspace{-15.4cm}} @{\hspace{1mm}}} 
& &   & Transf.\ & Input & Output
\\
$g$ & $j^{[g]}$ & $\tau^{[g-1]}$  
& matrix & $j^{[g-1]}$ &  $\tau^{[g]}$ and $\mathcal{A}$
\\[1mm]  \hline & & & & & \\[-2mm]  
6 & 2 & \TTab{1 \, \,, \,, \,}
&
\begin{adjustbox}{valign=t}
$
\begin{bmatrix}
1
\end{bmatrix}
$
\end{adjustbox}
&
2
&
\TTab{1 \, \,, \, 6, \,}
\\[5mm]
& & \TTab{1 \,, \, \,, \,}
&
\begin{adjustbox}{valign=t}
$
\begin{bmatrix}
1
\end{bmatrix}
$
\end{adjustbox}
&
1
&
\TTab{1 \, 6, \, \,, \,}
\\[5mm]
& 3 & \TTab{1 2 3 4 5}
&
\begin{adjustbox}{valign=t}
$
\begin{bmatrix}
1
\end{bmatrix}
$
\end{adjustbox}
&
3
&
\TTab{1 2 3 4 5, 6}
\\[2.5mm]
& & \TTab{1 \, \, \,, \,}
&
\begin{adjustbox}{valign=t}
$
\begin{bmatrix}
\tfrac{-\sqrt{80}}{\sqrt{112}} & \tfrac{\sqrt{7}}{\sqrt{112}} & \tfrac{-\sqrt{25}}{\sqrt{112}} \\[1mm]
\tfrac{\sqrt{80}}{\sqrt{336}} & \tfrac{\sqrt{175}}{\sqrt{336}} & \tfrac{-\sqrt{81}}{\sqrt{336}} \\[1mm]
\tfrac{-\sqrt{4}}{\sqrt{84}} & \tfrac{\sqrt{35}}{\sqrt{84}} & \tfrac{\sqrt{45}}{\sqrt{84}} 
\end{bmatrix}
$
\end{adjustbox}
&
2,3,4
&
\multirow{3.8}{\hsize}{\TTab{1 \, \, \, 6, \,},\TTab{1 \, \, \,, \, 6},\\[0.5mm] \TTab{1 \, \, \,, \,, 6}}
\\[13mm]
& & \TTab{1 \, \,, \, \,}
&
\begin{adjustbox}{valign=t}
$
\begin{bmatrix}
\tfrac{\sqrt{2}}{\sqrt{3}} & \tfrac{-1}{\sqrt{3}}  \\[1mm]
\tfrac{1}{\sqrt{3}} & \tfrac{\sqrt{2}}{\sqrt{3}} 
\end{bmatrix}
$
\end{adjustbox}
&
2,3
&
\TTab{1 \, \, 6, \, \,},\TTab{1 \, \,, \, \, 6}
\\[8mm]
& & \TTab{1 \, \,, \,, \,}
&
\begin{adjustbox}{valign=t}
$
\begin{bmatrix}
1
\end{bmatrix}
$
\end{adjustbox}
&
2
&
\TTab{1 \, \, 6, \,, \,}
\\[5mm]
\end{tabular} 
&
\begin{tabular}[t]{@{\hspace{2mm}} l @{\hspace{2mm}} l @{\hspace{2mm}} l
@{\hspace{1mm}} l @{\hspace{1mm}} l @{\hspace{2mm}} l @{\hspace{-16.3cm}} @{\hspace{1mm}}} 
& &   & Transf.\ & Input & Output
\\
$g$ & $j^{[g]}$ & $\tau^{[g-1]}$  
& matrix & $j^{[g-1]}$ &  $\tau^{[g]}$ and $\mathcal{A}$
\\[1mm]  \hline & & & & & \\[-2mm]  
6 & 4 & \TTab{1 2 3 4 5}
&
\begin{adjustbox}{valign=t}
$
\begin{bmatrix}
\tfrac{\sqrt{22}}{\sqrt{27}} & \tfrac{\sqrt{5}}{\sqrt{27}} \\[1mm]
\tfrac{\sqrt{5}}{\sqrt{27}} & \tfrac{-\sqrt{22}}{\sqrt{27}} 
\end{bmatrix}
$
\end{adjustbox}
&
3,5
&
\multirow{3}{\hsize}{\TTab{1 2 3 4 5 6},\\[0.5mm] \TTab{1 2 3 4 5, 6}}
\\[8mm]
& & \TTab{1 \, \, \,, \,}
&
\begin{adjustbox}{valign=t}
$
\begin{bmatrix}
\tfrac{\sqrt{15}}{4} & \tfrac{1}{4} \\[1mm]
\tfrac{1}{4} & \tfrac{-\sqrt{15}}{4} 
\end{bmatrix}
$
\end{adjustbox}
&
3,4
&
\multirow{3.5}{\hsize}{\TTab{1 \, \, \, 6, \,},\\[0.5mm] \TTab{1 \, \, \,, \, 6}}
\\[9.5mm]
& & \TTab{1 \, \,, \, \,}
&
\begin{adjustbox}{valign=t}
$
\begin{bmatrix}
1
\end{bmatrix}
$
\end{adjustbox}
&
3
&
\TTab{1 \, \, 6, \, \,}
\\[5mm]
& 5 & \TTab{1 2 3 4 5}
&
\begin{adjustbox}{valign=t}
$
\begin{bmatrix}
1
\end{bmatrix}
$
\end{adjustbox}
&
5
&
\TTab{1 2 3 4 5, 6}
\\[2.5mm]
& & \TTab{1 \, \, \,, \,}
&
\begin{adjustbox}{valign=t}
$
\begin{bmatrix}
1
\end{bmatrix}
$
\end{adjustbox}
&
4
&
\TTab{1 \, \, \, 6, \,}
\\[2.5mm]
& 6 & \TTab{1 2 3 4 5}
&
\begin{adjustbox}{valign=t}
$
\begin{bmatrix}
1
\end{bmatrix}
$
\end{adjustbox}
&
5
&
\TTab{1 2 3 4 5 6}
\\[2.5mm]
\end{tabular} 
\\
\hline \hline  \\[-2mm]
\end{tabular}
\ytableausetup{smalltableaux,centertableaux}
\end{table}
}

Last, we show the visualization of the dynamics of a polarization transfer from spin one to spin two in 
a system consisting of four coupled spins $1/2$ under isotropic mixing conditions.\cite{Luy20001999}
Isotropic mixing is one of the most important
methods to transfer polarization in high-resolution
NMR spectroscopy and is frequently used in homonuclear and heteronuclear
experiments to maximize polarization transfer.
Its efficiency depends extremely on the mixing time duration.
For  four coupled spins $1/2$, the ideal
isotropic mixing Hamiltonian has the form
$H=2 \pi \sum_{i<j}^4 \mathrm{J}_{ij}(I_{ix}I_{jx}+I_{iy}I_{jy}+I_{iz}I_{jz})$.
For the model system consisting of the $^1$H nuclear spins of trans-phenylcyclopropane carboxylic acid, 
the coupling constants are given by
$\mathrm{J}_{12}=4.1$ Hz, $\mathrm{J}_{13}=9.4$ Hz, $\mathrm{J}_{14}=6.8$ Hz, 
$\mathrm{J}_{23}=5.3$ Hz, $\mathrm{J}_{24}=8.2$ Hz, and $\mathrm{J}_{34}=-4.6$ Hz. 
Starting with the initial density density operator $\rho(0)=I_{1z}$,
Fig.~\ref{fig:exps_spins1d2}(c1)-(c4) shows the DROPS representation of the states for different mixing 
times: (c1) 0 ms, (c2) 20 ms, (c3) 40 ms, and (c4) 133 ms. 
Again, nodes represent the particles and edges their couplings.
Note that for simplicity, here we only plotted 
the linear and bilinear tensor components.
During the course of the experiment, the free evolution under isotropic mixing conditions results in the generation of coherences, which is reflected by the occurrence of non-vanishing bilinear tensors ($g=2$) and visualized by droplets located 
on the edges. Also small amounts of polarization occurs on the other spins depicted by
the droplet functions on these nodes. After 133 ms [panel (c4)], almost all polarization has been transferred from spin one to spin two.

\subsection{Representing systems consisting of five and six coupled spins $1/2$ systems\label{sec:app_randomMat}}
We also show the droplet functions for a (complex) random matrix 
$A \in \mathbb{C}^{32 \times 32}$ for a five-spin-$1/2$ system. Although such systems are quite complex, with our approach, we can
conveniently partition the information in different subsystems given by the panels
and subpanels in Fig.~\ref{fig:5spin_1d2}. 
We find one zero-linear subsystem $G = \emptyset$ with one droplet, five linear subsystems with 
one droplet in each $G \in \{\{1\},\{2\},\{3\},\{4\},\{5\}\}$, and ten bilinear subsystems 
with also one droplet in each $G \in \{\{1,\!2\},\{1,\!3\},\{1,\!4\},\{1,\!5\},\{2,\!3\},\{2,\!4\},\{2,\!5\},\{3,\!4\},\{3,\!5\},\{4,\!5\}\}$. 
They can be plotted together as shown in the upper left panel of Fig.~\ref{fig:5spin_1d2}, 
where the droplets visualizing the linear subsystems are plotted on the corresponding 
nodes representing the spins and the droplet functions for the bilinear subsystems are placed on the edges between two spins.
The identity part (Id) is placed beneath this scheme.
In the upper right panel, the
ten trilinear subsystems  
$G \in \{\{1,\!2,\!3\},\{1,\!2,\!4\},\{1,\!2,\!5\},\{1,\!3,\!4\},\{1,\!3,\!5\},\{1,\!4,\!5\},\{2,\!3,\!4\},\{2,\!3,\!5\},\{2,\!4,\!5\},\{3,\!4,\!5\}\}$, each represented by four droplets.
The topology of the involved spins of each subsystem are symbolized by the sketch at the top of each subpanel.  
The bottom left panel shows the five four-linear subsystems 
with nine droplets for each $G \in \{\{1,\!2,\!3,\!4\},\{1,\!2,\!3,\!5\},\{1,\!2,\!4,\!5\},\{1,\!3,\!4,\!5\},\{2,\!3,\!4,\!5\}\}$.
Again, the subsystem is graphically given at the beginning of each subpanel. 
Finally, the $5$-linear subsystem with 21 droplets is illustrated in the bottom right panel of Fig.~\ref{fig:5spin_1d2}. In total we 
have 122 droplet functions uniquely representing the matrix $A$. Note that
 we omit the superscript $[g]$ for $\tau^{[G]}$ in this figure, since $g=|G|$ holds and $G$ is clear from the context.

We conclude this section by visualizing a (complex) random matrix 
$A \in \mathbb{C}^{64 \times 64}$ for a six coupled spins $1/2$ as given in Fig.~\ref{fig:6spin_1d2}.
The information can be analogously partitioned and presented as given in Fig.~\ref{fig:5spin_1d2}.
In the upper left panel, the topology of the system is sketched. The
zero-linear subsystem containing 
one droplet can be located beneath the scheme. The droplet functions representing
the six linear subsystems 
$G \in \{ \{1\},\{2\},\{3\},\{4\},\{5\},\{6\} \}$ are plotted on the nodes, the droplet functions of the fifteen bilinear subsystems 
with also one droplet in each 
$G \in \binom{\{1,\ldots,6\}}{2}$
are plotted on the
edges. The droplet functions of the six-linear system 
$G=\{1,\!2,\!3,\!4,\!5,\!6\}$ are shown in the upper right panel. 
The panel in the center presents the twenty trilinear subsystems
$G \in \binom{\{1,\ldots,6\}}{3}$
with the related droplet functions. The droplets of the 5-linear systems 
$G \in \binom{\{1,\ldots,6\}}{5}$
are plotted in the bottom panels of Fig.~\ref{fig:6spin_1d2}. Finally, the
4-linear subsystems $G \in \binom{\{1,\ldots,6\}}{4}$
with the nine corresponding droplets for each $G$ are given in Fig.~\ref{fig:6spin_1d2_B}.
In total we find 423 droplets, which uniquely represents 
the information contained in the $64^2=4096$ complex matrix elements of
such an operator. 
Again,
 we omit the superscript $[g]$ for $\tau$ in this figure and we use the additional {\it ad hoc} sublabels only when required, i.e., for $g=6$ and $\tau_i^{[6]}$ with $i \in \{7,\dots ,15\}$.
The standard Young tableaux for
$g=5$ and $g=6$ are 
summarized in Table~\ref{tab:tableaux}.

\section{Explicit values of the fractional parentage coefficients\label{CFP_app}}
The explicit values of the fractional parentage coefficients for up to $g=6$ 
are given in 
Tables~\ref{tab_cfp_a} and \ref{tab_cfp_a}. We have used the fractional parentage 
coefficients as defined in Ref.~\onlinecite{JvW51}, and we have \emph{not} applied the phase amendments
from the footnote on p.~241 of Ref.~\onlinecite{EHJ53}.


\begin{thebibliography}{129}%
\makeatletter
\providecommand \@ifxundefined [1]{%
 \@ifx{#1\undefined}
}%
\providecommand \@ifnum [1]{%
 \ifnum #1\expandafter \@firstoftwo
 \else \expandafter \@secondoftwo
 \fi
}%
\providecommand \@ifx [1]{%
 \ifx #1\expandafter \@firstoftwo
 \else \expandafter \@secondoftwo
 \fi
}%
\providecommand \natexlab [1]{#1}%
\providecommand \enquote  [1]{``#1''}%
\providecommand \bibnamefont  [1]{#1}%
\providecommand \bibfnamefont [1]{#1}%
\providecommand \citenamefont [1]{#1}%
\providecommand \href@noop [0]{\@secondoftwo}%
\providecommand \href [0]{\begingroup \@sanitize@url \@href}%
\providecommand \@href[1]{\@@startlink{#1}\@@href}%
\providecommand \@@href[1]{\endgroup#1\@@endlink}%
\providecommand \@sanitize@url [0]{\catcode `\\12\catcode `\$12\catcode
  `\&12\catcode `\#12\catcode `\^12\catcode `\_12\catcode `\%12\relax}%
\providecommand \@@startlink[1]{}%
\providecommand \@@endlink[0]{}%
\providecommand \url  [0]{\begingroup\@sanitize@url \@url }%
\providecommand \@url [1]{\endgroup\@href {#1}{\urlprefix }}%
\providecommand \urlprefix  [0]{URL }%
\providecommand \Eprint [0]{\href }%
\providecommand \doibase [0]{http://dx.doi.org/}%
\providecommand \selectlanguage [0]{\@gobble}%
\providecommand \bibinfo  [0]{\@secondoftwo}%
\providecommand \bibfield  [0]{\@secondoftwo}%
\providecommand \translation [1]{[#1]}%
\providecommand \BibitemOpen [0]{}%
\providecommand \bibitemStop [0]{}%
\providecommand \bibitemNoStop [0]{.\EOS\space}%
\providecommand \EOS [0]{\spacefactor3000\relax}%
\providecommand \BibitemShut  [1]{\csname bibitem#1\endcsname}%
\let\auto@bib@innerbib\@empty
\bibitem [{\citenamefont {Feynman}, \citenamefont {Vernon},\ and\ \citenamefont
  {Hellwarth}(1957)}]{Feynman_Vernon_57}%
  \BibitemOpen
  \bibfield  {author} {\bibinfo {author} {\bibfnamefont {R.~P.}\ \bibnamefont
  {Feynman}}, \bibinfo {author} {\bibfnamefont {F.~L.}\ \bibnamefont {Vernon},
  \bibfnamefont {Jr.}}, \ and\ \bibinfo {author} {\bibfnamefont {R.~W.}\
  \bibnamefont {Hellwarth}},\ }\bibfield  {title} {\enquote {\bibinfo {title}
  {{Geometrical Representation of the Schr{\"o}dinger Equation for Solving
  Maser Problems}},}\ }\href@noop {} {\bibfield  {journal} {\bibinfo  {journal}
  {J. Appl. Phys.}\ }\textbf {\bibinfo {volume} {28}},\ \bibinfo {pages}
  {49--52} (\bibinfo {year} {1957})}\BibitemShut {NoStop}%
\bibitem [{\citenamefont {Bernstein}, \citenamefont {King},\ and\ \citenamefont
  {Zhou}(2004)}]{handbook_04}%
  \BibitemOpen
  \bibfield  {author} {\bibinfo {author} {\bibfnamefont {M.~A.}\ \bibnamefont
  {Bernstein}}, \bibinfo {author} {\bibfnamefont {K.~F.}\ \bibnamefont {King}},
  \ and\ \bibinfo {author} {\bibfnamefont {X.~J.}\ \bibnamefont {Zhou}},\
  }\href@noop {} {\emph {\bibinfo {title} {{Handbook of MRI Pulse
  Sequences}}}}\ (\bibinfo  {publisher} {Elsevier, Burlington-San
  Diego-London},\ \bibinfo {year} {2004})\BibitemShut {NoStop}%
\bibitem [{\citenamefont {Ernst}, \citenamefont {Bodenhausen},\ and\
  \citenamefont {Wokaun}(1987)}]{EBW87}%
  \BibitemOpen
  \bibfield  {author} {\bibinfo {author} {\bibfnamefont {R.~R.}\ \bibnamefont
  {Ernst}}, \bibinfo {author} {\bibfnamefont {G.}~\bibnamefont {Bodenhausen}},
  \ and\ \bibinfo {author} {\bibfnamefont {A.}~\bibnamefont {Wokaun}},\
  }\href@noop {} {\emph {\bibinfo {title} {{Principles of Nuclear Magnetic
  Resonance in One and Two Dimensions}}}}\ (\bibinfo  {publisher} {Clarendon
  Press, Oxford},\ \bibinfo {year} {1987})\BibitemShut {NoStop}%
\bibitem [{\citenamefont {Schleich}(2001)}]{SchleichBook}%
  \BibitemOpen
  \bibfield  {author} {\bibinfo {author} {\bibfnamefont {W.~P.}\ \bibnamefont
  {Schleich}},\ }\href@noop {} {\emph {\bibinfo {title} {{Quantum Optics in
  Phase Space}}}}\ (\bibinfo  {publisher} {Wiley-VCH, Weinheim},\ \bibinfo
  {year} {2001})\BibitemShut {NoStop}%
\bibitem [{\citenamefont {Nielsen}\ and\ \citenamefont {Chuang}(2000)}]{NC00}%
  \BibitemOpen
  \bibfield  {author} {\bibinfo {author} {\bibfnamefont {M.~A.}\ \bibnamefont
  {Nielsen}}\ and\ \bibinfo {author} {\bibfnamefont {I.~L.}\ \bibnamefont
  {Chuang}},\ }\href@noop {} {\emph {\bibinfo {title} {{Quantum Computation and
  Quantum Information}}}}\ (\bibinfo  {publisher} {Cambridge University Press,
  Cambridge (UK)},\ \bibinfo {year} {2000})\BibitemShut {NoStop}%
\bibitem [{\citenamefont {S{\o}rensen}\ \emph {et~al.}(1983)\citenamefont
  {S{\o}rensen}, \citenamefont {Eich}, \citenamefont {Levitt}, \citenamefont
  {Bodenhausen},\ and\ \citenamefont {Ernst}}]{SEL:1983}%
  \BibitemOpen
  \bibfield  {author} {\bibinfo {author} {\bibfnamefont {O.~W.}\ \bibnamefont
  {S{\o}rensen}}, \bibinfo {author} {\bibfnamefont {G.~W.}\ \bibnamefont
  {Eich}}, \bibinfo {author} {\bibfnamefont {M.~H.}\ \bibnamefont {Levitt}},
  \bibinfo {author} {\bibfnamefont {G.}~\bibnamefont {Bodenhausen}}, \ and\
  \bibinfo {author} {\bibfnamefont {R.~R.}\ \bibnamefont {Ernst}},\ }\bibfield
  {title} {\enquote {\bibinfo {title} {Product operator formalism for the
  description of nmr pulse experiments},}\ }\href@noop {} {\bibfield  {journal}
  {\bibinfo  {journal} {Progr. NMR Spectrosc.}\ }\textbf {\bibinfo {volume}
  {16}},\ \bibinfo {pages} {163--192} (\bibinfo {year} {1983})}\BibitemShut
  {NoStop}%
\bibitem [{\citenamefont {Donne}\ and\ \citenamefont
  {Gorenstein}(1997)}]{Donne_Gorenstein}%
  \BibitemOpen
  \bibfield  {author} {\bibinfo {author} {\bibfnamefont {D.~G.}\ \bibnamefont
  {Donne}}\ and\ \bibinfo {author} {\bibfnamefont {D.~G.}\ \bibnamefont
  {Gorenstein}},\ }\bibfield  {title} {\enquote {\bibinfo {title} {{A Pictorial
  Representation of Product Operator Formalism: Nonclassical Vector Diagrams
  for Multidimensional NMR}},}\ }\href@noop {} {\bibfield  {journal} {\bibinfo
  {journal} {Concepts Magn. Reson.}\ }\textbf {\bibinfo {volume} {9}},\
  \bibinfo {pages} {95--111} (\bibinfo {year} {1997})}\BibitemShut {NoStop}%
\bibitem [{\citenamefont {Freeman}(1997)}]{Freeman97}%
  \BibitemOpen
  \bibfield  {author} {\bibinfo {author} {\bibfnamefont {R.}~\bibnamefont
  {Freeman}},\ }\href@noop {} {\emph {\bibinfo {title} {{A Handbook of Nuclear
  Magnetic Resonance}}}},\ \bibinfo {edition} {2nd}\ ed.\ (\bibinfo
  {publisher} {Addison Wesley Longman, Harlow},\ \bibinfo {year}
  {1997})\BibitemShut {NoStop}%
\bibitem [{\citenamefont {Curtright}, \citenamefont {Fairlie},\ and\
  \citenamefont {Zachos}(2014)}]{Curtright-review}%
  \BibitemOpen
  \bibfield  {author} {\bibinfo {author} {\bibfnamefont {T.~L.}\ \bibnamefont
  {Curtright}}, \bibinfo {author} {\bibfnamefont {D.~B.}\ \bibnamefont
  {Fairlie}}, \ and\ \bibinfo {author} {\bibfnamefont {C.~K.}\ \bibnamefont
  {Zachos}},\ }\href@noop {} {\emph {\bibinfo {title} {{A Concise Treatise on
  Quantum Mechanics in Phase Space}}}}\ (\bibinfo  {publisher} {World
  Scientific, Singapore},\ \bibinfo {year} {2014})\BibitemShut {NoStop}%
\bibitem [{\citenamefont {Zachos}, \citenamefont {Fairlie},\ and\ \citenamefont
  {Curtright}(2005)}]{Zachos2005}%
  \BibitemOpen
  \bibfield  {author} {\bibinfo {author} {\bibfnamefont {C.~K.}\ \bibnamefont
  {Zachos}}, \bibinfo {author} {\bibfnamefont {D.~B.}\ \bibnamefont {Fairlie}},
  \ and\ \bibinfo {author} {\bibfnamefont {T.~L.}\ \bibnamefont {Curtright}},\
  }\href@noop {} {\emph {\bibinfo {title} {Quantum Mechanics in Phase Space: An
  Overview with Selected Papers}}}\ (\bibinfo  {publisher} {World Scientific},\
  \bibinfo {address} {Singapore},\ \bibinfo {year} {2005})\BibitemShut
  {NoStop}%
\bibitem [{\citenamefont {Schroeck~Jr.}(2013)}]{Schroeck2013}%
  \BibitemOpen
  \bibfield  {author} {\bibinfo {author} {\bibfnamefont {F.~E.}\ \bibnamefont
  {Schroeck~Jr.}},\ }\href@noop {} {\emph {\bibinfo {title} {Quantum Mechanics
  on Phase Space}}}\ (\bibinfo  {publisher} {Springer},\ \bibinfo {address}
  {Dordrecht},\ \bibinfo {year} {2013})\BibitemShut {NoStop}%
\bibitem [{\citenamefont {Wigner}(1932)}]{Wig32}%
  \BibitemOpen
  \bibfield  {author} {\bibinfo {author} {\bibfnamefont {E.}~\bibnamefont
  {Wigner}},\ }\bibfield  {title} {\enquote {\bibinfo {title} {{On the Quantum
  Correction For Thermodynamic Equilibrium}},}\ }\href@noop {} {\bibfield
  {journal} {\bibinfo  {journal} {Phys. Rev.}\ }\textbf {\bibinfo {volume}
  {40}},\ \bibinfo {pages} {749--759} (\bibinfo {year} {1932})}\BibitemShut
  {NoStop}%
\bibitem [{\citenamefont {Smithey}\ \emph
  {et~al.}(1993{\natexlab{a}})\citenamefont {Smithey}, \citenamefont {Beck},
  \citenamefont {Raymer},\ and\ \citenamefont {Faridani}}]{smithey1}%
  \BibitemOpen
  \bibfield  {author} {\bibinfo {author} {\bibfnamefont {D.~T.}\ \bibnamefont
  {Smithey}}, \bibinfo {author} {\bibfnamefont {M.}~\bibnamefont {Beck}},
  \bibinfo {author} {\bibfnamefont {M.~G.}\ \bibnamefont {Raymer}}, \ and\
  \bibinfo {author} {\bibfnamefont {A.}~\bibnamefont {Faridani}},\ }\bibfield
  {title} {\enquote {\bibinfo {title} {Measurement of the {Wigner} distribution
  and the density matrix of a light mode using optical homodyne tomography:
  Application to squeezed states and the vacuum},}\ }\href@noop {} {\bibfield
  {journal} {\bibinfo  {journal} {Phys. Rev. Lett.}\ }\textbf {\bibinfo
  {volume} {70}},\ \bibinfo {pages} {1244--1247} (\bibinfo {year}
  {1993}{\natexlab{a}})}\BibitemShut {NoStop}%
\bibitem [{\citenamefont {Smithey}\ \emph
  {et~al.}(1993{\natexlab{b}})\citenamefont {Smithey}, \citenamefont {Beck},
  \citenamefont {Cooper}, \citenamefont {Raymer},\ and\ \citenamefont
  {Faridani}}]{smithey2}%
  \BibitemOpen
  \bibfield  {author} {\bibinfo {author} {\bibfnamefont {D.~T.}\ \bibnamefont
  {Smithey}}, \bibinfo {author} {\bibfnamefont {M.}~\bibnamefont {Beck}},
  \bibinfo {author} {\bibfnamefont {J.}~\bibnamefont {Cooper}}, \bibinfo
  {author} {\bibfnamefont {M.~G.}\ \bibnamefont {Raymer}}, \ and\ \bibinfo
  {author} {\bibfnamefont {A.}~\bibnamefont {Faridani}},\ }\bibfield  {title}
  {\enquote {\bibinfo {title} {Complete experimental characterization of the
  quantum state of a light mode via the wigner function and the density matrix:
  application to quantum phase distributions of vacuum and squeezed-vacuum
  states},}\ }\href@noop {} {\bibfield  {journal} {\bibinfo  {journal} {Phys.
  Scripta}\ }\textbf {\bibinfo {volume} {1993}},\ \bibinfo {pages} {35}
  (\bibinfo {year} {1993}{\natexlab{b}})}\BibitemShut {NoStop}%
\bibitem [{\citenamefont {Smithey}\ \emph
  {et~al.}(1993{\natexlab{c}})\citenamefont {Smithey}, \citenamefont {Beck},
  \citenamefont {Cooper},\ and\ \citenamefont {Raymer}}]{smithey3}%
  \BibitemOpen
  \bibfield  {author} {\bibinfo {author} {\bibfnamefont {D.~T.}\ \bibnamefont
  {Smithey}}, \bibinfo {author} {\bibfnamefont {M.}~\bibnamefont {Beck}},
  \bibinfo {author} {\bibfnamefont {J.}~\bibnamefont {Cooper}}, \ and\ \bibinfo
  {author} {\bibfnamefont {M.~G.}\ \bibnamefont {Raymer}},\ }\bibfield  {title}
  {\enquote {\bibinfo {title} {Measurement of number-phase uncertainty
  relations of optical fields},}\ }\href@noop {} {\bibfield  {journal}
  {\bibinfo  {journal} {Phys. Rev. A}\ }\textbf {\bibinfo {volume} {48}},\
  \bibinfo {pages} {3159--3167} (\bibinfo {year}
  {1993}{\natexlab{c}})}\BibitemShut {NoStop}%
\bibitem [{\citenamefont {Leonhardt}(1997)}]{leonhardt}%
  \BibitemOpen
  \bibfield  {author} {\bibinfo {author} {\bibfnamefont {U.}~\bibnamefont
  {Leonhardt}},\ }\href@noop {} {\emph {\bibinfo {title} {Measuring the Quantum
  State of Light}}}\ (\bibinfo  {publisher} {Cambridge University Press,
  Cambridge},\ \bibinfo {year} {1997})\BibitemShut {NoStop}%
\bibitem [{\citenamefont {Paris}\ and\ \citenamefont
  {Rehacek}(2004)}]{paris_rehacek}%
  \BibitemOpen
  \bibinfo {editor} {\bibfnamefont {M.}~\bibnamefont {Paris}}\ and\ \bibinfo
  {editor} {\bibfnamefont {J.}~\bibnamefont {Rehacek}},\ eds.,\ \href@noop {}
  {\emph {\bibinfo {title} {Quantum State Estimation}}}\ (\bibinfo  {publisher}
  {Springer, Berlin},\ \bibinfo {year} {2004})\BibitemShut {NoStop}%
\bibitem [{\citenamefont {Wooters}(1987)}]{Wooters87}%
  \BibitemOpen
  \bibfield  {author} {\bibinfo {author} {\bibfnamefont {W.~K.}\ \bibnamefont
  {Wooters}},\ }\bibfield  {title} {\enquote {\bibinfo {title} {{A
  Wigner-Function Formulation of Finite-State Quantum Mechanics}},}\
  }\href@noop {} {\bibfield  {journal} {\bibinfo  {journal} {Ann. Phys.}\
  }\textbf {\bibinfo {volume} {176}},\ \bibinfo {pages} {1--21} (\bibinfo
  {year} {1987})}\BibitemShut {NoStop}%
\bibitem [{\citenamefont {Leonhardt}(1996)}]{leonhardt1996}%
  \BibitemOpen
  \bibfield  {author} {\bibinfo {author} {\bibfnamefont {U.}~\bibnamefont
  {Leonhardt}},\ }\bibfield  {title} {\enquote {\bibinfo {title} {Discrete
  {W}igner function and quantum-state tomography},}\ }\href@noop {} {\bibfield
  {journal} {\bibinfo  {journal} {Phys. Rev. A}\ }\textbf {\bibinfo {volume}
  {53}},\ \bibinfo {pages} {2998} (\bibinfo {year} {1996})}\BibitemShut
  {NoStop}%
\bibitem [{\citenamefont {Miquel}, \citenamefont {Paz},\ and\ \citenamefont
  {Saraceno}(2002)}]{Miquel}%
  \BibitemOpen
  \bibfield  {author} {\bibinfo {author} {\bibfnamefont {C.}~\bibnamefont
  {Miquel}}, \bibinfo {author} {\bibfnamefont {J.~P.}\ \bibnamefont {Paz}}, \
  and\ \bibinfo {author} {\bibfnamefont {M.}~\bibnamefont {Saraceno}},\
  }\bibfield  {title} {\enquote {\bibinfo {title} {Quantum computers in phase
  space},}\ }\href@noop {} {\bibfield  {journal} {\bibinfo  {journal} {Phys.
  Rev. A}\ }\textbf {\bibinfo {volume} {65}},\ \bibinfo {pages} {062309}
  (\bibinfo {year} {2002})}\BibitemShut {NoStop}%
\bibitem [{\citenamefont {Miquel}\ \emph {et~al.}(2002)\citenamefont {Miquel},
  \citenamefont {Paz}, \citenamefont {Saraceno}, \citenamefont {Knill},
  \citenamefont {Laflamme},\ and\ \citenamefont {Negrevergne}}]{Miquel2}%
  \BibitemOpen
  \bibfield  {author} {\bibinfo {author} {\bibfnamefont {C.}~\bibnamefont
  {Miquel}}, \bibinfo {author} {\bibfnamefont {J.~P.}\ \bibnamefont {Paz}},
  \bibinfo {author} {\bibfnamefont {M.}~\bibnamefont {Saraceno}}, \bibinfo
  {author} {\bibfnamefont {E.}~\bibnamefont {Knill}}, \bibinfo {author}
  {\bibfnamefont {R.}~\bibnamefont {Laflamme}}, \ and\ \bibinfo {author}
  {\bibfnamefont {C.}~\bibnamefont {Negrevergne}},\ }\bibfield  {title}
  {\enquote {\bibinfo {title} {Interpretation of tomography and spectroscopy as
  dual forms of quantum computation},}\ }\href@noop {} {\bibfield  {journal}
  {\bibinfo  {journal} {Nature}\ }\textbf {\bibinfo {volume} {418}},\ \bibinfo
  {pages} {59--62} (\bibinfo {year} {2002})}\BibitemShut {NoStop}%
\bibitem [{\citenamefont {Gibbons}, \citenamefont {Hoffman},\ and\
  \citenamefont {Wootters}(2004)}]{gibbons2004}%
  \BibitemOpen
  \bibfield  {author} {\bibinfo {author} {\bibfnamefont {K.~S.}\ \bibnamefont
  {Gibbons}}, \bibinfo {author} {\bibfnamefont {M.~J.}\ \bibnamefont
  {Hoffman}}, \ and\ \bibinfo {author} {\bibfnamefont {W.~K.}\ \bibnamefont
  {Wootters}},\ }\bibfield  {title} {\enquote {\bibinfo {title} {Discrete phase
  space based on finite fields},}\ }\href@noop {} {\bibfield  {journal}
  {\bibinfo  {journal} {Phys. Rev. A}\ }\textbf {\bibinfo {volume} {70}},\
  \bibinfo {pages} {062101} (\bibinfo {year} {2004})}\BibitemShut {NoStop}%
\bibitem [{\citenamefont {Ferrie}\ and\ \citenamefont {Emerson}(2009)}]{FE09}%
  \BibitemOpen
  \bibfield  {author} {\bibinfo {author} {\bibfnamefont {C.}~\bibnamefont
  {Ferrie}}\ and\ \bibinfo {author} {\bibfnamefont {J.}~\bibnamefont
  {Emerson}},\ }\bibfield  {title} {\enquote {\bibinfo {title} {{Framed Hilbert
  space: hanging the quasi-probability pictures of quantum theory}},}\
  }\href@noop {} {\bibfield  {journal} {\bibinfo  {journal} {New J. Phys.}\
  }\textbf {\bibinfo {volume} {11}},\ \bibinfo {pages} {063040} (\bibinfo
  {year} {2009})}\BibitemShut {NoStop}%
\bibitem [{\citenamefont {Stratonovich}(1956)}]{Stratonovich}%
  \BibitemOpen
  \bibfield  {author} {\bibinfo {author} {\bibfnamefont {R.~L.}\ \bibnamefont
  {Stratonovich}},\ }\bibfield  {title} {\enquote {\bibinfo {title} {{On
  distributions in Representation Space}},}\ }\href@noop {} {\bibfield
  {journal} {\bibinfo  {journal} {JETP. (U.S.S.R.)}\ }\textbf {\bibinfo
  {volume} {31}},\ \bibinfo {pages} {1012--1020} (\bibinfo {year}
  {1956})}\BibitemShut {NoStop}%
\bibitem [{\citenamefont {Agarwal}(1981)}]{Agarwal81}%
  \BibitemOpen
  \bibfield  {author} {\bibinfo {author} {\bibfnamefont {G.~S.}\ \bibnamefont
  {Agarwal}},\ }\bibfield  {title} {\enquote {\bibinfo {title} {Relation
  between atomic coherent-state representation, state multipoles, and
  generalized phase-space distributions},}\ }\href@noop {} {\bibfield
  {journal} {\bibinfo  {journal} {Phys. Rev. A}\ }\textbf {\bibinfo {volume}
  {24}},\ \bibinfo {pages} {2889--2896} (\bibinfo {year} {1981})}\BibitemShut
  {NoStop}%
\bibitem [{\citenamefont {V{\'a}rrily}\ and\ \citenamefont
  {Garcia-Bond{\'i}a}(1989)}]{VGB89}%
  \BibitemOpen
  \bibfield  {author} {\bibinfo {author} {\bibfnamefont {J.~C.}\ \bibnamefont
  {V{\'a}rrily}}\ and\ \bibinfo {author} {\bibfnamefont {J.~M.}\ \bibnamefont
  {Garcia-Bond{\'i}a}},\ }\bibfield  {title} {\enquote {\bibinfo {title} {{The
  Moyal representation for spin}},}\ }\href@noop {} {\bibfield  {journal}
  {\bibinfo  {journal} {Ann. Phys.}\ }\textbf {\bibinfo {volume} {190}},\
  \bibinfo {pages} {107--148} (\bibinfo {year} {1989})}\BibitemShut {NoStop}%
\bibitem [{\citenamefont {Brif}\ and\ \citenamefont {Mann}(1999)}]{Brif98}%
  \BibitemOpen
  \bibfield  {author} {\bibinfo {author} {\bibfnamefont {C.}~\bibnamefont
  {Brif}}\ and\ \bibinfo {author} {\bibfnamefont {A.}~\bibnamefont {Mann}},\
  }\bibfield  {title} {\enquote {\bibinfo {title} {{Phase-space formulation of
  quantum mechanics and quantum-state reconstruction for physical systems with
  Lie-group symmetries}},}\ }\href@noop {} {\bibfield  {journal} {\bibinfo
  {journal} {Phys. Rev. A}\ }\textbf {\bibinfo {volume} {59}},\ \bibinfo
  {pages} {971} (\bibinfo {year} {1999})}\BibitemShut {NoStop}%
\bibitem [{\citenamefont {Brif}\ and\ \citenamefont {Mann}(1997)}]{Brif97}%
  \BibitemOpen
  \bibfield  {author} {\bibinfo {author} {\bibfnamefont {C.}~\bibnamefont
  {Brif}}\ and\ \bibinfo {author} {\bibfnamefont {A.}~\bibnamefont {Mann}},\
  }\bibfield  {title} {\enquote {\bibinfo {title} {{A general theory of
  phase-space quasiprobability distributions}},}\ }\href@noop {} {\bibfield
  {journal} {\bibinfo  {journal} {J. Phys. A: Math. Gen.}\ }\textbf {\bibinfo
  {volume} {31}},\ \bibinfo {pages} {L9--L17} (\bibinfo {year}
  {1997})}\BibitemShut {NoStop}%
\bibitem [{\citenamefont {Heiss}\ and\ \citenamefont
  {Weigert}(2000)}]{heiss2000discrete}%
  \BibitemOpen
  \bibfield  {author} {\bibinfo {author} {\bibfnamefont {S.}~\bibnamefont
  {Heiss}}\ and\ \bibinfo {author} {\bibfnamefont {S.}~\bibnamefont
  {Weigert}},\ }\bibfield  {title} {\enquote {\bibinfo {title} {{Discrete
  Moyal-type representations for a spin}},}\ }\href@noop {} {\bibfield
  {journal} {\bibinfo  {journal} {Phys. Rev. A}\ }\textbf {\bibinfo {volume}
  {63}},\ \bibinfo {pages} {012105} (\bibinfo {year} {2000})}\BibitemShut
  {NoStop}%
\bibitem [{\citenamefont {Klimov}(2002)}]{klimov2002ExactEvolution}%
  \BibitemOpen
  \bibfield  {author} {\bibinfo {author} {\bibfnamefont {A.}~\bibnamefont
  {Klimov}},\ }\bibfield  {title} {\enquote {\bibinfo {title} {Exact evolution
  equations for $\mathrm{SU}(2)$ quasidistribution functions},}\ }\href@noop {}
  {\bibfield  {journal} {\bibinfo  {journal} {J. Math. Phys.}\ }\textbf
  {\bibinfo {volume} {43}},\ \bibinfo {pages} {2202--2213} (\bibinfo {year}
  {2002})}\BibitemShut {NoStop}%
\bibitem [{\citenamefont {Klimov}\ and\ \citenamefont
  {Espinoza}(2005)}]{klimov2005classical}%
  \BibitemOpen
  \bibfield  {author} {\bibinfo {author} {\bibfnamefont {A.}~\bibnamefont
  {Klimov}}\ and\ \bibinfo {author} {\bibfnamefont {P.}~\bibnamefont
  {Espinoza}},\ }\bibfield  {title} {\enquote {\bibinfo {title} {Classical
  evolution of quantum fluctuations in spin-like systems: squeezing and
  entanglement},}\ }\href@noop {} {\bibfield  {journal} {\bibinfo  {journal}
  {J. Opt. B}\ }\textbf {\bibinfo {volume} {7}},\ \bibinfo {pages} {183}
  (\bibinfo {year} {2005})}\BibitemShut {NoStop}%
\bibitem [{\citenamefont {Klimov}\ and\ \citenamefont
  {Espinoza}(2002)}]{StarProd}%
  \BibitemOpen
  \bibfield  {author} {\bibinfo {author} {\bibfnamefont {A.~B.}\ \bibnamefont
  {Klimov}}\ and\ \bibinfo {author} {\bibfnamefont {P.}~\bibnamefont
  {Espinoza}},\ }\bibfield  {title} {\enquote {\bibinfo {title} {{Moyal-like
  form of the star product for generalized $\mathrm{SU}(2)$ Stratonovich-Weyl
  symbols}},}\ }\href@noop {} {\bibfield  {journal} {\bibinfo  {journal} {J.
  Phys. A}\ }\textbf {\bibinfo {volume} {35}},\ \bibinfo {pages} {8435}
  (\bibinfo {year} {2002})}\BibitemShut {NoStop}%
\bibitem [{\citenamefont {Dowling}, \citenamefont {Agarwal},\ and\
  \citenamefont {Schleich}(1994)}]{DowlingAgarwalSchleich}%
  \BibitemOpen
  \bibfield  {author} {\bibinfo {author} {\bibfnamefont {J.~P.}\ \bibnamefont
  {Dowling}}, \bibinfo {author} {\bibfnamefont {G.~S.}\ \bibnamefont
  {Agarwal}}, \ and\ \bibinfo {author} {\bibfnamefont {W.~P.}\ \bibnamefont
  {Schleich}},\ }\bibfield  {title} {\enquote {\bibinfo {title} {{Wigner
  distribution of a general angular-momentum state: Applications to a
  collection of two-level atoms}},}\ }\href@noop {} {\bibfield  {journal}
  {\bibinfo  {journal} {Phys. Rev. A}\ }\textbf {\bibinfo {volume} {49}},\
  \bibinfo {pages} {4101--4109} (\bibinfo {year} {1994})}\BibitemShut {NoStop}%
\bibitem [{\citenamefont {Jessen}\ \emph {et~al.}(2001)\citenamefont {Jessen},
  \citenamefont {Haycock}, \citenamefont {Klose}, \citenamefont {Smith},
  \citenamefont {Deutsch},\ and\ \citenamefont {Brennen}}]{JHKS}%
  \BibitemOpen
  \bibfield  {author} {\bibinfo {author} {\bibfnamefont {P.~S.}\ \bibnamefont
  {Jessen}}, \bibinfo {author} {\bibfnamefont {D.~L.}\ \bibnamefont {Haycock}},
  \bibinfo {author} {\bibfnamefont {G.}~\bibnamefont {Klose}}, \bibinfo
  {author} {\bibfnamefont {G.~A.}\ \bibnamefont {Smith}}, \bibinfo {author}
  {\bibfnamefont {I.~H.}\ \bibnamefont {Deutsch}}, \ and\ \bibinfo {author}
  {\bibfnamefont {G.~K.}\ \bibnamefont {Brennen}},\ }\bibfield  {title}
  {\enquote {\bibinfo {title} {{Quantum control and information processing in
  optical lattices}},}\ }\href@noop {} {\bibfield  {journal} {\bibinfo
  {journal} {Quant. Inf. Computation}\ }\textbf {\bibinfo {volume} {1}},\
  \bibinfo {pages} {20--32} (\bibinfo {year} {2001})}\BibitemShut {NoStop}%
\bibitem [{\citenamefont {Philp}\ and\ \citenamefont
  {Kuchel}(2005)}]{PhilpKuchel}%
  \BibitemOpen
  \bibfield  {author} {\bibinfo {author} {\bibfnamefont {D.~J.}\ \bibnamefont
  {Philp}}\ and\ \bibinfo {author} {\bibfnamefont {P.~W.}\ \bibnamefont
  {Kuchel}},\ }\bibfield  {title} {\enquote {\bibinfo {title} {{A Way of
  Visualizing NMR Experiments on Quadrupolar Nuclei}},}\ }\href@noop {}
  {\bibfield  {journal} {\bibinfo  {journal} {Concepts Magn. Reson. A}\
  }\textbf {\bibinfo {volume} {25A}},\ \bibinfo {pages} {40--52} (\bibinfo
  {year} {2005})}\BibitemShut {NoStop}%
\bibitem [{\citenamefont {Harland}\ \emph {et~al.}(2012)\citenamefont
  {Harland}, \citenamefont {Everitt}, \citenamefont {Nemoto}, \citenamefont
  {Tilma},\ and\ \citenamefont {Spiller}}]{Harland}%
  \BibitemOpen
  \bibfield  {author} {\bibinfo {author} {\bibfnamefont {D.}~\bibnamefont
  {Harland}}, \bibinfo {author} {\bibfnamefont {M.~J.}\ \bibnamefont
  {Everitt}}, \bibinfo {author} {\bibfnamefont {K.}~\bibnamefont {Nemoto}},
  \bibinfo {author} {\bibfnamefont {T.}~\bibnamefont {Tilma}}, \ and\ \bibinfo
  {author} {\bibfnamefont {T.~P.}\ \bibnamefont {Spiller}},\ }\bibfield
  {title} {\enquote {\bibinfo {title} {{Towards a complete and continious
  Wigner function for an ensemble of spins or qubits}},}\ }\href@noop {}
  {\bibfield  {journal} {\bibinfo  {journal} {Phys. Rev. A}\ }\textbf {\bibinfo
  {volume} {86}},\ \bibinfo {pages} {062117} (\bibinfo {year}
  {2012})}\BibitemShut {NoStop}%
\bibitem [{\citenamefont {Garon}, \citenamefont {Zeier},\ and\ \citenamefont
  {Glaser}(2015)}]{Garon15}%
  \BibitemOpen
  \bibfield  {author} {\bibinfo {author} {\bibfnamefont {A.}~\bibnamefont
  {Garon}}, \bibinfo {author} {\bibfnamefont {R.}~\bibnamefont {Zeier}}, \ and\
  \bibinfo {author} {\bibfnamefont {S.~J.}\ \bibnamefont {Glaser}},\ }\bibfield
   {title} {\enquote {\bibinfo {title} {Visualizing operators of coupled spin
  systems},}\ }\href@noop {} {\bibfield  {journal} {\bibinfo  {journal} {Phys.
  Rev. A}\ }\textbf {\bibinfo {volume} {91}},\ \bibinfo {pages} {042122}
  (\bibinfo {year} {2015})}\BibitemShut {NoStop}%
\bibitem [{\citenamefont {Tilma}\ \emph {et~al.}(2016)\citenamefont {Tilma},
  \citenamefont {Everitt}, \citenamefont {Samson}, \citenamefont {Munro},\ and\
  \citenamefont {Nemoto}}]{tilma2016}%
  \BibitemOpen
  \bibfield  {author} {\bibinfo {author} {\bibfnamefont {T.}~\bibnamefont
  {Tilma}}, \bibinfo {author} {\bibfnamefont {M.~J.}\ \bibnamefont {Everitt}},
  \bibinfo {author} {\bibfnamefont {J.~H.}\ \bibnamefont {Samson}}, \bibinfo
  {author} {\bibfnamefont {W.~J.}\ \bibnamefont {Munro}}, \ and\ \bibinfo
  {author} {\bibfnamefont {K.}~\bibnamefont {Nemoto}},\ }\bibfield  {title}
  {\enquote {\bibinfo {title} {Wigner functions for arbitrary quantum
  systems},}\ }\href@noop {} {\bibfield  {journal} {\bibinfo  {journal} {Phys.
  Rev. Lett.}\ }\textbf {\bibinfo {volume} {117}},\ \bibinfo {pages} {180401}
  (\bibinfo {year} {2016})}\BibitemShut {NoStop}%
\bibitem [{\citenamefont {Koczor}, \citenamefont {Zeier},\ and\ \citenamefont
  {Glaser}(2016)}]{koczor2016}%
  \BibitemOpen
  \bibfield  {author} {\bibinfo {author} {\bibfnamefont {B.}~\bibnamefont
  {Koczor}}, \bibinfo {author} {\bibfnamefont {R.}~\bibnamefont {Zeier}}, \
  and\ \bibinfo {author} {\bibfnamefont {S.~J.}\ \bibnamefont {Glaser}},\
  }\href@noop {} {\enquote {\bibinfo {title} {{Time evolution of coupled spin
  systems in a generalized Wigner representation}},}\ } (\bibinfo {year}
  {2016}),\ \bibinfo {note} {(\emph{Preprint}
  \href{https://arxiv.org/abs/1612.06777v2}{\tt arXiv:1612.06777v2})},\ \Eprint
  {http://arxiv.org/abs/1808.02697} {arXiv:1808.02697} \BibitemShut {NoStop}%
\bibitem [{\citenamefont {Rundle}\ \emph
  {et~al.}(2017{\natexlab{a}})\citenamefont {Rundle}, \citenamefont {Mills},
  \citenamefont {Tilma}, \citenamefont {Samson},\ and\ \citenamefont
  {Everitt}}]{rundle2017}%
  \BibitemOpen
  \bibfield  {author} {\bibinfo {author} {\bibfnamefont {R.~P.}\ \bibnamefont
  {Rundle}}, \bibinfo {author} {\bibfnamefont {P.~W.}\ \bibnamefont {Mills}},
  \bibinfo {author} {\bibfnamefont {T.}~\bibnamefont {Tilma}}, \bibinfo
  {author} {\bibfnamefont {J.~H.}\ \bibnamefont {Samson}}, \ and\ \bibinfo
  {author} {\bibfnamefont {M.~J.}\ \bibnamefont {Everitt}},\ }\bibfield
  {title} {\enquote {\bibinfo {title} {Simple procedure for phase-space
  measurement and entanglement validation},}\ }\href@noop {} {\bibfield
  {journal} {\bibinfo  {journal} {Phys. Rev. A}\ }\textbf {\bibinfo {volume}
  {96}},\ \bibinfo {pages} {022117} (\bibinfo {year}
  {2017}{\natexlab{a}})}\BibitemShut {NoStop}%
\bibitem [{\citenamefont {Rundle}\ \emph
  {et~al.}(2017{\natexlab{b}})\citenamefont {Rundle}, \citenamefont {Tilma},
  \citenamefont {Samson}, \citenamefont {Dwyer}, \citenamefont {Bishop},\ and\
  \citenamefont {Everitt}}]{RTD17}%
  \BibitemOpen
  \bibfield  {author} {\bibinfo {author} {\bibfnamefont {R.~P.}\ \bibnamefont
  {Rundle}}, \bibinfo {author} {\bibfnamefont {T.}~\bibnamefont {Tilma}},
  \bibinfo {author} {\bibfnamefont {J.~H.}\ \bibnamefont {Samson}}, \bibinfo
  {author} {\bibfnamefont {V.~M.}\ \bibnamefont {Dwyer}}, \bibinfo {author}
  {\bibfnamefont {R.~F.}\ \bibnamefont {Bishop}}, \ and\ \bibinfo {author}
  {\bibfnamefont {M.~J.}\ \bibnamefont {Everitt}},\ }\href@noop {} {\enquote
  {\bibinfo {title} {{A general approach to quantum mechanics as a statistical
  theory}},}\ } (\bibinfo {year} {2017}{\natexlab{b}}),\ \Eprint
  {http://arxiv.org/abs/1708.03814v3} {arXiv:1708.03814v3} \BibitemShut
  {NoStop}%
\bibitem [{\citenamefont {Koczor}, \citenamefont {Zeier},\ and\ \citenamefont
  {Glaser}(2017)}]{koczor2017}%
  \BibitemOpen
  \bibfield  {author} {\bibinfo {author} {\bibfnamefont {B.}~\bibnamefont
  {Koczor}}, \bibinfo {author} {\bibfnamefont {R.}~\bibnamefont {Zeier}}, \
  and\ \bibinfo {author} {\bibfnamefont {S.~J.}\ \bibnamefont {Glaser}},\
  }\href@noop {} {\enquote {\bibinfo {title} {{Continuous phase-space
  representations for finite-dimensional quantum states and their
  tomography}},}\ } (\bibinfo {year} {2017}),\ \Eprint
  {http://arxiv.org/abs/1711.07994v2} {arXiv:1711.07994v2} \BibitemShut
  {NoStop}%
\bibitem [{\citenamefont {Koczor}, \citenamefont {Zeier},\ and\ \citenamefont
  {Glaser}(2018)}]{koczor2018}%
  \BibitemOpen
  \bibfield  {author} {\bibinfo {author} {\bibfnamefont {B.}~\bibnamefont
  {Koczor}}, \bibinfo {author} {\bibfnamefont {R.}~\bibnamefont {Zeier}}, \
  and\ \bibinfo {author} {\bibfnamefont {S.~J.}\ \bibnamefont {Glaser}},\
  }\href@noop {} {\enquote {\bibinfo {title} {{Continuous phase spaces and the
  time evolution of spins: star products and spin-weighted spherical
  harmonics}},}\ } (\bibinfo {year} {2018}),\ \Eprint
  {http://arxiv.org/abs/1808.02697} {arXiv:1808.02697} \BibitemShut {NoStop}%
\bibitem [{\citenamefont {Leiner}, \citenamefont {Zeier},\ and\ \citenamefont
  {Glaser}(2017)}]{davidtomo}%
  \BibitemOpen
  \bibfield  {author} {\bibinfo {author} {\bibfnamefont {D.}~\bibnamefont
  {Leiner}}, \bibinfo {author} {\bibfnamefont {R.}~\bibnamefont {Zeier}}, \
  and\ \bibinfo {author} {\bibfnamefont {S.~J.}\ \bibnamefont {Glaser}},\
  }\bibfield  {title} {\enquote {\bibinfo {title} {Wigner tomography of
  multispin quantum states},}\ }\href@noop {} {\bibfield  {journal} {\bibinfo
  {journal} {Phys. Rev. A}\ }\textbf {\bibinfo {volume} {96}},\ \bibinfo
  {pages} {063413} (\bibinfo {year} {2017})}\BibitemShut {NoStop}%
\bibitem [{\citenamefont {Leiner}\ and\ \citenamefont {Glaser}(2018)}]{LG18}%
  \BibitemOpen
  \bibfield  {author} {\bibinfo {author} {\bibfnamefont {D.}~\bibnamefont
  {Leiner}}\ and\ \bibinfo {author} {\bibfnamefont {S.~J.}\ \bibnamefont
  {Glaser}},\ }\bibfield  {title} {\enquote {\bibinfo {title} {Wigner process
  tomography: Visualization of spin propagators and their spinor properties},}\
  }\href@noop {} {\bibfield  {journal} {\bibinfo  {journal} {Phys. Rev. A}\
  }\textbf {\bibinfo {volume} {98}},\ \bibinfo {pages} {012112} (\bibinfo
  {year} {2018})}\BibitemShut {NoStop}%
\bibitem [{\citenamefont {Jackson}(1999)}]{Jac99}%
  \BibitemOpen
  \bibfield  {author} {\bibinfo {author} {\bibfnamefont {J.~D.}\ \bibnamefont
  {Jackson}},\ }\href@noop {} {\emph {\bibinfo {title} {{Classical
  Electrodynamics}}}},\ \bibinfo {edition} {3rd}\ ed.\ (\bibinfo  {publisher}
  {John Wiley \& Sons, New York},\ \bibinfo {year} {1999})\BibitemShut
  {NoStop}%
\bibitem [{\citenamefont {James}(1978)}]{James78}%
  \BibitemOpen
  \bibfield  {author} {\bibinfo {author} {\bibfnamefont {G.~D.}\ \bibnamefont
  {James}},\ }\href@noop {} {\emph {\bibinfo {title} {{The Representation
  Theory of the Symmetric Group}}}}\ (\bibinfo  {publisher} {Springer,
  Berlin},\ \bibinfo {year} {1978})\BibitemShut {NoStop}%
\bibitem [{\citenamefont {James}\ and\ \citenamefont {Kerber}(1981)}]{JK81}%
  \BibitemOpen
  \bibfield  {author} {\bibinfo {author} {\bibfnamefont {G.}~\bibnamefont
  {James}}\ and\ \bibinfo {author} {\bibfnamefont {A.}~\bibnamefont {Kerber}},\
  }\href@noop {} {\emph {\bibinfo {title} {{The Representation Theory of the
  Symmetric Group}}}}\ (\bibinfo  {publisher} {Addison-Wesley, Reading, MA},\
  \bibinfo {year} {1981})\BibitemShut {NoStop}%
\bibitem [{\citenamefont {Ceccherini-Silberstein}, \citenamefont {Scarabotti},\
  and\ \citenamefont {Tolli}(2010)}]{CSST10}%
  \BibitemOpen
  \bibfield  {author} {\bibinfo {author} {\bibfnamefont {T.}~\bibnamefont
  {Ceccherini-Silberstein}}, \bibinfo {author} {\bibfnamefont {F.}~\bibnamefont
  {Scarabotti}}, \ and\ \bibinfo {author} {\bibfnamefont {F.}~\bibnamefont
  {Tolli}},\ }\href@noop {} {\emph {\bibinfo {title} {{Representation Theory of
  the Symmetric Groups}}}}\ (\bibinfo  {publisher} {Cambridge University Press,
  Cambridge},\ \bibinfo {year} {2010})\BibitemShut {NoStop}%
\bibitem [{\citenamefont {Boerner}(1967)}]{Boerner67}%
  \BibitemOpen
  \bibfield  {author} {\bibinfo {author} {\bibfnamefont {H.}~\bibnamefont
  {Boerner}},\ }\href@noop {} {\emph {\bibinfo {title} {{Darstellungen von
  Gruppen}}}},\ \bibinfo {edition} {2nd}\ ed.\ (\bibinfo  {publisher}
  {Sprin\-ger, Berlin},\ \bibinfo {year} {1967})\BibitemShut {NoStop}%
\bibitem [{\citenamefont {Hamermesh}(1962)}]{Hamermesh62}%
  \BibitemOpen
  \bibfield  {author} {\bibinfo {author} {\bibfnamefont {M.}~\bibnamefont
  {Hamermesh}},\ }\href@noop {} {\emph {\bibinfo {title} {{Group Theory}}}}\
  (\bibinfo  {publisher} {Addison-Wesley, Reading, MA},\ \bibinfo {year}
  {1962})\BibitemShut {NoStop}%
\bibitem [{\citenamefont {Sagan}(2001)}]{Sagan01}%
  \BibitemOpen
  \bibfield  {author} {\bibinfo {author} {\bibfnamefont {B.~E.}\ \bibnamefont
  {Sagan}},\ }\href@noop {} {\emph {\bibinfo {title} {{The Symmetric
  Group}}}},\ \bibinfo {edition} {2nd}\ ed.\ (\bibinfo  {publisher} {Springer,
  New York},\ \bibinfo {year} {2001})\BibitemShut {NoStop}%
\bibitem [{\citenamefont {Tung}(1985)}]{tung1985group}%
  \BibitemOpen
  \bibfield  {author} {\bibinfo {author} {\bibfnamefont {W.~K.}\ \bibnamefont
  {Tung}},\ }\href@noop {} {\emph {\bibinfo {title} {Group Theory in
  Physics}}}\ (\bibinfo  {publisher} {World Scientific Publishing Company,
  Incorporated},\ \bibinfo {year} {1985})\BibitemShut {NoStop}%
\bibitem [{\citenamefont {Racah}(1965)}]{Racah65}%
  \BibitemOpen
  \bibfield  {author} {\bibinfo {author} {\bibfnamefont {G.}~\bibnamefont
  {Racah}},\ }\bibfield  {title} {\enquote {\bibinfo {title} {{Group Theory and
  Spectroscopy}},}\ }in\ \href@noop {} {\emph {\bibinfo {booktitle}
  {{Ergebnisse der exakten Natur\-wis\-sen\-schaf\-ten, 37. Band}}}}\ (\bibinfo
   {publisher} {Springer, Berlin},\ \bibinfo {year} {1965})\ pp.\ \bibinfo
  {pages} {28--84}\BibitemShut {NoStop}%
\bibitem [{\citenamefont {Elliott}\ and\ \citenamefont {Lane}(1957)}]{EL57}%
  \BibitemOpen
  \bibfield  {author} {\bibinfo {author} {\bibfnamefont {J.~P.}\ \bibnamefont
  {Elliott}}\ and\ \bibinfo {author} {\bibfnamefont {A.~M.}\ \bibnamefont
  {Lane}},\ }\bibfield  {title} {\enquote {\bibinfo {title} {{The Nuclear
  Shell-Model}},}\ }in\ \href@noop {} {\emph {\bibinfo {booktitle}
  {{Encyclopedia of Physics, Volume XXXIX, Structure of Atomic Nuclei}}}}\
  (\bibinfo  {publisher} {Springer, Berlin},\ \bibinfo {year} {1957})\ pp.\
  \bibinfo {pages} {241--410}\BibitemShut {NoStop}%
\bibitem [{\citenamefont {Kaplan}(1975)}]{Kaplan75}%
  \BibitemOpen
  \bibfield  {author} {\bibinfo {author} {\bibfnamefont {I.~G.}\ \bibnamefont
  {Kaplan}},\ }\href@noop {} {\emph {\bibinfo {title} {{Symmetry of
  Many-Electron Systems}}}}\ (\bibinfo  {publisher} {Academic Press, New
  York},\ \bibinfo {year} {1975})\BibitemShut {NoStop}%
\bibitem [{\citenamefont {Silver}(1976)}]{Silver76}%
  \BibitemOpen
  \bibfield  {author} {\bibinfo {author} {\bibfnamefont {B.~L.}\ \bibnamefont
  {Silver}},\ }\href@noop {} {\emph {\bibinfo {title} {{Irreducible Tensor
  Methods}}}}\ (\bibinfo  {publisher} {Academic Press, New York},\ \bibinfo
  {year} {1976})\BibitemShut {NoStop}%
\bibitem [{\citenamefont {Chisholm}(1976)}]{Chisholm76}%
  \BibitemOpen
  \bibfield  {author} {\bibinfo {author} {\bibfnamefont {C.~D.~H.}\
  \bibnamefont {Chisholm}},\ }\href@noop {} {\emph {\bibinfo {title} {{Group
  Theoretical Techniques in Quantum Chemistry}}}}\ (\bibinfo  {publisher}
  {Academic Press, London},\ \bibinfo {year} {1976})\BibitemShut {NoStop}%
\bibitem [{\citenamefont {Kramer}, \citenamefont {John},\ and\ \citenamefont
  {Schenzle}(1981)}]{KJS81}%
  \BibitemOpen
  \bibfield  {author} {\bibinfo {author} {\bibfnamefont {P.}~\bibnamefont
  {Kramer}}, \bibinfo {author} {\bibfnamefont {G.}~\bibnamefont {John}}, \ and\
  \bibinfo {author} {\bibfnamefont {D.}~\bibnamefont {Schenzle}},\ }\href@noop
  {} {\emph {\bibinfo {title} {{Group Theory and the Interaction of Composite
  Nucleon Systems}}}}\ (\bibinfo  {publisher} {Vieweg, Braunschweig},\ \bibinfo
  {year} {1981})\BibitemShut {NoStop}%
\bibitem [{\citenamefont {Jahn}\ and\ \citenamefont {van
  Wieringen}(1951)}]{JvW51}%
  \BibitemOpen
  \bibfield  {author} {\bibinfo {author} {\bibfnamefont {H.~A.}\ \bibnamefont
  {Jahn}}\ and\ \bibinfo {author} {\bibfnamefont {H.}~\bibnamefont {van
  Wieringen}},\ }\bibfield  {title} {\enquote {\bibinfo {title} {{Theoretical
  studies in nuclear structure IV. Wave functions for the nuclear $p$-shell
  Part A. $\langle p^n | p^{n{-}1}p \rangle$ fractional parentage
  coefficients}},}\ }\href@noop {} {\bibfield  {journal} {\bibinfo  {journal}
  {Proc. R. Soc. Lond. A}\ }\textbf {\bibinfo {volume} {209}},\ \bibinfo
  {pages} {502--524} (\bibinfo {year} {1951})}\BibitemShut {NoStop}%
\bibitem [{\citenamefont {Weyl}(1927)}]{Wey27}%
  \BibitemOpen
  \bibfield  {author} {\bibinfo {author} {\bibfnamefont {H.}~\bibnamefont
  {Weyl}},\ }\bibfield  {title} {\enquote {\bibinfo {title} {{Quantenmechanik
  und Gruppentheorie}},}\ }\href@noop {} {\bibfield  {journal} {\bibinfo
  {journal} {Z. Phys.}\ }\textbf {\bibinfo {volume} {46}},\ \bibinfo {pages}
  {1--33} (\bibinfo {year} {1927})}\BibitemShut {NoStop}%
\bibitem [{\citenamefont {Weyl}(1931)}]{Weyl31}%
  \BibitemOpen
  \bibfield  {author} {\bibinfo {author} {\bibfnamefont {H.}~\bibnamefont
  {Weyl}},\ }\href@noop {} {\emph {\bibinfo {title} {{Gruppentheorie und
  Quantenmechanik}}}},\ \bibinfo {edition} {2nd}\ ed.\ (\bibinfo  {publisher}
  {Hirzel, Leipzig},\ \bibinfo {year} {1931})\ \bibinfo {note} {english
  translation in \cite{Weyl50}}\BibitemShut {NoStop}%
\bibitem [{\citenamefont {Weyl}(1950)}]{Weyl50}%
  \BibitemOpen
  \bibfield  {author} {\bibinfo {author} {\bibfnamefont {H.}~\bibnamefont
  {Weyl}},\ }\href@noop {} {\emph {\bibinfo {title} {{The Theory of Groups \&
  Quantum Mechanics}}}},\ \bibinfo {edition} {2nd}\ ed.\ (\bibinfo  {publisher}
  {Dover Publ., New York},\ \bibinfo {year} {1950})\BibitemShut {NoStop}%
\bibitem [{\citenamefont {Weyl}(1953)}]{Weyl46}%
  \BibitemOpen
  \bibfield  {author} {\bibinfo {author} {\bibfnamefont {H.}~\bibnamefont
  {Weyl}},\ }\href@noop {} {\emph {\bibinfo {title} {{The Classical Groups:
  Their Invariants and Representations}}}},\ \bibinfo {edition} {2nd}\ ed.\
  (\bibinfo  {publisher} {Princeton University Press, Princeton},\ \bibinfo
  {year} {1953})\BibitemShut {NoStop}%
\bibitem [{\citenamefont {Wigner}(1931)}]{Wigner31}%
  \BibitemOpen
  \bibfield  {author} {\bibinfo {author} {\bibfnamefont {E.}~\bibnamefont
  {Wigner}},\ }\href@noop {} {\emph {\bibinfo {title} {{Gruppentheorie und ihre
  Anwendung auf die Quantenmechanik der Atomspektren}}}}\ (\bibinfo
  {publisher} {Friedrich Vieweg \& Sohn, Braunschweig},\ \bibinfo {year}
  {1931})\ \bibinfo {note} {({E}nglish translation in
  \cite{Wigner59})}\BibitemShut {NoStop}%
\bibitem [{\citenamefont {Wigner}(1959)}]{Wigner59}%
  \BibitemOpen
  \bibfield  {author} {\bibinfo {author} {\bibfnamefont {E.~P.}\ \bibnamefont
  {Wigner}},\ }\href@noop {} {\emph {\bibinfo {title} {{Group Theory and its
  Application to the Quantum Mechanics of Atomic Spectra}}}}\ (\bibinfo
  {publisher} {Academic Press, London},\ \bibinfo {year} {1959})\BibitemShut
  {NoStop}%
\bibitem [{\citenamefont {Condon}\ and\ \citenamefont
  {Shortley}(1935)}]{CondonShortley}%
  \BibitemOpen
  \bibfield  {author} {\bibinfo {author} {\bibfnamefont {E.~U.}\ \bibnamefont
  {Condon}}\ and\ \bibinfo {author} {\bibfnamefont {G.~H.}\ \bibnamefont
  {Shortley}},\ }\href@noop {} {\emph {\bibinfo {title} {{The Theory of Atomic
  Spectra}}}}\ (\bibinfo  {publisher} {Cambridge University Press, Cambridge},\
  \bibinfo {year} {1935})\BibitemShut {NoStop}%
\bibitem [{\citenamefont {Racah}(1941)}]{Racah41}%
  \BibitemOpen
  \bibfield  {author} {\bibinfo {author} {\bibfnamefont {G.}~\bibnamefont
  {Racah}},\ }\bibfield  {title} {\enquote {\bibinfo {title} {{Theory of
  Complex Spectra I}},}\ }\href@noop {} {\bibfield  {journal} {\bibinfo
  {journal} {Phys. Rev.}\ }\textbf {\bibinfo {volume} {61}},\ \bibinfo {pages}
  {186--197} (\bibinfo {year} {1941})}\BibitemShut {NoStop}%
\bibitem [{\citenamefont {Racah}(1942)}]{Racah42}%
  \BibitemOpen
  \bibfield  {author} {\bibinfo {author} {\bibfnamefont {G.}~\bibnamefont
  {Racah}},\ }\bibfield  {title} {\enquote {\bibinfo {title} {{Theory of
  Complex Spectra II}},}\ }\href@noop {} {\bibfield  {journal} {\bibinfo
  {journal} {Phys. Rev.}\ }\textbf {\bibinfo {volume} {62}},\ \bibinfo {pages}
  {438--462} (\bibinfo {year} {1942})}\BibitemShut {NoStop}%
\bibitem [{\citenamefont {Racah}(1943)}]{Racah43}%
  \BibitemOpen
  \bibfield  {author} {\bibinfo {author} {\bibfnamefont {G.}~\bibnamefont
  {Racah}},\ }\bibfield  {title} {\enquote {\bibinfo {title} {{Theory of
  Complex Spectra III}},}\ }\href@noop {} {\bibfield  {journal} {\bibinfo
  {journal} {Phys. Rev.}\ }\textbf {\bibinfo {volume} {63}},\ \bibinfo {pages}
  {367--382} (\bibinfo {year} {1943})}\BibitemShut {NoStop}%
\bibitem [{\citenamefont {Racah}(1949)}]{Racah49}%
  \BibitemOpen
  \bibfield  {author} {\bibinfo {author} {\bibfnamefont {G.}~\bibnamefont
  {Racah}},\ }\bibfield  {title} {\enquote {\bibinfo {title} {{Theory of
  Complex Spectra IV}},}\ }\href@noop {} {\bibfield  {journal} {\bibinfo
  {journal} {Phys. Rev.}\ }\textbf {\bibinfo {volume} {76}},\ \bibinfo {pages}
  {1353--1365} (\bibinfo {year} {1949})}\BibitemShut {NoStop}%
\bibitem [{\citenamefont {Fano}\ and\ \citenamefont {Racah}(1959)}]{Fano59}%
  \BibitemOpen
  \bibfield  {author} {\bibinfo {author} {\bibfnamefont {U.}~\bibnamefont
  {Fano}}\ and\ \bibinfo {author} {\bibfnamefont {G.}~\bibnamefont {Racah}},\
  }\href@noop {} {\emph {\bibinfo {title} {{Irreducible Tensorial Sets}}}}\
  (\bibinfo  {publisher} {Academic Press, New York},\ \bibinfo {year}
  {1959})\BibitemShut {NoStop}%
\bibitem [{\citenamefont {Edmonds}(1960)}]{edmonds1960}%
  \BibitemOpen
  \bibfield  {author} {\bibinfo {author} {\bibfnamefont {A.~R.}\ \bibnamefont
  {Edmonds}},\ }\href@noop {} {\emph {\bibinfo {title} {{Angular momentum in
  quantum mechanics}}}}\ (\bibinfo  {publisher} {Princeton University Press,
  Princeton},\ \bibinfo {year} {1960})\BibitemShut {NoStop}%
\bibitem [{\citenamefont {Griffith}(2006)}]{Griffith06}%
  \BibitemOpen
  \bibfield  {author} {\bibinfo {author} {\bibfnamefont {J.~S.}\ \bibnamefont
  {Griffith}},\ }\href@noop {} {\emph {\bibinfo {title} {{The Irreducible
  Tensor Method for Molecular Symmetry Groups}}}}\ (\bibinfo  {publisher}
  {Dover Publications, Mineola, NY},\ \bibinfo {year} {2006})\BibitemShut
  {NoStop}%
\bibitem [{\citenamefont {Judd}(1998)}]{Judd98}%
  \BibitemOpen
  \bibfield  {author} {\bibinfo {author} {\bibfnamefont {B.~R.}\ \bibnamefont
  {Judd}},\ }\href@noop {} {\emph {\bibinfo {title} {{Operator Techniques in
  Atomic Spectroscopy}}}}\ (\bibinfo  {publisher} {Princeton University Press,
  Princeton},\ \bibinfo {year} {1998})\BibitemShut {NoStop}%
\bibitem [{\citenamefont {Miller}(1972)}]{Miller72}%
  \BibitemOpen
  \bibfield  {author} {\bibinfo {author} {\bibfnamefont {W.}~\bibnamefont
  {Miller}},\ }\href@noop {} {\emph {\bibinfo {title} {{Symmetry Groups and
  Their Applications}}}},\ Pure and Applied Mathematics\ (\bibinfo  {publisher}
  {Academic Press, London},\ \bibinfo {year} {1972})\BibitemShut {NoStop}%
\bibitem [{\citenamefont {Ludwig}\ and\ \citenamefont
  {Falter}(1996)}]{Ludwig96}%
  \BibitemOpen
  \bibfield  {author} {\bibinfo {author} {\bibfnamefont {W.}~\bibnamefont
  {Ludwig}}\ and\ \bibinfo {author} {\bibfnamefont {C.}~\bibnamefont
  {Falter}},\ }\href@noop {} {\emph {\bibinfo {title} {{Symmetries in
  Physics}}}},\ \bibinfo {edition} {2nd}\ ed.\ (\bibinfo  {publisher}
  {Springer, Berlin},\ \bibinfo {year} {1996})\BibitemShut {NoStop}%
\bibitem [{\citenamefont {Slater}(1960)}]{Slater1960}%
  \BibitemOpen
  \bibfield  {author} {\bibinfo {author} {\bibfnamefont {J.~C.}\ \bibnamefont
  {Slater}},\ }\href@noop {} {\emph {\bibinfo {title} {{Quantum Theory of
  Atomic Structure, Vol. II}}}}\ (\bibinfo  {publisher} {McGraw-Hill, New
  York},\ \bibinfo {year} {1960})\BibitemShut {NoStop}%
\bibitem [{\citenamefont {{de-Shalit}}\ and\ \citenamefont
  {Talmi}(1963)}]{ShalitTalmi63}%
  \BibitemOpen
  \bibfield  {author} {\bibinfo {author} {\bibfnamefont {A.}~\bibnamefont
  {{de-Shalit}}}\ and\ \bibinfo {author} {\bibfnamefont {I.}~\bibnamefont
  {Talmi}},\ }\href@noop {} {\emph {\bibinfo {title} {{Nuclear Shell
  Theory}}}}\ (\bibinfo  {publisher} {Academic Press, New York},\ \bibinfo
  {year} {1963})\BibitemShut {NoStop}%
\bibitem [{\citenamefont {Pauncz}(1967)}]{Pauncz67}%
  \BibitemOpen
  \bibfield  {author} {\bibinfo {author} {\bibfnamefont {R.}~\bibnamefont
  {Pauncz}},\ }\href@noop {} {\emph {\bibinfo {title} {{Alternant Molecular
  Orbital Method}}}}\ (\bibinfo  {publisher} {W. B. Saunders Company,
  Philadelphia},\ \bibinfo {year} {1967})\BibitemShut {NoStop}%
\bibitem [{\citenamefont {Wybourne}(1970)}]{Wybourne70}%
  \BibitemOpen
  \bibfield  {author} {\bibinfo {author} {\bibfnamefont {B.~G.}\ \bibnamefont
  {Wybourne}},\ }\href@noop {} {\emph {\bibinfo {title} {{Symmetry Principles
  and Atomic Spectroscopy}}}}\ (\bibinfo  {publisher} {Wiley-Interscience, New
  York},\ \bibinfo {year} {1970})\BibitemShut {NoStop}%
\bibitem [{\citenamefont {Elliott}\ and\ \citenamefont {Dawber}(1979)}]{ED79}%
  \BibitemOpen
  \bibfield  {author} {\bibinfo {author} {\bibfnamefont {J.~P.}\ \bibnamefont
  {Elliott}}\ and\ \bibinfo {author} {\bibfnamefont {P.~G.}\ \bibnamefont
  {Dawber}},\ }\href@noop {} {\emph {\bibinfo {title} {{Symmetry in Physics, 2
  Volumes}}}}\ (\bibinfo  {publisher} {Macmillan, Hampshire},\ \bibinfo {year}
  {1979})\BibitemShut {NoStop}%
\bibitem [{\citenamefont {Condon}\ and\ \citenamefont
  {Odaba{\c{s}}i}(1980)}]{Condon1980}%
  \BibitemOpen
  \bibfield  {author} {\bibinfo {author} {\bibfnamefont {E.~U.}\ \bibnamefont
  {Condon}}\ and\ \bibinfo {author} {\bibfnamefont {H.}~\bibnamefont
  {Odaba{\c{s}}i}},\ }\href@noop {} {\emph {\bibinfo {title} {{Atomic
  Structure}}}}\ (\bibinfo  {publisher} {Cambridge University Press,
  Cambridge},\ \bibinfo {year} {1980})\BibitemShut {NoStop}%
\bibitem [{\citenamefont {Rudzikas}(1997)}]{Rudzikas97}%
  \BibitemOpen
  \bibfield  {author} {\bibinfo {author} {\bibfnamefont {Z.}~\bibnamefont
  {Rudzikas}},\ }\href@noop {} {\emph {\bibinfo {title} {{Theoretical Atomic
  Spectroscopy}}}}\ (\bibinfo  {publisher} {Cambrdige University Press,
  Cambridge},\ \bibinfo {year} {1997})\BibitemShut {NoStop}%
\bibitem [{\citenamefont {Chaichian}\ and\ \citenamefont
  {Hagedorn}(1998)}]{CH98}%
  \BibitemOpen
  \bibfield  {author} {\bibinfo {author} {\bibfnamefont {M.}~\bibnamefont
  {Chaichian}}\ and\ \bibinfo {author} {\bibfnamefont {R.}~\bibnamefont
  {Hagedorn}},\ }\href@noop {} {\emph {\bibinfo {title} {{Symmetries in Quantum
  Mechanics: From Angular Momentum to Supersymmetry}}}}\ (\bibinfo  {publisher}
  {Institute of Physics, Bristol},\ \bibinfo {year} {1998})\BibitemShut
  {NoStop}%
\bibitem [{\citenamefont {Rowe}\ and\ \citenamefont {Wood}(2010)}]{RW10}%
  \BibitemOpen
  \bibfield  {author} {\bibinfo {author} {\bibfnamefont {D.~J.}\ \bibnamefont
  {Rowe}}\ and\ \bibinfo {author} {\bibfnamefont {J.~L.}\ \bibnamefont
  {Wood}},\ }\href@noop {} {\emph {\bibinfo {title} {{Fundamentals of Nuclear
  Models: Foundational Models}}}}\ (\bibinfo  {publisher} {World Scientific
  Publ., Singapore},\ \bibinfo {year} {2010})\BibitemShut {NoStop}%
\bibitem [{\citenamefont {Listerud}(1987)}]{Listerud_thesis}%
  \BibitemOpen
  \bibfield  {author} {\bibinfo {author} {\bibfnamefont {J.}~\bibnamefont
  {Listerud}},\ }\emph {\bibinfo {title} {{Techniques in Solid State NMR}}},\
  \href@noop {} {Ph.D. thesis},\ \bibinfo  {school} {University of Washington}
  (\bibinfo {year} {1987})\BibitemShut {NoStop}%
\bibitem [{\citenamefont {Listerud}, \citenamefont {Glaser},\ and\
  \citenamefont {Drobny}(1993)}]{LGD93}%
  \BibitemOpen
  \bibfield  {author} {\bibinfo {author} {\bibfnamefont {J.}~\bibnamefont
  {Listerud}}, \bibinfo {author} {\bibfnamefont {S.~J.}\ \bibnamefont
  {Glaser}}, \ and\ \bibinfo {author} {\bibfnamefont {G.~P.}\ \bibnamefont
  {Drobny}},\ }\bibfield  {title} {\enquote {\bibinfo {title} {{Symmetry and
  Isotropic Coherence Transfer. II. Three Spin Calculations Using a Young
  Tableaux Formulation}},}\ }\href@noop {} {\bibfield  {journal} {\bibinfo
  {journal} {Mol. Phys.}\ }\textbf {\bibinfo {volume} {78}},\ \bibinfo {pages}
  {629--658} (\bibinfo {year} {1993})}\BibitemShut {NoStop}%
\bibitem [{\citenamefont {Biedenharn}\ and\ \citenamefont
  {Louck}(1981)}]{BL81}%
  \BibitemOpen
  \bibfield  {author} {\bibinfo {author} {\bibfnamefont {L.~C.}\ \bibnamefont
  {Biedenharn}}\ and\ \bibinfo {author} {\bibfnamefont {J.~D.}\ \bibnamefont
  {Louck}},\ }\href@noop {} {\emph {\bibinfo {title} {{Angular Momentum in
  Quantum Physics}}}}\ (\bibinfo  {publisher} {Addison-Wesley, Reading, MA},\
  \bibinfo {year} {1981})\BibitemShut {NoStop}%
\bibitem [{Note1()}]{Note1}%
  \BibitemOpen
  \bibinfo {note} {Spherical harmonics $Y_{jm}(\theta ,\phi )=r(\theta ,\phi )
  \protect \qopname \relax o{exp}[i\eta (\theta ,\phi )]$ (and droplet
  functions) are plotted throughout this work by mapping their spherical
  coordinates $\theta $ and $\phi $ to the radial part $r(\theta ,\phi )$ and
  phase $\eta (\theta ,\phi )$.}\BibitemShut {Stop}%
\bibitem [{\citenamefont {Judd}\ \emph {et~al.}(1974)\citenamefont {Judd},
  \citenamefont {Miller}, \citenamefont {Patera},\ and\ \citenamefont
  {Winternitz}}]{JMPW74}%
  \BibitemOpen
  \bibfield  {author} {\bibinfo {author} {\bibfnamefont {B.~R.}\ \bibnamefont
  {Judd}}, \bibinfo {author} {\bibfnamefont {W.}~\bibnamefont {Miller},
  \bibfnamefont {Jr.}}, \bibinfo {author} {\bibfnamefont {J.}~\bibnamefont
  {Patera}}, \ and\ \bibinfo {author} {\bibfnamefont {P.}~\bibnamefont
  {Winternitz}},\ }\bibfield  {title} {\enquote {\bibinfo {title} {{Complete
  sets of commuting operators and $\mathrm{O}(3)$ scalars in the enveloping
  algebra of $\mathrm{SU}(3)$}},}\ }\href@noop {} {\bibfield  {journal}
  {\bibinfo  {journal} {J. Math. Phys.}\ }\textbf {\bibinfo {volume} {15}},\
  \bibinfo {pages} {1787--1799} (\bibinfo {year} {1974})}\BibitemShut {NoStop}%
\bibitem [{\citenamefont {Sharp}(1975)}]{Sharp75}%
  \BibitemOpen
  \bibfield  {author} {\bibinfo {author} {\bibfnamefont {R.~T.}\ \bibnamefont
  {Sharp}},\ }\bibfield  {title} {\enquote {\bibinfo {title}
  {{Internal-labeling operators}},}\ }\href@noop {} {\bibfield  {journal}
  {\bibinfo  {journal} {J. Math. Phys.}\ }\textbf {\bibinfo {volume} {16}},\
  \bibinfo {pages} {2050--2053} (\bibinfo {year} {1975})}\BibitemShut {NoStop}%
\bibitem [{\citenamefont {Iachello}\ and\ \citenamefont {Levine}(1995)}]{IL95}%
  \BibitemOpen
  \bibfield  {author} {\bibinfo {author} {\bibfnamefont {F.}~\bibnamefont
  {Iachello}}\ and\ \bibinfo {author} {\bibfnamefont {R.~D.}\ \bibnamefont
  {Levine}},\ }\href@noop {} {\emph {\bibinfo {title} {{Algebraic Theory of
  Molecules}}}}\ (\bibinfo  {publisher} {Oxford University Press, New York},\
  \bibinfo {year} {1995})\BibitemShut {NoStop}%
\bibitem [{\citenamefont {Merzbacher}(1998)}]{Merzbacher98}%
  \BibitemOpen
  \bibfield  {author} {\bibinfo {author} {\bibfnamefont {E.}~\bibnamefont
  {Merzbacher}},\ }\href@noop {} {\emph {\bibinfo {title} {{Quantum
  Mechanics}}}},\ \bibinfo {edition} {3rd}\ ed.\ (\bibinfo  {publisher} {John
  Wiley \& Sons, New York},\ \bibinfo {year} {1998})\BibitemShut {NoStop}%
\bibitem [{\citenamefont {Pauncz}(1995)}]{Pauncz95}%
  \BibitemOpen
  \bibfield  {author} {\bibinfo {author} {\bibfnamefont {R.}~\bibnamefont
  {Pauncz}},\ }\href@noop {} {\emph {\bibinfo {title} {{The Symmetric Group in
  Quantum Chemistry}}}}\ (\bibinfo  {publisher} {CRC Press, Boca Raton},\
  \bibinfo {year} {1995})\BibitemShut {NoStop}%
\bibitem [{\citenamefont {Feenberg}\ and\ \citenamefont
  {Phillips}(1937)}]{FP37}%
  \BibitemOpen
  \bibfield  {author} {\bibinfo {author} {\bibfnamefont {E.}~\bibnamefont
  {Feenberg}}\ and\ \bibinfo {author} {\bibfnamefont {M.}~\bibnamefont
  {Phillips}},\ }\bibfield  {title} {\enquote {\bibinfo {title} {{On the
  Structure of Light Nuclei}},}\ }\href@noop {} {\bibfield  {journal} {\bibinfo
   {journal} {Phys. Rev.}\ }\textbf {\bibinfo {volume} {51}},\ \bibinfo {pages}
  {597--608} (\bibinfo {year} {1937})}\BibitemShut {NoStop}%
\bibitem [{\citenamefont {Zare}(1988)}]{Zare88}%
  \BibitemOpen
  \bibfield  {author} {\bibinfo {author} {\bibfnamefont {R.~N.}\ \bibnamefont
  {Zare}},\ }\href@noop {} {\emph {\bibinfo {title} {{Angular Momentum}}}}\
  (\bibinfo  {publisher} {John Wiley \& Sons, New York},\ \bibinfo {year}
  {1988})\BibitemShut {NoStop}%
\bibitem [{\citenamefont {{Beringer, J.\ et al.\ (Particle Data
  Group)}}(2012)}]{PDG12}%
  \BibitemOpen
  \bibfield  {author} {\bibinfo {author} {\bibnamefont {{Beringer, J.\ et al.\
  (Particle Data Group)}}},\ }\bibfield  {title} {\enquote {\bibinfo {title}
  {{The Review of Particle Physics}},}\ }\href@noop {} {\bibfield  {journal}
  {\bibinfo  {journal} {Phys. Rev. D}\ }\textbf {\bibinfo {volume} {86}},\
  \bibinfo {pages} {010001} (\bibinfo {year} {2012})}\BibitemShut {NoStop}%
\bibitem [{\citenamefont {Sanctuary}\ and\ \citenamefont
  {Temme}(1985)}]{SancTemMNMRXIII}%
  \BibitemOpen
  \bibfield  {author} {\bibinfo {author} {\bibfnamefont {B.~C.}\ \bibnamefont
  {Sanctuary}}\ and\ \bibinfo {author} {\bibfnamefont {F.~P.}\ \bibnamefont
  {Temme}},\ }\bibfield  {title} {\enquote {\bibinfo {title} {{Multipole N.M.R.
  XIII. multispin interactions and symmetry in Liouville space}},}\ }\href@noop
  {} {\bibfield  {journal} {\bibinfo  {journal} {Mol. Phys.}\ }\textbf
  {\bibinfo {volume} {55}},\ \bibinfo {pages} {1049--1062} (\bibinfo {year}
  {1985})}\BibitemShut {NoStop}%
\bibitem [{\citenamefont {Fano}(1953)}]{Fano53}%
  \BibitemOpen
  \bibfield  {author} {\bibinfo {author} {\bibfnamefont {U.}~\bibnamefont
  {Fano}},\ }\bibfield  {title} {\enquote {\bibinfo {title} {Geometrical
  characterization of nuclear states and the theory of angular correlations},}\
  }\href@noop {} {\bibfield  {journal} {\bibinfo  {journal} {Phys. Rev.}\
  }\textbf {\bibinfo {volume} {90}},\ \bibinfo {pages} {577--579} (\bibinfo
  {year} {1953})}\BibitemShut {NoStop}%
\bibitem [{\citenamefont {Messiah}(1962)}]{MesII:1962}%
  \BibitemOpen
  \bibfield  {author} {\bibinfo {author} {\bibfnamefont {A.}~\bibnamefont
  {Messiah}},\ }\href@noop {} {\emph {\bibinfo {title} {Quantum mechanics}}},\
  Vol.~\bibinfo {volume} {II}\ (\bibinfo  {publisher} {North-Holland},\
  \bibinfo {address} {Amsterdam},\ \bibinfo {year} {1962})\BibitemShut
  {NoStop}%
\bibitem [{\citenamefont {Landau}\ and\ \citenamefont
  {Lifshitz}(1977)}]{Landau_Lifshitz:1977}%
  \BibitemOpen
  \bibfield  {author} {\bibinfo {author} {\bibfnamefont {L.~D.}\ \bibnamefont
  {Landau}}\ and\ \bibinfo {author} {\bibfnamefont {E.~M.}\ \bibnamefont
  {Lifshitz}},\ }\href@noop {} {\emph {\bibinfo {title} {Quantum mechanics}}},\
  Vol.~\bibinfo {volume} {3}\ (\bibinfo  {publisher} {Pergamon Press},\
  \bibinfo {address} {Oxford},\ \bibinfo {year} {1977})\BibitemShut {NoStop}%
\bibitem [{\citenamefont {D{\"u}r}, \citenamefont {Vidal},\ and\ \citenamefont
  {Cirac}(2000)}]{DVC00}%
  \BibitemOpen
  \bibfield  {author} {\bibinfo {author} {\bibfnamefont {W.}~\bibnamefont
  {D{\"u}r}}, \bibinfo {author} {\bibfnamefont {G.}~\bibnamefont {Vidal}}, \
  and\ \bibinfo {author} {\bibfnamefont {J.~I.}\ \bibnamefont {Cirac}},\
  }\bibfield  {title} {\enquote {\bibinfo {title} {{Three qubits can be
  entangled in two inequivalent ways}},}\ }\href@noop {} {\bibfield  {journal}
  {\bibinfo  {journal} {Phys. Rev. A}\ }\textbf {\bibinfo {volume} {62}},\
  \bibinfo {pages} {062314} (\bibinfo {year} {2000})}\BibitemShut {NoStop}%
\bibitem [{\citenamefont {Briegel}\ and\ \citenamefont
  {Raussendorf}(2001)}]{BR01}%
  \BibitemOpen
  \bibfield  {author} {\bibinfo {author} {\bibfnamefont {H.~J.}\ \bibnamefont
  {Briegel}}\ and\ \bibinfo {author} {\bibfnamefont {R.}~\bibnamefont
  {Raussendorf}},\ }\bibfield  {title} {\enquote {\bibinfo {title} {{Persistent
  Entanglement in Arrays of Interacting Particles}},}\ }\href@noop {}
  {\bibfield  {journal} {\bibinfo  {journal} {Phys. Rev. Lett.}\ }\textbf
  {\bibinfo {volume} {86}},\ \bibinfo {pages} {910--913} (\bibinfo {year}
  {2001})}\BibitemShut {NoStop}%
\bibitem [{\citenamefont {Verstraete}\ \emph {et~al.}(2002)\citenamefont
  {Verstraete}, \citenamefont {Dehaene}, \citenamefont {De~Moor},\ and\
  \citenamefont {Verschelde}}]{PhysRevA.65.052112}%
  \BibitemOpen
  \bibfield  {author} {\bibinfo {author} {\bibfnamefont {F.}~\bibnamefont
  {Verstraete}}, \bibinfo {author} {\bibfnamefont {J.}~\bibnamefont {Dehaene}},
  \bibinfo {author} {\bibfnamefont {B.}~\bibnamefont {De~Moor}}, \ and\
  \bibinfo {author} {\bibfnamefont {H.}~\bibnamefont {Verschelde}},\ }\bibfield
   {title} {\enquote {\bibinfo {title} {Four qubits can be entangled in nine
  different ways},}\ }\href@noop {} {\bibfield  {journal} {\bibinfo  {journal}
  {Phys. Rev. A}\ }\textbf {\bibinfo {volume} {65}},\ \bibinfo {pages} {052112}
  (\bibinfo {year} {2002})}\BibitemShut {NoStop}%
\bibitem [{\citenamefont {Dicke}(1954)}]{Dicke1954}%
  \BibitemOpen
  \bibfield  {author} {\bibinfo {author} {\bibfnamefont {R.~H.}\ \bibnamefont
  {Dicke}},\ }\bibfield  {title} {\enquote {\bibinfo {title} {{Coherence in
  spontaneous radiation processes}},}\ }\href@noop {} {\bibfield  {journal}
  {\bibinfo  {journal} {Phys. Rev.}\ }\textbf {\bibinfo {volume} {93}},\
  \bibinfo {pages} {99--110} (\bibinfo {year} {1954})}\BibitemShut {NoStop}%
\bibitem [{\citenamefont {Stockton}\ \emph {et~al.}(2003)\citenamefont
  {Stockton}, \citenamefont {Geremia}, \citenamefont {Doherty},\ and\
  \citenamefont {Mabuchi}}]{stockton2003}%
  \BibitemOpen
  \bibfield  {author} {\bibinfo {author} {\bibfnamefont {J.~K.}\ \bibnamefont
  {Stockton}}, \bibinfo {author} {\bibfnamefont {J.~M.}\ \bibnamefont
  {Geremia}}, \bibinfo {author} {\bibfnamefont {A.~C.}\ \bibnamefont
  {Doherty}}, \ and\ \bibinfo {author} {\bibfnamefont {H.}~\bibnamefont
  {Mabuchi}},\ }\bibfield  {title} {\enquote {\bibinfo {title} {{Characterizing
  the entanglement of symmetric many-particle spin-$1/2$ systems}},}\
  }\href@noop {} {\bibfield  {journal} {\bibinfo  {journal} {Phys. Rev. A}\
  }\textbf {\bibinfo {volume} {67}},\ \bibinfo {pages} {022112} (\bibinfo
  {year} {2003})}\BibitemShut {NoStop}%
\bibitem [{\citenamefont {K\"{o}cher}\ \emph {et~al.}(2016)\citenamefont
  {K\"{o}cher}, \citenamefont {Heydenreich}, \citenamefont {Zhang},
  \citenamefont {Reddy}, \citenamefont {Caldarelli}, \citenamefont {Yuan},\
  and\ \citenamefont {Glaser}}]{maxq}%
  \BibitemOpen
  \bibfield  {author} {\bibinfo {author} {\bibfnamefont {S.}~\bibnamefont
  {K\"{o}cher}}, \bibinfo {author} {\bibfnamefont {T.}~\bibnamefont
  {Heydenreich}}, \bibinfo {author} {\bibfnamefont {Y.}~\bibnamefont {Zhang}},
  \bibinfo {author} {\bibfnamefont {G.~N.}\ \bibnamefont {Reddy}}, \bibinfo
  {author} {\bibfnamefont {S.}~\bibnamefont {Caldarelli}}, \bibinfo {author}
  {\bibfnamefont {H.}~\bibnamefont {Yuan}}, \ and\ \bibinfo {author}
  {\bibfnamefont {S.~J.}\ \bibnamefont {Glaser}},\ }\bibfield  {title}
  {\enquote {\bibinfo {title} {Time-optimal excitation of maximum quantum
  coherence: Physical limits and pulse sequences},}\ }\href@noop {} {\bibfield
  {journal} {\bibinfo  {journal} {J. Chem. Phys.}\ }\textbf {\bibinfo {volume}
  {144}},\ \bibinfo {pages} {164103} (\bibinfo {year} {2016})}\BibitemShut
  {NoStop}%
\bibitem [{Note2()}]{Note2}%
  \BibitemOpen
  \bibinfo {note} {Hermitian operators lead to positive and negative values,
  which are shown in red (dark gray) and green (light gray).}\BibitemShut
  {Stop}%
\bibitem [{\citenamefont {Glaser}\ \emph {et~al.}(2015)\citenamefont {Glaser},
  \citenamefont {Boscain}, \citenamefont {Calarco}, \citenamefont {Koch},
  \citenamefont {K{\"o}ckenberger}, \citenamefont {Kosloff}, \citenamefont
  {Kuprov}, \citenamefont {Luy}, \citenamefont {Schirmer}, \citenamefont
  {Schulte-Herb{\"u}ggen}, \citenamefont {Sugny},\ and\ \citenamefont
  {Wilhelm}}]{Roadmap2015}%
  \BibitemOpen
  \bibfield  {author} {\bibinfo {author} {\bibfnamefont {S.~J.}\ \bibnamefont
  {Glaser}}, \bibinfo {author} {\bibfnamefont {U.}~\bibnamefont {Boscain}},
  \bibinfo {author} {\bibfnamefont {T.}~\bibnamefont {Calarco}}, \bibinfo
  {author} {\bibfnamefont {C.~P.}\ \bibnamefont {Koch}}, \bibinfo {author}
  {\bibfnamefont {W.}~\bibnamefont {K{\"o}ckenberger}}, \bibinfo {author}
  {\bibfnamefont {R.}~\bibnamefont {Kosloff}}, \bibinfo {author} {\bibfnamefont
  {I.}~\bibnamefont {Kuprov}}, \bibinfo {author} {\bibfnamefont
  {B.}~\bibnamefont {Luy}}, \bibinfo {author} {\bibfnamefont {S.}~\bibnamefont
  {Schirmer}}, \bibinfo {author} {\bibfnamefont {T.}~\bibnamefont
  {Schulte-Herb{\"u}ggen}}, \bibinfo {author} {\bibfnamefont {D.}~\bibnamefont
  {Sugny}}, \ and\ \bibinfo {author} {\bibfnamefont {F.~K.}\ \bibnamefont
  {Wilhelm}},\ }\bibfield  {title} {\enquote {\bibinfo {title} {{Training
  Schr{\"o}dinger's Cat: Quantum Optimal Control}},}\ }\href@noop {} {\bibfield
   {journal} {\bibinfo  {journal} {Eur. Phys. J. D}\ }\textbf {\bibinfo
  {volume} {69}},\ \bibinfo {pages} {279} (\bibinfo {year} {2015})}\BibitemShut
  {NoStop}%
\bibitem [{\citenamefont {Glaser}, \citenamefont {Tesch},\ and\ \citenamefont
  {Glaser}(2015)}]{ipad_app}%
  \BibitemOpen
  \bibfield  {author} {\bibinfo {author} {\bibfnamefont {N.~J.}\ \bibnamefont
  {Glaser}}, \bibinfo {author} {\bibfnamefont {M.}~\bibnamefont {Tesch}}, \
  and\ \bibinfo {author} {\bibfnamefont {S.~J.}\ \bibnamefont {Glaser}},\
  }\href@noop {} {\enquote {\bibinfo {title} {{SpinDrops (Version 1.2.2)
  [Mobile application]}},}\ } (\bibinfo {year} {2015}),\ \bibinfo {note}
  {\href{http://itunes.apple.com}{itunes.apple.com}}\BibitemShut {NoStop}%
\bibitem [{\citenamefont {Tesch}, \citenamefont {Glaser},\ and\ \citenamefont
  {Glaser}(2018)}]{app}%
  \BibitemOpen
  \bibfield  {author} {\bibinfo {author} {\bibfnamefont {M.}~\bibnamefont
  {Tesch}}, \bibinfo {author} {\bibfnamefont {N.~J.}\ \bibnamefont {Glaser}}, \
  and\ \bibinfo {author} {\bibfnamefont {S.~J.}\ \bibnamefont {Glaser}},\
  }\href@noop {} {\enquote {\bibinfo {title} {Spindrops 2.0},}\ } (\bibinfo
  {year} {2018}),\ \bibinfo {note}
  {\href{https://spindrops.org}{https://spindrops.org}}\BibitemShut {NoStop}%
\bibitem [{\citenamefont {Luy}\ and\ \citenamefont {Glaser}(2000)}]{Luy2000}%
  \BibitemOpen
  \bibfield  {author} {\bibinfo {author} {\bibfnamefont {B.}~\bibnamefont
  {Luy}}\ and\ \bibinfo {author} {\bibfnamefont {S.~J.}\ \bibnamefont
  {Glaser}},\ }\bibfield  {title} {\enquote {\bibinfo {title} {{Negative
  polarization transfer between a spin 1/2 and a spin 1}},}\ }\href@noop {}
  {\bibfield  {journal} {\bibinfo  {journal} {Chem. Phys. Lett.}\ }\textbf
  {\bibinfo {volume} {323}},\ \bibinfo {pages} {377--381} (\bibinfo {year}
  {2000})}\BibitemShut {NoStop}%
\bibitem [{\citenamefont {Horn}\ and\ \citenamefont {Johnson}(1991)}]{HJ2}%
  \BibitemOpen
  \bibfield  {author} {\bibinfo {author} {\bibfnamefont {R.~A.}\ \bibnamefont
  {Horn}}\ and\ \bibinfo {author} {\bibfnamefont {C.~R.}\ \bibnamefont
  {Johnson}},\ }\href@noop {} {\emph {\bibinfo {title} {{Topics in Matrix
  Analysis}}}}\ (\bibinfo  {publisher} {Cambridge University Press,
  Cambridge},\ \bibinfo {year} {1991})\BibitemShut {NoStop}%
\bibitem [{\citenamefont {Henderson}\ and\ \citenamefont
  {Searle}(1981)}]{HS81}%
  \BibitemOpen
  \bibfield  {author} {\bibinfo {author} {\bibfnamefont {H.~V.}\ \bibnamefont
  {Henderson}}\ and\ \bibinfo {author} {\bibfnamefont {S.~R.}\ \bibnamefont
  {Searle}},\ }\bibfield  {title} {\enquote {\bibinfo {title} {{The
  Vec-Permutation Matrix, the Vec Operator, and Kronecker Products: A
  Review}},}\ }\href@noop {} {\bibfield  {journal} {\bibinfo  {journal} {Lin.
  Multilin. Alg.}\ }\textbf {\bibinfo {volume} {9}},\ \bibinfo {pages}
  {271--288} (\bibinfo {year} {1981})}\BibitemShut {NoStop}%
\bibitem [{\citenamefont {Fuhrmann}(2012)}]{Fuhrmann12}%
  \BibitemOpen
  \bibfield  {author} {\bibinfo {author} {\bibfnamefont {P.~A.}\ \bibnamefont
  {Fuhrmann}},\ }\href@noop {} {\emph {\bibinfo {title} {{A Polynomial Approach
  to Linear Algebra}}}},\ \bibinfo {edition} {2nd}\ ed.\ (\bibinfo  {publisher}
  {Springer, New York},\ \bibinfo {year} {2012})\BibitemShut {NoStop}%
\bibitem [{\citenamefont {Huizenga}()}]{MO115067}%
  \BibitemOpen
  \bibfield  {author} {\bibinfo {author} {\bibfnamefont {J.}~\bibnamefont
  {Huizenga}},\ }\href@noop {} {\enquote {\bibinfo {title} {Necessary and
  sufficient conditions for a sum of idempotents to be idempotent},}\ }\bibinfo
  {howpublished} {MathOverflow},\ \bibinfo {note}
  {\url{https://mathoverflow.net/q/115255} (2012-12-03)}\BibitemShut {NoStop}%
\bibitem [{\citenamefont {Simon}(1995)}]{Simon1995}%
  \BibitemOpen
  \bibfield  {author} {\bibinfo {author} {\bibfnamefont {B.}~\bibnamefont
  {Simon}},\ }\href@noop {} {\emph {\bibinfo {title} {Representations of Finite
  and Compact Groups}}}\ (\bibinfo  {publisher} {Oxford University Press},\
  \bibinfo {address} {Oxford},\ \bibinfo {year} {1995})\BibitemShut {NoStop}%
\bibitem [{\citenamefont {Keppeler}\ and\ \citenamefont
  {Sj{\"o}dahl}(2014)}]{Keppeler2014}%
  \BibitemOpen
  \bibfield  {author} {\bibinfo {author} {\bibfnamefont {S.}~\bibnamefont
  {Keppeler}}\ and\ \bibinfo {author} {\bibfnamefont {M.}~\bibnamefont
  {Sj{\"o}dahl}},\ }\bibfield  {title} {\enquote {\bibinfo {title} {{Hermitian
  Young operators}},}\ }\href@noop {} {\bibfield  {journal} {\bibinfo
  {journal} {J. Math. Phys.}\ }\textbf {\bibinfo {volume} {55}},\ \bibinfo
  {pages} {021702} (\bibinfo {year} {2014})}\BibitemShut {NoStop}%
\bibitem [{\citenamefont {Alcock-Zeilinger}\ and\ \citenamefont
  {Weigert}(2017{\natexlab{a}})}]{Alcock2017b}%
  \BibitemOpen
  \bibfield  {author} {\bibinfo {author} {\bibfnamefont {J.}~\bibnamefont
  {Alcock-Zeilinger}}\ and\ \bibinfo {author} {\bibfnamefont {H.}~\bibnamefont
  {Weigert}},\ }\bibfield  {title} {\enquote {\bibinfo {title} {Compact
  {H}ermitian {Y}oung projection operators},}\ }\href@noop {} {\bibfield
  {journal} {\bibinfo  {journal} {J. Math. Phys.}\ }\textbf {\bibinfo {volume}
  {58}},\ \bibinfo {pages} {051702} (\bibinfo {year}
  {2017}{\natexlab{a}})}\BibitemShut {NoStop}%
\bibitem [{\citenamefont {Alcock-Zeilinger}\ and\ \citenamefont
  {Weigert}(2017{\natexlab{b}})}]{Alcock2017c}%
  \BibitemOpen
  \bibfield  {author} {\bibinfo {author} {\bibfnamefont {J.}~\bibnamefont
  {Alcock-Zeilinger}}\ and\ \bibinfo {author} {\bibfnamefont {H.}~\bibnamefont
  {Weigert}},\ }\bibfield  {title} {\enquote {\bibinfo {title} {Transition
  operators},}\ }\href@noop {} {\bibfield  {journal} {\bibinfo  {journal} {J.
  Math. Phys.}\ }\textbf {\bibinfo {volume} {58}},\ \bibinfo {pages} {051703}
  (\bibinfo {year} {2017}{\natexlab{b}})}\BibitemShut {NoStop}%
\bibitem [{\citenamefont {Stembridge}(2011)}]{Stembridge11}%
  \BibitemOpen
  \bibfield  {author} {\bibinfo {author} {\bibfnamefont {J.~R.}\ \bibnamefont
  {Stembridge}},\ }\bibfield  {title} {\enquote {\bibinfo {title} {{Orthogonal
  sets of Young symmetrizers}},}\ }\href@noop {} {\bibfield  {journal}
  {\bibinfo  {journal} {Adv. Appl. Math.}\ }\textbf {\bibinfo {volume} {46}},\
  \bibinfo {pages} {576--582} (\bibinfo {year} {2011})}\BibitemShut {NoStop}%
\bibitem [{\citenamefont {Acin}\ \emph {et~al.}(2018)\citenamefont {Acin},
  \citenamefont {Bloch}, \citenamefont {Buhrman}, \citenamefont {Calarco},
  \citenamefont {Eichler}, \citenamefont {Eisert}, \citenamefont {Esteve},
  \citenamefont {Gisin}, \citenamefont {Glaser}, \citenamefont {Jelezko},
  \citenamefont {Kuhr}, \citenamefont {Lewenstein}, \citenamefont {Riedel},
  \citenamefont {Schmidt}, \citenamefont {Thew}, \citenamefont {Wallraff},
  \citenamefont {Walmsley},\ and\ \citenamefont {Wilhelm}}]{Acin2018}%
  \BibitemOpen
  \bibfield  {author} {\bibinfo {author} {\bibfnamefont {A.}~\bibnamefont
  {Acin}}, \bibinfo {author} {\bibfnamefont {I.}~\bibnamefont {Bloch}},
  \bibinfo {author} {\bibfnamefont {H.}~\bibnamefont {Buhrman}}, \bibinfo
  {author} {\bibfnamefont {T.}~\bibnamefont {Calarco}}, \bibinfo {author}
  {\bibfnamefont {C.}~\bibnamefont {Eichler}}, \bibinfo {author} {\bibfnamefont
  {J.}~\bibnamefont {Eisert}}, \bibinfo {author} {\bibfnamefont
  {D.}~\bibnamefont {Esteve}}, \bibinfo {author} {\bibfnamefont
  {N.}~\bibnamefont {Gisin}}, \bibinfo {author} {\bibfnamefont {S.~J.}\
  \bibnamefont {Glaser}}, \bibinfo {author} {\bibfnamefont {F.}~\bibnamefont
  {Jelezko}}, \bibinfo {author} {\bibfnamefont {S.}~\bibnamefont {Kuhr}},
  \bibinfo {author} {\bibfnamefont {M.}~\bibnamefont {Lewenstein}}, \bibinfo
  {author} {\bibfnamefont {M.~F.}\ \bibnamefont {Riedel}}, \bibinfo {author}
  {\bibfnamefont {P.~O.}\ \bibnamefont {Schmidt}}, \bibinfo {author}
  {\bibfnamefont {R.}~\bibnamefont {Thew}}, \bibinfo {author} {\bibfnamefont
  {A.}~\bibnamefont {Wallraff}}, \bibinfo {author} {\bibfnamefont
  {I.}~\bibnamefont {Walmsley}}, \ and\ \bibinfo {author} {\bibfnamefont
  {F.~K.}\ \bibnamefont {Wilhelm}},\ }\bibfield  {title} {\enquote {\bibinfo
  {title} {The quantum technologies roadmap: a {European} community view},}\
  }\href@noop {} {\bibfield  {journal} {\bibinfo  {journal} {New J. Phys.}\
  }\textbf {\bibinfo {volume} {20}},\ \bibinfo {pages} {080201} (\bibinfo
  {year} {2018})}\BibitemShut {NoStop}%
\bibitem [{\citenamefont {Bosma}, \citenamefont {Cannon},\ and\ \citenamefont
  {Playoust}(1997)}]{MAGMA}%
  \BibitemOpen
  \bibfield  {author} {\bibinfo {author} {\bibfnamefont {W.}~\bibnamefont
  {Bosma}}, \bibinfo {author} {\bibfnamefont {J.~J.}\ \bibnamefont {Cannon}}, \
  and\ \bibinfo {author} {\bibfnamefont {C.}~\bibnamefont {Playoust}},\
  }\bibfield  {title} {\enquote {\bibinfo {title} {{The {\sf MAGMA} Algebra
  System~I: The User Language}},}\ }\href@noop {} {\bibfield  {journal}
  {\bibinfo  {journal} {J. Symbolic Comput.}\ }\textbf {\bibinfo {volume}
  {24}},\ \bibinfo {pages} {235--265} (\bibinfo {year} {1997})}\BibitemShut
  {NoStop}%
\bibitem [{\citenamefont {{The MathWorks Inc.}}(2017)}]{matlab}%
  \BibitemOpen
  \bibfield  {author} {\bibinfo {author} {\bibnamefont {{The MathWorks
  Inc.}}},\ }\href@noop {} {\emph {\bibinfo {title} {MATLAB, version 9.2
  (R2017a)}}},\ \bibinfo {address} {Natick, Massachusetts} (\bibinfo {year}
  {2017})\BibitemShut {NoStop}%
\bibitem [{\citenamefont {{The Sage Developers}}(2018)}]{sagemath}%
  \BibitemOpen
  \bibfield  {author} {\bibinfo {author} {\bibnamefont {{The Sage
  Developers}}},\ }\href@noop {} {\emph {\bibinfo {title} {{S}ageMath, the
  {S}age {M}athematics {S}oftware {S}ystem ({V}ersion 8.3)}}} (\bibinfo {year}
  {2018}),\ \bibinfo {note} {\url{http://www.sagemath.org}}\BibitemShut
  {NoStop}%
\bibitem [{\citenamefont {Khaneja}\ and\ \citenamefont {Glaser}(2002)}]{KG02}%
  \BibitemOpen
  \bibfield  {author} {\bibinfo {author} {\bibfnamefont {N.}~\bibnamefont
  {Khaneja}}\ and\ \bibinfo {author} {\bibfnamefont {S.~J.}\ \bibnamefont
  {Glaser}},\ }\bibfield  {title} {\enquote {\bibinfo {title} {{Efficient
  Transfer of Coherence through Ising Spin Chains}},}\ }\href@noop {}
  {\bibfield  {journal} {\bibinfo  {journal} {Phys. Rev. A}\ }\textbf {\bibinfo
  {volume} {66}},\ \bibinfo {pages} {060301(R)} (\bibinfo {year}
  {2002})}\BibitemShut {NoStop}%
\bibitem [{\citenamefont {Luy}, \citenamefont {Schedletzky},\ and\
  \citenamefont {Glaser}(1999)}]{Luy20001999}%
  \BibitemOpen
  \bibfield  {author} {\bibinfo {author} {\bibfnamefont {B.}~\bibnamefont
  {Luy}}, \bibinfo {author} {\bibfnamefont {O.}~\bibnamefont {Schedletzky}}, \
  and\ \bibinfo {author} {\bibfnamefont {S.~J.}\ \bibnamefont {Glaser}},\
  }\bibfield  {title} {\enquote {\bibinfo {title} {{Analytical polarization
  transfer functions for four coupled spins 1/2 under isotropic mixing
  conditions}},}\ }\href@noop {} {\bibfield  {journal} {\bibinfo  {journal} {J.
  Magn. Reson.}\ }\textbf {\bibinfo {volume} {138}},\ \bibinfo {pages} {19--27}
  (\bibinfo {year} {1999})}\BibitemShut {NoStop}%
\bibitem [{\citenamefont {Elliott}, \citenamefont {Hope},\ and\ \citenamefont
  {Jahn}(1953)}]{EHJ53}%
  \BibitemOpen
  \bibfield  {author} {\bibinfo {author} {\bibfnamefont {J.~P.}\ \bibnamefont
  {Elliott}}, \bibinfo {author} {\bibfnamefont {J.}~\bibnamefont {Hope}}, \
  and\ \bibinfo {author} {\bibfnamefont {H.~A.}\ \bibnamefont {Jahn}},\
  }\bibfield  {title} {\enquote {\bibinfo {title} {{Theoretical studies in
  nuclear structure IV. Wave functions for the nuclear $p$-shell Part B.
  $\langle p^n | p^{n{-}2}p^2 \rangle$ fractional parentage coefficients}},}\
  }\href@noop {} {\bibfield  {journal} {\bibinfo  {journal} {Philos. Trans.
  Royal Soc. A}\ }\textbf {\bibinfo {volume} {246}},\ \bibinfo {pages}
  {241--279} (\bibinfo {year} {1953})}\BibitemShut {NoStop}%
\end{thebibliography}

%

\end{document}